\documentclass[12pt]{article}
\pdfoutput=1
\usepackage[utf8]{inputenc}
\usepackage{amsmath,amssymb,graphicx,epsfig,color,float,cancel,cite}
\usepackage[titletoc,title]{appendix} 
\usepackage[export]{adjustbox}
\usepackage{multicol,tabu,multirow,bm}
\usepackage{hyperref}
\usepackage{slashed}
                                              
\numberwithin{equation}{section}

\def\nn{\nonumber}

\def\nn{\nonumber}
\def\bea{\begin{eqnarray}}
\def\eea{\end{eqnarray}}
\def\hedt{\bar c \, \sigma^{\mu\nu} \,  b}
\def\bal#1\eal{\begin{align}#1\end{align}}

\def\Jps{{J/\psi}}

\newcommand{\beq}{\begin{equation}}
\newcommand{\eeq}{\end{equation}}
\newcommand{\bseq}{\begin{subequations}}
\newcommand{\eseq}{\end{subequations}}

\newcommand{\del}{\partial}

\def\vn#1#2{\textcolor{black}{(V_{#1}^\nu)_{#2'#2}}}
\def\vnd#1#2{\textcolor{black}{(V_{#1}^\nu)^\dagger_{#2 #2'}}}
\def\ve#1#2{\textcolor{black}{(V_{#1}^e)_{#2'#2}}}
\def\ved#1#2{\textcolor{black}{(V_{#1}^e)^\dagger_{#2 #2'}}}
\def\vu#1#2{\textcolor{black}{(V_{#1}^u)_{#2'#2}}}
\def\vud#1#2{\textcolor{black}{(V_{#1}^u)^\dagger_{#2 #2'}}}
\def\vd#1#2{\textcolor{black}{(V_{#1}^d)_{#2'#2}}}
\def\vdd#1#2{\textcolor{black}{(V_{#1}^d)^\dagger_{#2 #2'}}}
\def\clqT#1#2#3#4{ \textcolor{black}{[\tilde C_{lq}^{(3) #1 #2 #3 
#4}]_{prst}}}
\def\cledq#1#2#3#4{ \textcolor{black}{[\tilde C_{ledq}^{ #1 #2 #3 
#4}]_{prst}}}
\def\cledqO#1#2#3#4{ \textcolor{black}{[\tilde C_{lequ}^{ (1) #1 #2 #3 
#4}]_{prst}}}
\def\cledqT#1#2#3#4{ \textcolor{black}{[\tilde C_{lequ}^{ (3) #1 #2 #3 
#4}]_{prst}}}
\def\cplT#1#2{ \textcolor{black}{[\tilde C_{\phi l}^{ (3) #1 #2 }]_{pr}}}

\def\bra{\langle}
\def\ket{\rangle}
\def\PDSQ{|p_D|^2}
\def\PDd{|p_D|}
\def\pds{\left| p_{D^\ast}\right|^2}
\def\mds{M_{D^\ast}^2}

\def\CSLSq{|{\bf C_{\rm SL}^{\ell}|^2}}

\def\CVLSq{|{\bf C_{\rm VL}^{\ell}|^2}}

\def\CTLSq{|{\bf C_{\rm TL}^{\ell}|^2}}

\def\FzeroSQ{{\bf F_0^2}}
\def\FoneSQ{{\bf  F_+^2}}
\def\FtwoSQ{{\bf  F_T^2}}

\def\re{\mathcal{R}}
\def\cvl{C_{\rm VL}^\ell}
\def\call{C_{\rm AL}^\ell}
\def\calconj{C_{\rm AL}^{\ell\ast}}
\def\cpl{C_{\rm PL}^\ell}
\def\cplconj{C_{\rm PL}^{\ell\ast}}

\def\ctl{C_{\rm TL}^\ell}
\def\ctlconj{C_{\rm TL}^{\ell\ast}}
\def\ctlsq{\left|C_{\rm TL}^\ell\right|^2}
\def\cvlsq{\left|C_{\rm VL}^\ell\right|^2}
\def\calsq{\left|C_{\rm AL}^\ell\right|^2}
\def\cplsq{\left|C_{\rm PL}^\ell\right|^2}

\def\vsq{V^2}
\def\azsq{A_0^2}
\def\aosq{A_1^2}
\def\atsq{A_2^2}
\def\vsq{V^2}
\def\mbmd{\left(M_B + M_{D^\ast}\right)}
\def\mbmc{\left(m_b + m_c\right)}
\def\mbmdq2{\left(M_B^2 - M_{D^\ast}^2 - q^2\right)}
\def\mbmdsq{\left(M_B^2 - M_{D^\ast}^2 \right)}
\def\mbmdqsq{\left(M_B^2 - M_{D^\ast}^2 - q^2\right)}
\def\mbmdqthsq{\left(M_B^2 + 3 M_{D^\ast}^2 - q^2\right)}
\def\re{\mathcal{R}}
\def\nn{\nonumber}
\def\llx{\ell_{\hspace{-0.3mm}0}}
\def\dblone{\hbox{$1\hskip -1.2pt\vrule depth 0pt height 1.6ex
width 0.7pt 
\vrule depth 0pt height 0.3pt width 0.12em$}}
\def\slash#1{\mathord{\not\mathrel{{\mathrel{#1}}}}}
\newcommand{\be}{\begin{equation}}
\newcommand{\ee}{\end{equation}}
\newcommand{\bc}{\begin{center}}
\newcommand{\ec}{\end{center}}

\allowdisplaybreaks
\begin{document}
\thispagestyle{empty}
\hspace{13cm} SISSA  20/2018/FISI
\begin{center}
\vspace*{0.5cm}
{\LARGE\bf Anatomy of $b \to c \, \tau \, \nu$ anomalies \\}
\bigskip
{\large Aleksandr Azatov}\,$^{a,d,1}$, \, \, 
{\large Debjyoti Bardhan}\,$^{b,2}$, \, \, 
{\large Diptimoy Ghosh}\,$^{c,d,3}$, \, \, 
{\large Francesco Sgarlata}\,$^{a,d,4}$, \, \, 
{\large Elena Venturini}\,$^{a,d,5}$
\\
\bigskip 
\bigskip
{\small
$^a$ SISSA International School for Advanced Studies, Via Bonomea 265, 34136, Trieste, Italy \\ [2mm]
$^b$ Department of Physics, Ben-Gurion University, Beer-Sheva 8410501, Israel \\[2mm]
$^c$ International Centre for Theoretical Physics, Strada Costiera 11, 34014 Trieste, Italy \\[2mm]
$^d$ INFN Trieste, Via Bonomea 265, 34136, Trieste, Italy
}
\end{center}
\bigskip 
\vspace*{0cm}
\begin{center} 
{\Large\bf Abstract} 
\end{center}
\vspace*{-0.35in}
\begin{quotation}
\noindent 
In recent times, one of the strongest hints of Physics Beyond the Standard Model (BSM) has been the anomaly in the ratios
$R_D$ and $R_{D^\ast}$ measured in the charged current decays of $B$-mesons. In this work, we perform a comprehensive
analysis of these decay modes, first in a model independent way and subsequently, in the context of composite Higgs models.
We discuss in depth as to how linearly realised $\rm SU(2)_L \times U(1)_Y$ symmetry imposes severe constraint on the various
scenarios because of correlations with other $\Delta F =1$ processes and  $Z\,\tau\,\tau$ and $Z\,\nu\,\nu$ couplings. In the
composite Higgs paradigm with partial compositeness, we show that, irrespective of the flavour structure of the composite sector,
constraints from $\Delta F =2$ processes bring the compositeness scale down to $\sim 650$ GeV which is in tension with
electroweak precision observables. In the presence of composite leptoquarks, the situation improves only marginally (a factor of $\sqrt{2}$
in the compositeness scale), thus making the new states soon discoverable by direct searches at the LHC.
We also comment on the possible explanation of the $R_{K,K^*}$ anomalies within the composite Higgs framework.
\end{quotation}
\bigskip
%
%
\vfill
%
%
\bigskip
\hrule
\vspace*{-0.1in}
\hspace*{8mm}
\hspace{-1cm} $^1$	aleksandr.azatov@sissa.it  \hspace{1.2cm} $^2$ bardhan@post.bgu.ac.il  \hspace{1.2cm} $^3$ dghosh@ictp.it \\
\hspace*{2cm} $^4$ francesco.sgarlata@sissa.it  \hspace{3cm} $^5$ 	eventuri@sissa.it

\newpage 
\tableofcontents
\vspace*{-3mm}
\section{Introduction}

The Standard Model (SM) of particle physics has been remarkably successful in explaining almost all the measurements made till date in 
accelerator-based experiments, ranging from a few GeV in centre-of-mass energy to a few hundred GeV. 
However, deviations from the SM expectations approximately at the  $2\sigma - 4\sigma$ level have shown up in a number 
of recent measurements involving semi-leptonic $B$-meson decays, both in charged current and neutral current channels. 

In this work, we focus mainly on the analysis  of the charged current anomalies\footnote{While the statistical significance of these experimental results 
is not yet large enough to claim a discovery, we will call them `anomalies' by common usage of the word.}, namely $R_D$ and $R_{D^*}$ 
defined in the following way
\bea
R_{D^{(*)}} = \frac{\mathcal{B} \left(B \to D^{(\ast)} \, \tau \, \nu  \right)}{\mathcal{B} \left(B \to D^{(\ast)} \, \llx \, \nu \right)} \, ,
\label{def-rd}
\eea
where the  $\llx$ stands for either $e$ or $\mu$. 
The experimental results as well as the SM predictions for $R_D$ and $R_{D^*}$ are summarised in table~\ref{tab-exp}.  
\begin{table}[h!]
\begin{center}
\begin{tabular}{|c|cl|cc|}
\hline 
 Observable & SM prediction & & Measurement & \\ 
\hline
            &  $0.300 \pm 0.008$ &\cite{Aoki:2016frl}      &                                &                   \\
$R_D$ &  $0.299 \pm 0.011$ & \cite{Na:2015kha}     & $0.407 \pm 0.046$ & \cite{Amhis:2016xyh} \\
            &  $0.299 \pm 0.003$ & \cite{Bigi:2016mdz}  &                                &                   \\
\hline
$R_{D^\ast}$    &  $0.252 \pm 0.003$ &\cite{Fajfer:2012vx} & $0.304 \pm 0.015$ & \cite{Amhis:2016xyh} \\
                         &  $0.260 \pm 0.008$ &\cite{Bigi:2017jbd}    &                                &                   \\
\hline
$P_\tau ({D^*})$ & $-0.47 \pm 0.04$     &\cite{Bigi:2017jbd}   &    $-0.38 \pm 0.51 (\text{stat.}) \, ^{+0.21}_{-0.16} (\text{syst.}) $   
&  \cite{Hirose:2016wfn,Hirose:2017dxl}                 \\
\hline
$R_{\Jps}$ &  $  0 .290 $  &     & $0.71 \pm 0.25$ & \cite{Aaij:2017tyk} \\
\hline
\end{tabular}
\caption{\sf Observables, their SM predictions and the experimentally measured values. The experimental averages for $R_D$ and 
$R_{D^*}$ shown in the third column are based on \cite{Lees:2012xj, Lees:2013uzd, Huschle:2015rga, Aaij:2015yra, Sato:2016svk, Hirose:2016wfn, fpcp}. The SM prediction of $R_{\Jps}$ is based on the form-factors given in 
\cite{Wen-Fei:2013uea}, see appendix-\ref{Bc2Jpsi-FF} for more details. As the $B_c \to J/\psi$ form-factors  are not very reliably known, we do not 
show any uncertainty for $R_{\Jps}$. However, it is expected to be similar to that of $R_{D^\ast}$. \label{tab-exp}}
\end{center} 
\end{table}
We also show two other relevant recent measurements, which are however rather imprecise at the moment -- the $\tau$ polarisation, 
$P_\tau ({D^*})$, in the decay $B \to D^{(\ast)} \, \tau \, \nu$, and $R_{\Jps}$, a ratio similar to Eq.~\ref{def-rd} for the decay 
$B_c \to J/\psi \, \tau \, \nu$. 
It can be seen from Table \ref{tab-exp} that a successful explanation of the $R_{D, D^*}$ anomalies requires a new physics (NP)
contribution of the order of at least 10 - 20\% of the SM contribution to the branching ratio. As the SM contribution is generated at the
tree level by $W^\pm$ boson exchange, this is a rather large effect. Such a large effect puts any NP explanation under strong pressure 
arising from experimental measurements of other $\Delta F=1 $ and $\Delta F=2$ processes, electroweak precision observables
and direct searches at the LHC.

In the first part of this work (sections \ref{section2} - \ref{gauge-invariance}), we provide a comprehensive analysis of the possible 
explanations of these anomalies in as model independent way as possible. 
We discuss the various implications of (linearly realised) $\rm SU(2) \times U(1)$ symmetry. In particular, we focus on the correlations
among the observables which could be used in the future to decipher the physics behind these anomalies.

Some of the results which are presented in this part already exist in the literature in some form or another.
While we try to perform the analysis in a comprehensive and systematic way, and present our results
within a unified language, we will explicitly point out to the existing literature wherever appropriate.

In the second part of the paper (section \ref{comp-Higgs}), we apply these results to composite Higgs models with partial compositeness.
An explanation of the flavour anomalies within this framework has received a lot of attention in the recent  past
\cite{Barbieri:2015yvd,Gripaios:2014tna,Niehoff:2015iaa,Barbieri:2016las,Niehoff:2016zso,Megias:2016bde,Buttazzo:2016kid,
Barbieri:2017tuq, DAmbrosio:2017wis,Sannino:2017utc,Carmona:2017fsn,Marzocca:2018wcf}.
However, this scenario, motivated by the Higgs mass naturalness problem, generically predicts flavour violating
effects which are often too large to be compatible with experimetal measurements. This can be partially cured by introducing
additional flavour symmetries suppressing the undesirable flavour violating effects\cite{Barbieri:2015yvd,Barbieri:2016las,Barbieri:2017tuq}. 
In this work, instead of explicitly relying on flavour symmetries, we take an agnostic approach and rely only on the
correlations among the various flavour violating observables coming from partial compositness.
Interestingly, even without making any assumption on the flavour structure of the composite sector, we are able to
find strong correlations between $\Delta F=1$ and $\Delta F=2$ observables leading to an upper bound on the
scale of compositeness.


\section{Operators for $b \to c \, \ell \, \nu$ decay}
\label{section2}

The relevant Lagrangian for the quark level process  $b \to c \, \ell \, \nu$ can be written as, 
\begin{eqnarray}
{\cal L}^{b \to c \, \ell \, \nu}_{\rm eff} =  {\cal L}^{b \to c \, \ell \, \nu}_{\rm eff}|_{\rm SM} -  
\sum  \dfrac{g^{cb\ell\nu}_i}{\Lambda^2} {\cal O}^{cb\ell\nu}_{i}  + \rm h.c. + ...  ~~~(\ell = \tau, \mu, e)
\label{eff-lag}
\end{eqnarray} 
where the ellipses refer to terms which are suppressed by additional factors of $(\frac{\partial}{\Lambda})^2$. 
As $(\frac{\partial}{\Lambda})^2 \sim (\frac{M_B}{\Lambda})^2$ for the processes we are interested in,  these ellipses are 
completely negligible for new physics (NP) scales heavier than the weak scale. Here, 
\begin{eqnarray}
{\cal L}^{b \to c \, \ell \, \nu}_{\rm eff}|_{\rm SM}  =  
-\frac{2 G_F V_{cb}}{\sqrt{2}}  \left({\cal O}^{cb\ell\nu}_{\rm VL} -  {\cal O}^{cb\ell\nu}_{\rm AL}\right) \, , \, 
\label{eff-lag-sm}
\end{eqnarray} 
and the definition of the operators are
\vspace{2mm}
\begin{multicols}{2}
\noindent
\begin{align}
\label{b2c-basis}
{\cal O}^{cb\ell\nu}_{\rm VL}      &=    [\bar{c} \, \gamma^\mu \, b] [\bar \ell \, \gamma_\mu \, P_L \, \nu]  \nonumber \\ 
{\cal O}^{cb\ell\nu}_{\rm AL}       &=    [\bar{c} \, \gamma^\mu \, \gamma_5 \, b] [\bar \ell \, \gamma_\mu \, P_L \, \nu] \nonumber \\  
{\cal O}^{cb\ell\nu}_{\rm SL}       &=    [\bar{c} \, b] [\bar \ell  \, P_L \, \nu] \nonumber \\ 
{\cal O}^{cb\ell\nu}_{\rm PL}       &=    [\bar{c} \, \gamma_5 \, b] [[\bar \ell \, P_L \, \nu] \nonumber \\ 
{\cal O}^{cb\ell\nu}_{\rm TL}       &=   [\hedt] [\bar \ell \, \sigma_{\mu\nu} \, P_L \, \nu] \nonumber 
\end{align}
\begin{align}
{\cal O}^{cb\ell\nu}_{\rm VR}      &=    [\bar{c} \, \gamma^\mu \, b] [\bar \ell \, \gamma_\mu \, P_R \, \nu]  \nonumber \\ 
{\cal O}^{cb\ell\nu}_{\rm AR}       &=    [\bar{c} \, \gamma^\mu \, \gamma_5 \, b] [\bar \ell \, \gamma_\mu \, P_R \, \nu] \nonumber \\  
{\cal O}^{cb\ell\nu}_{\rm SR}       &=    [\bar{c} \, b] [\bar \ell  \, P_R \, \nu] \\ 
{\cal O}^{cb\ell\nu}_{\rm PR}       &=    [\bar{c} \, \gamma_5 \, b] [[\bar \ell \, P_R \, \nu] \nonumber \\ 
{\cal O}^{cb\ell\nu}_{\rm TR}       &=   [\hedt] [\bar \ell \, \sigma_{\mu\nu} \, P_R \, \nu] \, . \nonumber 
\end{align}
\end{multicols}

Note that the operators in the right hand side of Eq.~\eqref{b2c-basis} (referred to as right-chiral operators below) involve 
right-chiral neutrinos. 
In this work, we assume that light right-chiral neutrinos do not exist in nature\footnote{See \cite{Asadi:2018wea,Greljo:2018ogz} for
some recent proposals where the anomalies are explained by operators with right-handed neutrinos.}. Moreover, even in their presence, 
the operators involving them do not interfere with those involving left-chiral neutrinos (and hence not to the SM 
operators). This means, by naive power counting, that in order to explain the experimental data by the right-chiral operators, the required 
NP scale of these operators have to be lower than that for the left-chiral operators.

For notational convenience, we will normalise the new WCs by 
\bea
\dfrac{2 G_F V_{cb}}{\sqrt{2}} = \dfrac{1}{\Lambda^2_{\rm SM}}\approx \dfrac{1}{(1.2 \, \text{TeV})^2} \, .  \label{lambda-SM}
\eea 
Thus, we have
\begin{eqnarray}
{\cal L}^{b \to c \, \ell \, \nu}_{\rm eff} &=&  {\cal L}^{b \to c \, \ell \, \nu}_{\rm eff}|_{\rm SM} -  
\sum  \dfrac{g^{cb\ell\nu}_i}{\Lambda^2} {\cal O}^{cb\ell\nu}_{i}  + \rm h.c.  \nn \\
&=&-\frac{2 G_F V_{cb}}{\sqrt{2}}  \sum  C^{cb\ell\nu}_i  {\cal O}^{cb\ell\nu}_{i}  + \rm h.c.
\label{eff-lag-2}
\end{eqnarray} 
where, 
$ \dfrac{g_{\rm VL}^{cb\ell\nu}}{\Lambda^2} = \dfrac{2 G_F V_{cb}}{\sqrt{2}} (C_{\rm VL}^{cb\ell\nu} - 1), \,  
\dfrac{g_{\rm AL}^{cb\ell\nu}}{\Lambda^2} = \dfrac{2 G_F V_{cb}}{\sqrt{2}} (C_{\rm AL}^{cb\ell\nu} + 1), \, 
\dfrac{g_{\rm SL, PL, TL}^{cb\ell\nu}}{\Lambda^2} = \dfrac{2 G_F V_{cb}}{\sqrt{2}} C_{\rm SL, PL, TL}^{cb\ell\nu}$.

In the SM, 
$ C^{cb\ell\nu}_{\rm VL}  = 1, C^{cb\ell\nu}_{\rm AL}  = -1. $

Although there are five operators with left-chiral neutrinos, not all of them contribute to both $R_D$ and $R_{D^*}$. 
This is because the following matrix elements vanish identically:
\bea
\bra D(p_D, M_D)| \bar c \gamma^\mu \gamma_5 b |B(p_B, M_B)\ket &=& 0 \, , \\
\bra D(p_D, M_D)| \bar c \gamma_5 b |B(p_B, M_B)\ket &=& 0  \, , \\
\bra D^\ast(p_{D^\ast}, M_{D^\ast}, \epsilon_{D^*})| \bar c b |B(p_B, M_B)\ket &=&  0  \, .
\eea
Thus, the operators ${\cal O}^{cb\ell\nu}_{\rm AL}$ and ${\cal O}^{cb\ell\nu}_{\rm PL}$ do not contribute to $R_D$, 
and similarly, the operator ${\cal O}^{cb\ell\nu}_{\rm SL}$ does not contribute to $R_{D^*}$. 

In appendix~\ref{app-formulas}, we provide approximate semi-numerical formulas for $R_{D}$ and $R_{D^*}$ in
terms of the Wilson coefficients.

\section{Explaining $R_D$ and $R_{D^*}$}
\label{section3}
In this section, we systematically study the possible role of various dimension-6 operators in explaining the $R_D$ and $R_{D^*}$ 
anomalies. As mentioned before, some of the results that will be shown in this section are not new, and already exist in the literature
in some form or another \cite{Datta:2012qk,Tanaka:2012nw,Choudhury:2012hn,Sakaki:2013bfa,Bhattacharya:2014wla,Cata:2015lta,Freytsis:2015qca,
Choudhury:2016ulr,Celis:2016azn,Choudhury:2017qyt,Crivellin:2017zlb,Colangelo:2018cnj}.
Our aim would be to offer a coherent picture to the readers and add some important insights and aspects to the discussion.

For the calculation of $R_D$ we have used the vector and scalar form factors from\cite{Na:2015kha,Lattice:2015rga} and tensor form factors
from \cite{Melikhov:2000yu,Atoui:2013mqa}. As lattice QCD results at nonzero recoil are not yet available for $B \to D^*$, we have used the
HQET form factors parametrized by \cite{Caprini:1997mu} with the following numerical values of the relevant parameters\cite{Amhis:2014hma,Bailey:2014tva}
\bea
\label{bds-ff-param}
&&R_1(1) = 1.406 \pm 0.033,  ~ R_2(1) = 0.853 \pm 0.020,  ~ \rho_{D^*}^2 =1.207 \pm  0.026  \nonumber \\
&& \hspace{4cm}   h_{A_1}(1) = 0.906 \pm 0.013 \, .
\eea
In view of the absence of lattice calculations, to be conservative, we use two times larger uncertainties than those quoted above.

For state-of-the-art $B \to D^*$ form-factors we refer the readers to \cite{Bernlochner:2017jka} (see also \cite{Jung:2018lfu}). It should be noted that, since we have not used the state-of-the-art  form-factors, our results for the allowed
values of the Wilson Coefficients are correct only up to sub-leading terms in $\Lambda/m_{b,c}$.

\subsection{Vector and Axial-vector operators}
\label{res:VA}

Here we consider only the operators ${\cal O}_{\rm VL}^\tau$ and ${\cal O}_{\rm AL}^\tau$, and investigate whether they 
are capable of explaining $R_D$ and $R_{D^*}$ anomalies simultaneously. We also comment on the compatibility with 
the recent measurement of $R_{J/\psi}$.  

\begin{figure}[!h!]
\centering
\includegraphics[scale=0.6]{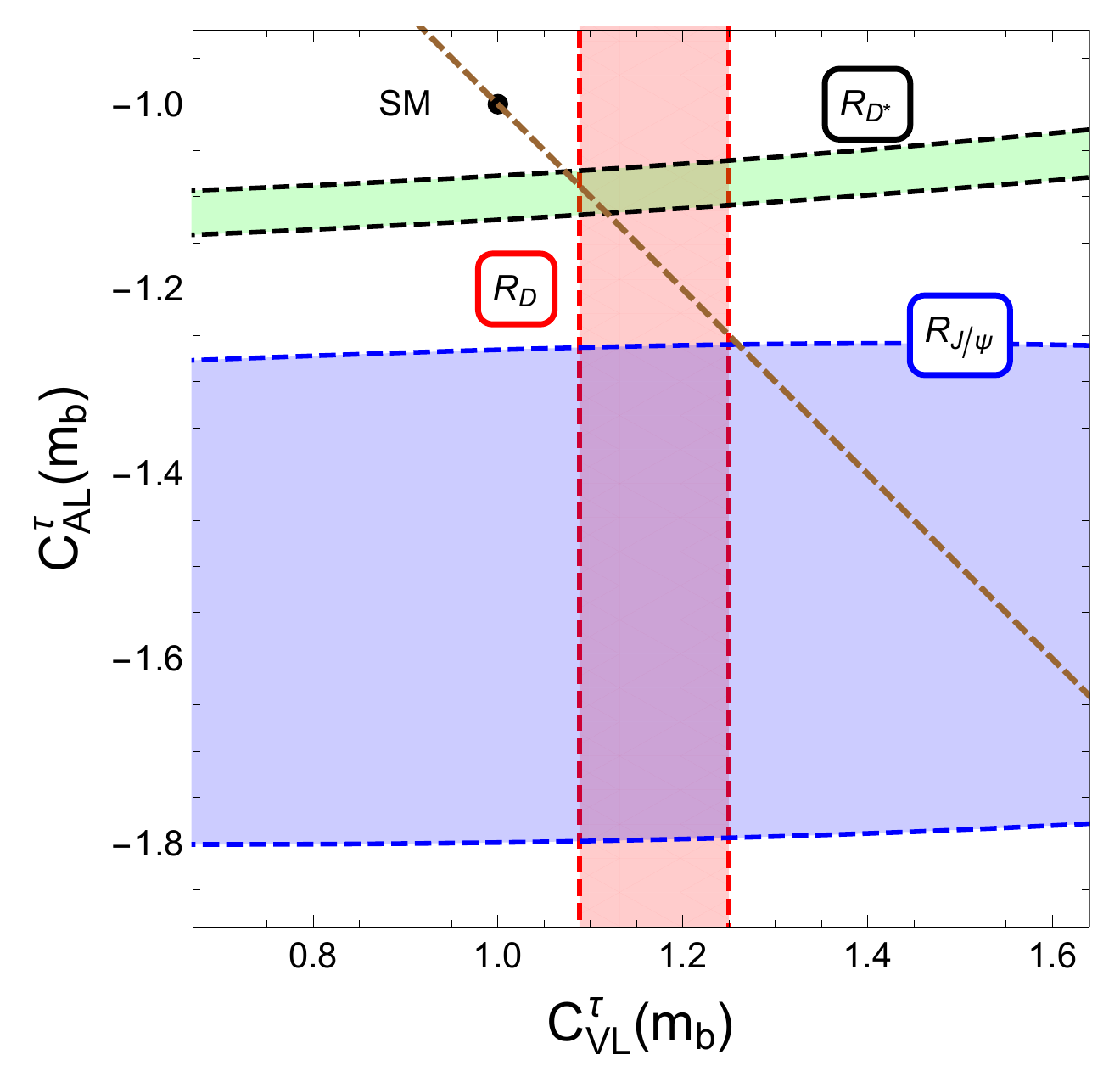}
\caption{\sf The vertical red band corresponds to the values of $C_{\rm VL}^\tau$ that satisfy the experimental measurement of 
$R_D$ within $1\sigma$. Similarly, the green (blue) region corresponds to the values of $C_{\rm VL}^\tau$ and $C_{\rm AL}^\tau$ that 
satisfy the experimental measurement of $R_{D^*}$ ($R_{J/\psi}$) within $1\sigma$. All the WCs are defines at the $m_b$ scale. 
The oblique dashed line is the locus of the equation $C_{\rm VL}^\tau= -C_{\rm AL}^\tau$. \label{fig:VL-AL}}
\end{figure}

In Fig.~\ref{fig:VL-AL}, we show the regions  in the $C_{\rm VL}^\tau$ - $C_{\rm AL}^\tau$ plane that satisfy the experimental data on 
$R_D$, $R_{D^*}$ and $R_{J/\psi}$ within $1\sigma$. Note that the uncertainties in the form-factors have been carefully taken 
into account in obtaining the various allowed regions. However, the semi-numerical formulas given in the previous section 
can be used to qualitatively understand the results.   
It can be seen that there is an overlap region (the overlap between the red and green bands) that successfully explains both 
$R_D$ and $R_{D^*}$. This overlap region is outside the $1\sigma$ 
experimental measurement of $R_{J/\psi}$, but consistent with $R_{J/\psi}$ at $\approx 1.5 \, \sigma$.

It is interesting that $C_{\rm VL}^\tau = -C_{\rm AL}^\tau \approx 1.1 $ falls in the overlap region mentioned above. As we will 
see in the next section, the relation $C_{\rm VL}^\tau= -C_{\rm AL}^\tau$ is expected if SU(2)$_{\rm L}$ $\times$ U(1)$_{\rm Y}$ 
gauge invariance in linearly realised at the dimension-6 level. 
Note that the vector and axial-vector operators do not have anomalous dimensions if only QCD interactions are considered 
(see, for example, appendix-E of \cite{Bardhan:2016uhr} and also \cite{Alonso:2013hga}).
Hence, we take $C_{\rm VL,AL}^\tau (\Lambda) = C_{\rm VL,AL}^\tau (m_b)$.

\subsection{Scalar and Pseudo-scalar operators}
\label{res:SP}

\begin{figure}[h!]
\centering
\begin{tabular}{cc}
\hspace*{-12mm} \includegraphics[width=7.7cm, height=7cm]{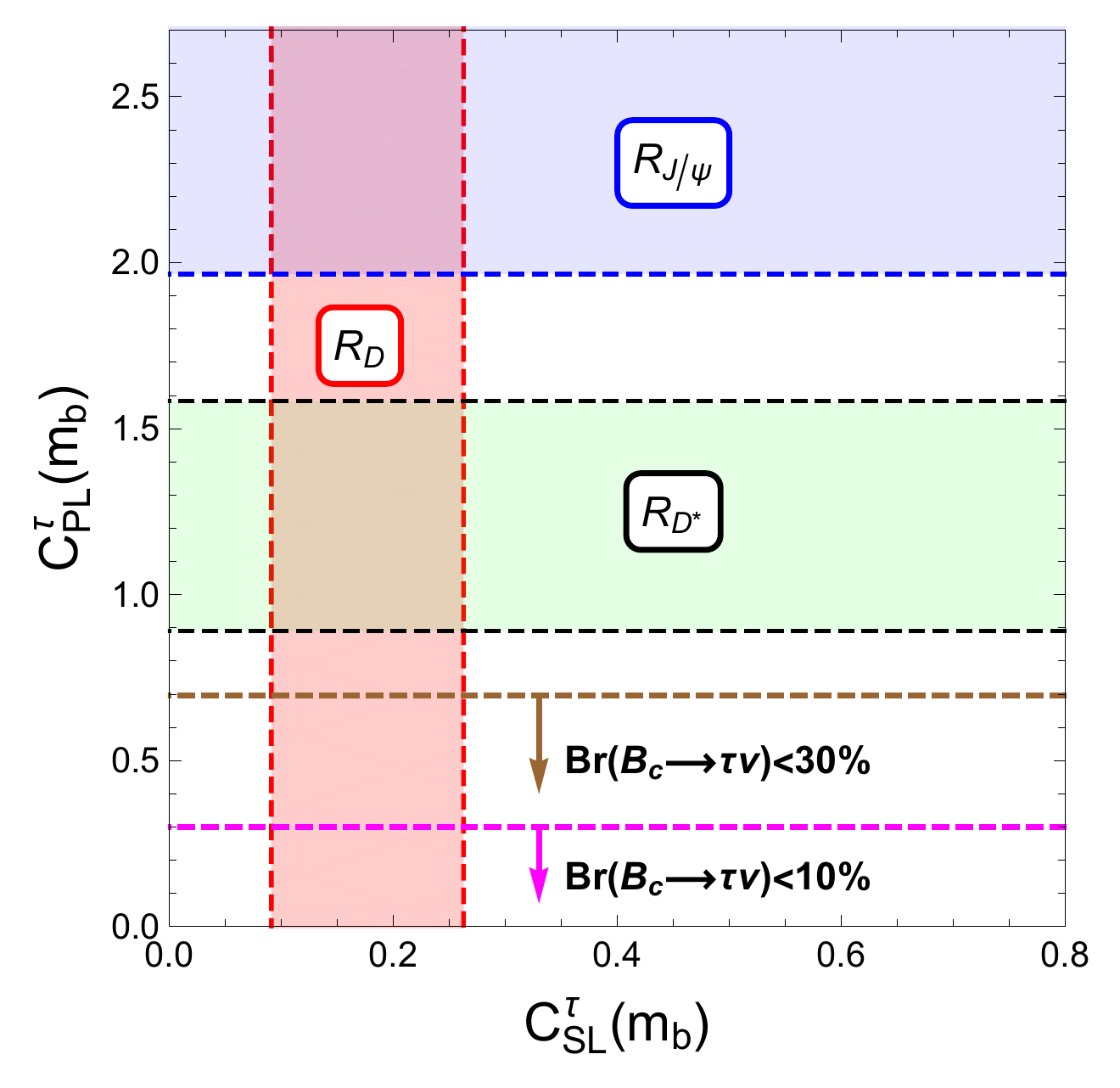} & \hspace*{-10mm}  \includegraphics[width=7.2cm, height=7.cm]{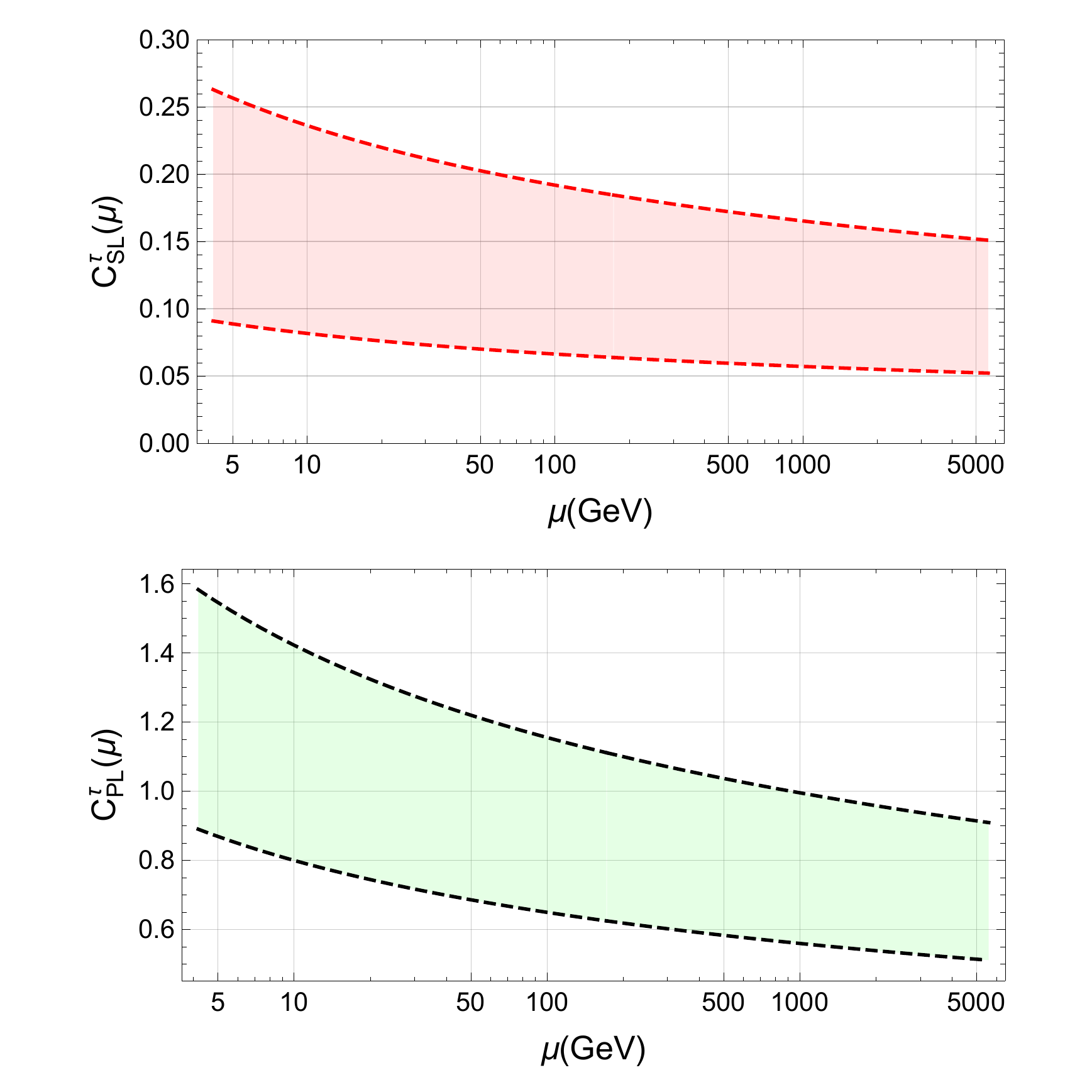}
\end{tabular}
\caption{\sf {\bf Left panel }: the red and green (blue) bands correspond to the values of $C_{\rm SL}^\tau$ and $C_{\rm PL}^\tau$ that satisfy the 
experimental measurement of $R_D$ and $R_{D^*}$ ($R_{J/\psi}$) within $1\sigma$ respectively. The values of $C_{\rm PL}^\tau$ 
that correspond to Br($B_c \to \tau \nu$) $< 30\%$ and $<10\%$ are also shown. 
{\bf Right panel} : renormalisation group running of the WCs $C_{\rm SL}^\tau$ and $C_{\rm PL}^\tau$.  \label{fig:SL-PL}}
\end{figure}

Here we consider the scalar and pseudo-scalar operators, ${\cal O}_{\rm SL}^\tau$ and ${\cal O}_{\rm PL}^\tau$ respectively. 
In the left panel of Fig.~\ref{fig:SL-PL}, we show the parameter space that satisfies the individual experimental data on 
$R_D$, $R_{D^*}$ and $R_{J/\psi}$ within $1\sigma$. 
As discussed before, while the operator ${\cal O}_{\rm SL}^\tau$ contributes to $R_D$ only, the operator ${\cal O}_{\rm PL}^\tau$ 
contributes only to $R_{D^*}$. This explains the vertical and horizontal nature of the allowed regions for $R_D$ and $R_{D^*}$ respectively.

Note that the operator ${\cal O}_{\rm PL}^\tau$ directly contributes to the decay $B_c \to \tau \nu$ also (refer to 
appendix-\ref{app:Bctaunu} for more details). The regions below the two horizontal dashed lines correspond to 
Br$(B_c \to \tau \nu) < 30\%$ and $<10\%$, which were claimed to be the indirect experimental upper bounds by 
the authors of \cite{Alonso:2016oyd} and \cite{Akeroyd:2017mhr} respectively.  
Thus, an explanation of $R_{D^*}$ by the operator ${\cal O}_{\rm PL}^\tau$ is in serious tension with the upper bound 
on Br$(B_c \to \tau \nu)$.

The right panel of Fig.~\ref{fig:SL-PL} shows the renormalisation group (RG) running (considering only QCD interactions)
of the WCs $C_{\rm SL}^\tau$ and $C_{\rm PL}^\tau$ 
from the $m_b$ scale to 5 TeV using the following equation \cite{Bardhan:2016uhr}, 
\begin{equation}
C(m_b)  = \left[\frac{\alpha_s(m_t)}{\alpha_s(m_b)}\right]^{\frac{\gamma}{2\beta_0^{(5)}}}
\left[\frac{\alpha_s(\Lambda)}{\alpha_s(m_t)}\right]^{\frac{\gamma}{2\beta_0^{(6)}}} C(\Lambda) \, , \label{RGE}
\end{equation}
where, $\gamma = -8$.
The values at the $m_b$ scale are taken from the allowed bands in the left panel.

\subsection{Tensor operators}

\begin{figure}[h!]
\centering
\begin{tabular}{cc}
\includegraphics[scale=0.55]{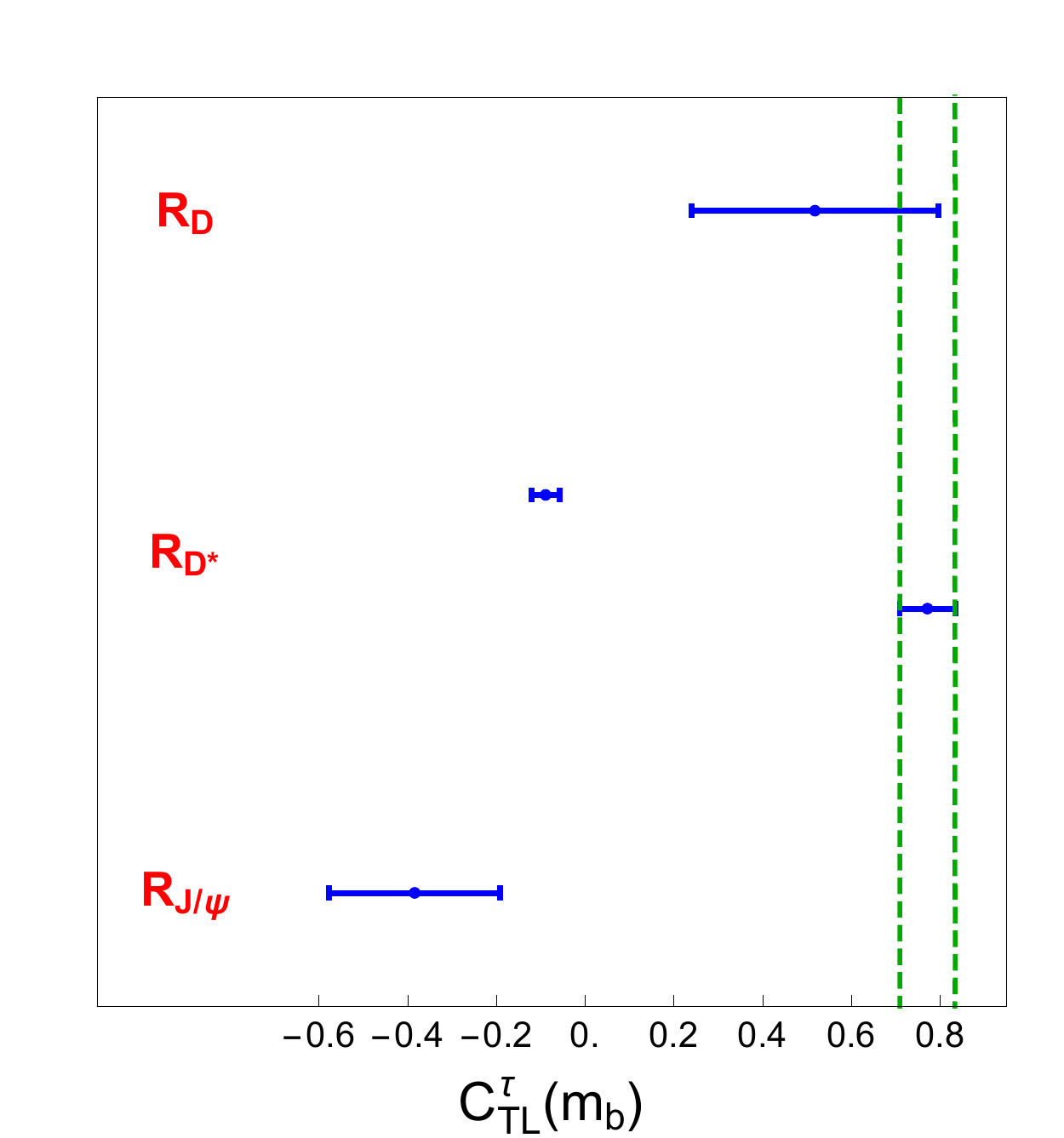} &  \hspace*{-10mm} \includegraphics[scale=0.55]{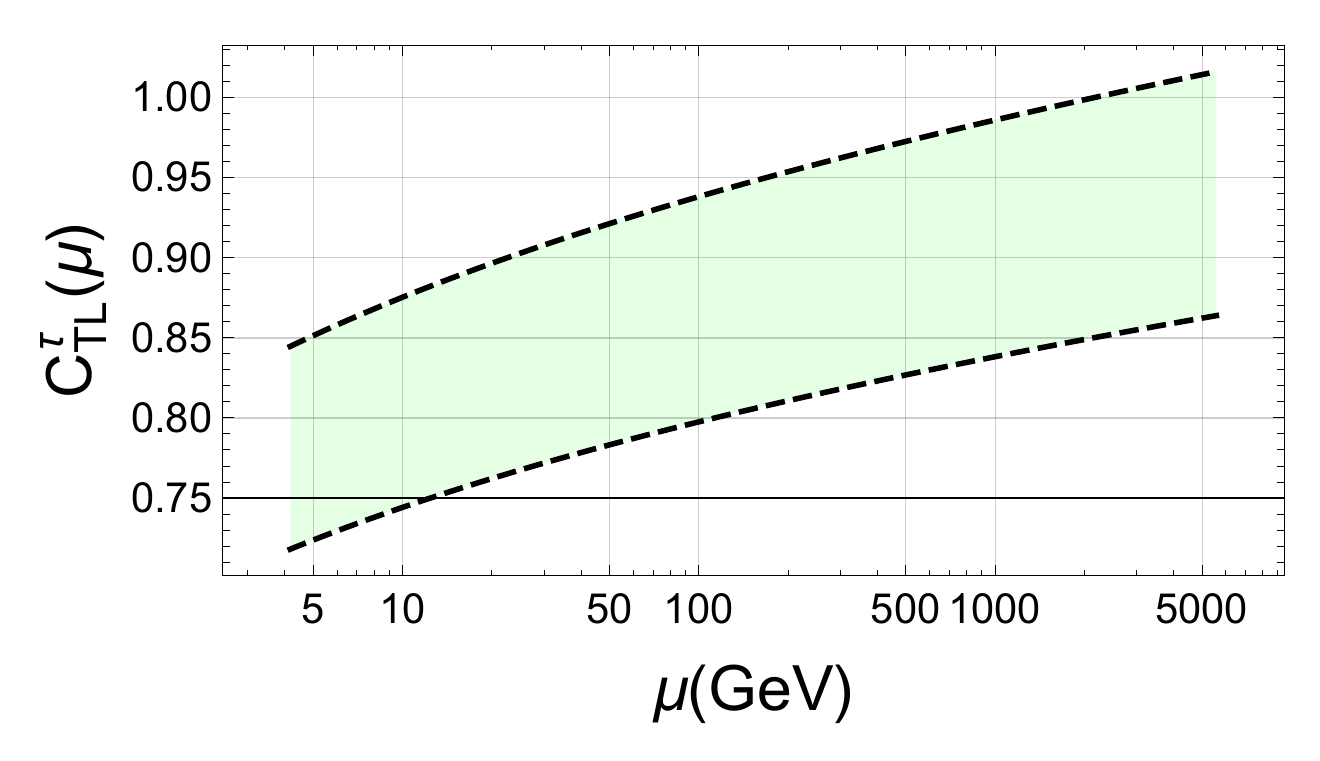}
\end{tabular}
\caption{\sf{\bf Left panel }: the horizontal lines correspond to the values of $C_{\rm TL}^\tau$ that satisfy the 
experimental measurement of $R_D$, $R_{D^*}$ and $R_{J/\psi}$ within $1\sigma$. The green band corresponds to 
values of $C_{\rm TL}^\tau$ that explains $R_D$ and $R_{D^*}$ simultaneously. 
{\bf Right panel} : renormalisation group running of  $C_{\rm TL}^\tau$. \label{fig:TL}}
\end{figure}

We now turn to the discussion of the tensor operator. In Fig.~\ref{fig:TL}, we show the allowed values of $C_{\rm TL}^\tau$ that are 
consistent with the $1\sigma$ experimental measurements of $R_D$, $R_{D^*}$ and $R_{J/\psi}$. The values enclosed by the 
green vertical dashed lines correspond to simultaneous explanation of $R_D$ and $R_{D^*}$ anomalies. Note however that the 
prediction for $R_{J/\psi}$ in this $C_{\rm TL}^\tau$ region is $\approx 0.17 - 0.23$, which is below the SM prediction and quite far 
from the current experimental central value.  The RG running of $C_{\rm TL}^\tau$ is shown in the right panel of Fig.~\ref{fig:TL} 
(using Eq.~\ref{RGE} with $\gamma=8/3$ \cite{Bardhan:2016uhr}) where the initial values of $C_{\rm TL}^\tau$ at the $m_b$ scale
correspond to the range enclosed by the two vertical dashed lines in the left panel. 

Note that the tensor operator does not contribute to the decay $B_c \to \tau \nu$ because the matrix element
$\langle 0 | \hedt | \bar{B}_c \rangle$ identically vanishes. Hence, there is no constraint on 
$C_{\rm TL}^\tau$ from the process $B_c \to \tau \nu$.

\subsection{Combination of Tensor, Scalar and Pseudo-scalar operators}

\begin{figure}[h!]
\centering
\begin{tabular}{ccc}
\hspace*{-5mm} \includegraphics[scale=0.4]{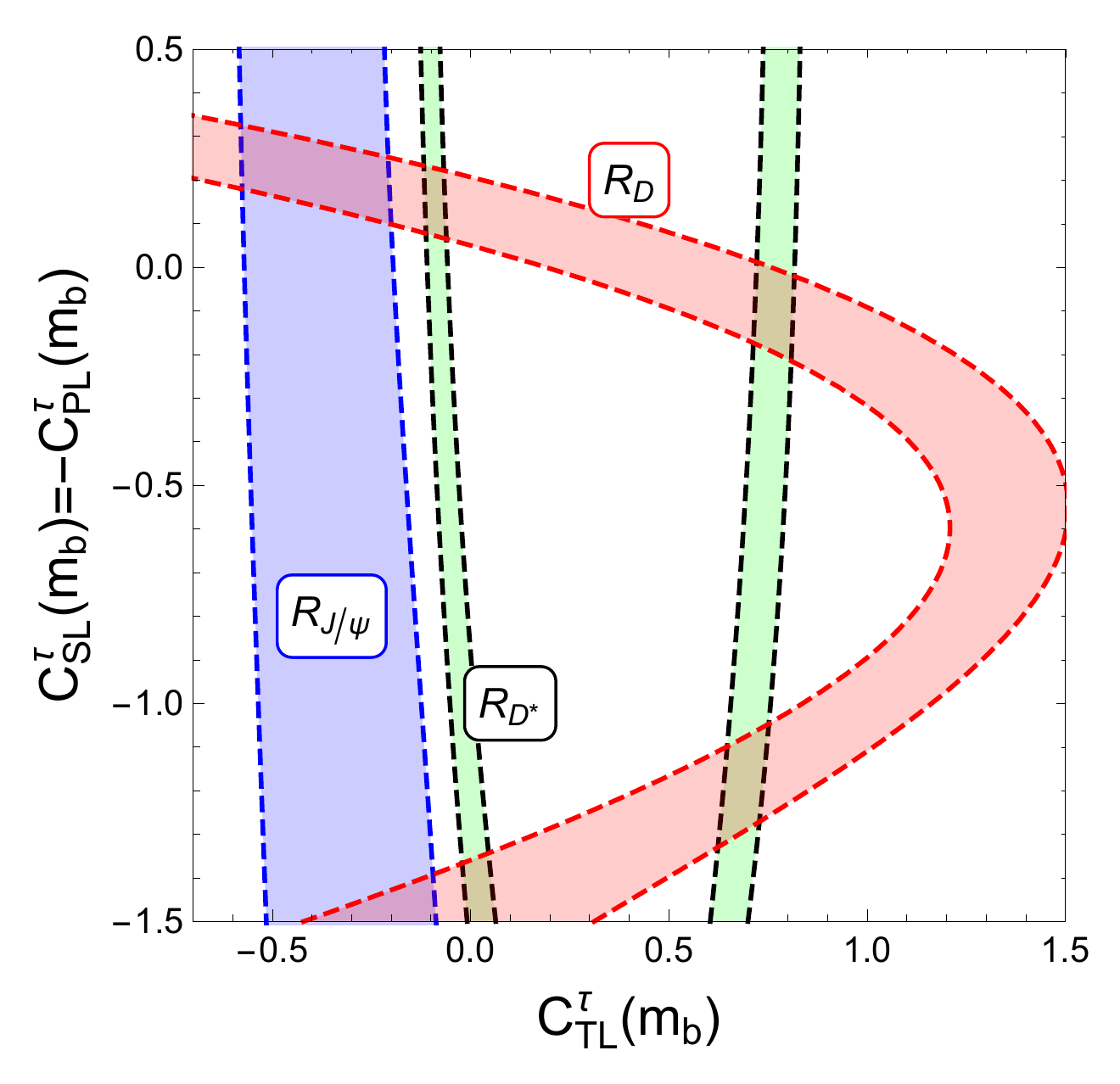} &  \hspace*{-6mm}  \includegraphics[scale=0.4]{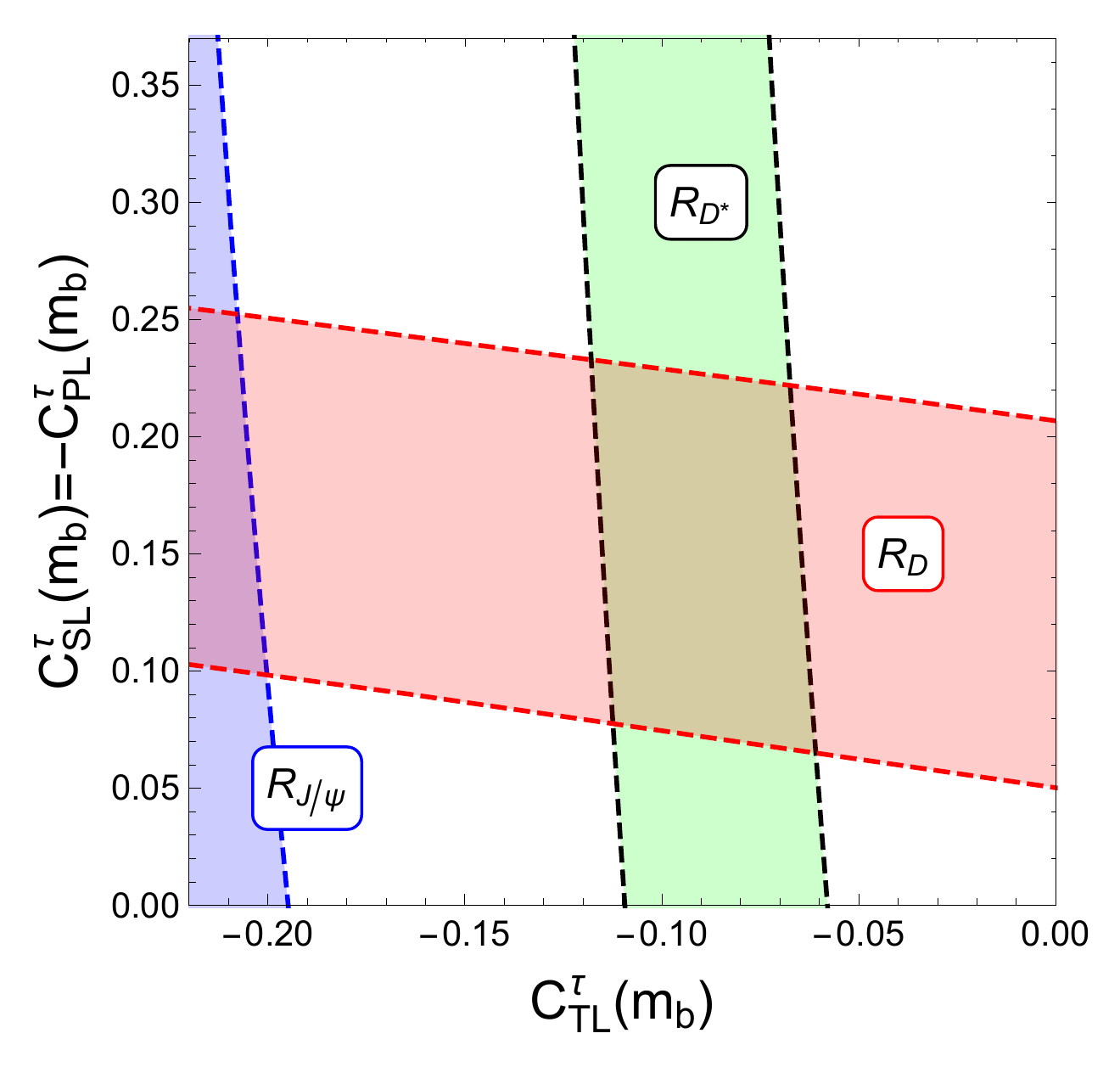} 
& \hspace*{-6mm} \includegraphics[scale=0.38]{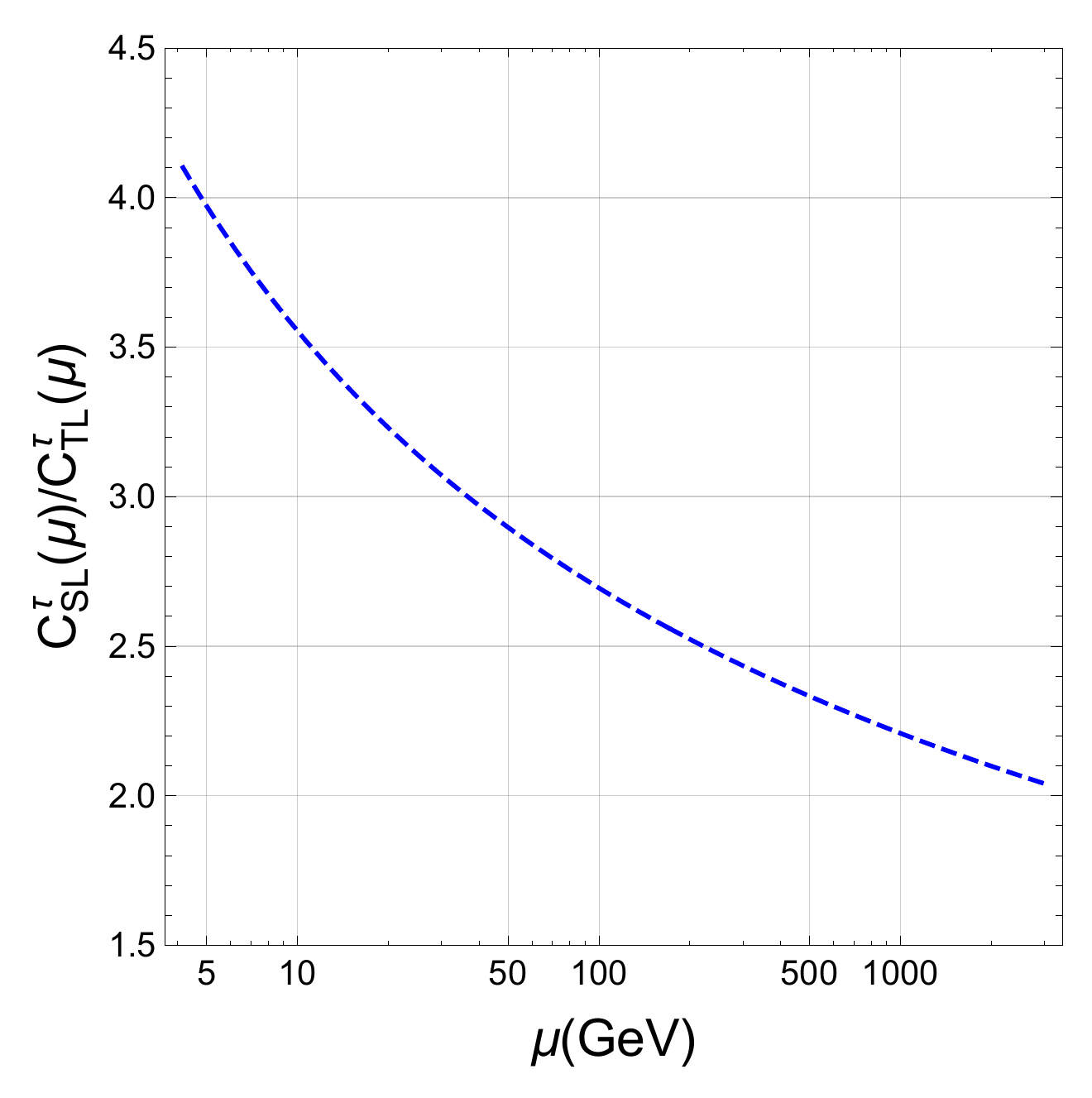}
\end{tabular}
\caption{\sf The red and green (blue) shaded regions in the left panel correspond to the values of $C_{\rm SL}^\tau=-C_{\rm PL}^\tau$ 
and $C_{\rm TL}^\tau$ that satisfy the experimental measurement of $R_D$ and $R_{D^*}$ ($R_{J/\psi}$) within $1\sigma$ respectively.  
The small overlap of the red and green regions for positive (negative) values of $C_{\rm SL}^\tau$ ($C_{\rm TL}^\tau$) is magnified separately
in the middle panel. The right panel shows the RG evolution of the coupling ratio $C_{\rm SL}^\tau / C_{\rm TL}^\tau$ assuming
$C_{\rm SL}^\tau / C_{\rm TL}^\tau = 2$ at 3 TeV.  See text for more details.  
\label{fig:SL-PL-TL} }
\end{figure}

In this section, we consider the scenario in which the scalar, pseudo-scalar and tensor operators are present
simultaneously\footnote{The combination of vector and scalar operators is discussed in appendix~\ref{sec:scalar-vector}.}. 
In the upper left panel of Fig.~\ref{fig:SL-PL-TL}, we show the various allowed regions in the $C_{\rm TL}^\tau$ - $C_{\rm SL}^\tau$ plane 
assuming the relation $C_{\rm SL}^\tau= -C_{\rm PL}^\tau$. From the upper panel of Fig.~\ref{fig:SL-PL-TL}, it can be seen that a
simultaneous explanation of the $R_D$ and $R_{D^*}$ anomalies requires $C_{\rm SL}^\tau (m_b)= - C_{\rm PL}^\tau (m_b) \in
[0.08, 0.23]$ and $C_{\rm TL}^\tau (m_b) \in [-0.11, -0.06]$ (the small overlap of the red and green regions for positive
values of $C_{\rm SL}^\tau$ and negative values of $C_{\rm TL}^\tau$). We are ignoring the overlap regions with $C_{\rm PL}^\tau > 1$ because of the bound from
Br($B_c \to \tau \nu$). There is also an overlap region enclosing $C_{\rm SL}^\tau = - C_{\rm PL}^\tau = 0$ and for non-zero $C_{\rm TL}^\tau$ which
corresponds to the tensor solution discussed in the previous section.

We would like to comment in passing that there exist scalar leptoquark models that generate the operator $\left(\bar{c}  P_L \nu \right) 
\left(\bar{\tau}  P_L b \right)$ at the matching
scale\footnote{This operator 
arises from a SU(2)$_{\rm L}$ $\times$ U(1)$_{\rm Y}$ gauge invariant operator 
$\left(\bar{l^\prime}^k  u^\prime\right)  \epsilon_{jk} \left( \bar{q^\prime}^j e^\prime  \right)$ which, by using Fierz transformation, gives 
$$
\left(\bar{l^\prime}^k  u^\prime\right)  \epsilon_{jk} \left( \bar{q^\prime}^j e^\prime  \right) = -\frac{1}{8} \Bigg[
4 \left(\bar{l^\prime}^j e^\prime \right) \epsilon_{jk} \left( \bar{q^\prime}^k u^\prime \right)  + 
\left(\bar{l^\prime}^j \sigma_{\mu\nu} e^\prime \right) \epsilon_{jk} \left( \bar{q^\prime}^k \sigma^{\mu\nu} u^\prime \right) \Bigg] \, .
$$ 
See section \ref{gauge-invariance} below for the notations.} $\Lambda$, see e.g., \cite{Dorsner:2013tla}.
This operator can be written in terms of the operators in Eq.~\eqref{b2c-basis} after performing the Fierz
transformation\footnote{Note that vector leptoquarks, after Fierz transformation, generate vector operators only in
the basis of section \ref{section2}. A scenario with vector leptoquarks will be discussed in section \ref{Leptoquark}.}, 
\bal
\left(\bar{c}  P_L \nu \right)  \left(\bar{\tau}  P_L b \right)  
&= -\frac{1}{8} \Bigg[2 ({\cal O}_{\rm SL}^\tau  - {\cal O}_{\rm PL}^\tau)  + {\cal O}_{\rm TL}^\tau \Bigg] \, .
\eal
Thus, at the matching scale one gets
\bal
C_{\rm SL}^\tau (\Lambda)= - C_{\rm PL}^\tau (\Lambda) = 2 C_{\rm TL}^\tau (\Lambda) \, .
\eal
This was our motivation to consider $C_{\rm SL}^\tau = - C_{\rm PL}^\tau$ in Fig.~\ref{fig:SL-PL-TL}.
The ratio $C_{\rm SL}^\tau / C_{\rm TL}^\tau$, however, increases with the decreasing RG scale as shown in the right panel of
Fig.~\ref{fig:SL-PL-TL}. Assuming $C_{\rm SL}^\tau (\Lambda)/ C_{\rm TL}^\tau (\Lambda)= 2$ for $\Lambda = 3$ TeV,
we get $C_{\rm SL}^\tau (m_b)/C_{\rm TL}^\tau (m_b) \approx 4$.

Note that, in the above discussion we have considered only real values of the Wilson coefficients. Allowing for complex Wilson
coefficients may lead to new possibilities, see for example \cite{Becirevic:2018afm}. 

\subsection{Distinguishing the various explanations}

In the previous subsections we  saw that simultaneous explanations of the $R_D$ and $R_{D^*}$ anomalies are possible by 
\begin{enumerate}
\item a combination of vector and axial-vector operators (the overlap of red and green regions in Fig.~\ref{fig:VL-AL})
\item a combination of scalar and pseudo-scalar operators (the overlap of red and green regions in Fig.~\ref{fig:SL-PL})
\item tensor operator only (the region between the two dashed vertical lines in Fig.~\ref{fig:TL})
\item a combination of scalar, pseudo-scalar and tensor operators (the overlap of red and green regions in Fig.~\ref{fig:SL-PL-TL},
in particular, the region with positive values of $C_{\rm SL}^\tau$ and negative values of $C_{\rm TL}^\tau$.)
\end{enumerate}

\begin{figure}[h!]
\centering
\begin{tabular}{cc}
\hspace*{-13mm}\includegraphics[width=7cm, height=7cm]{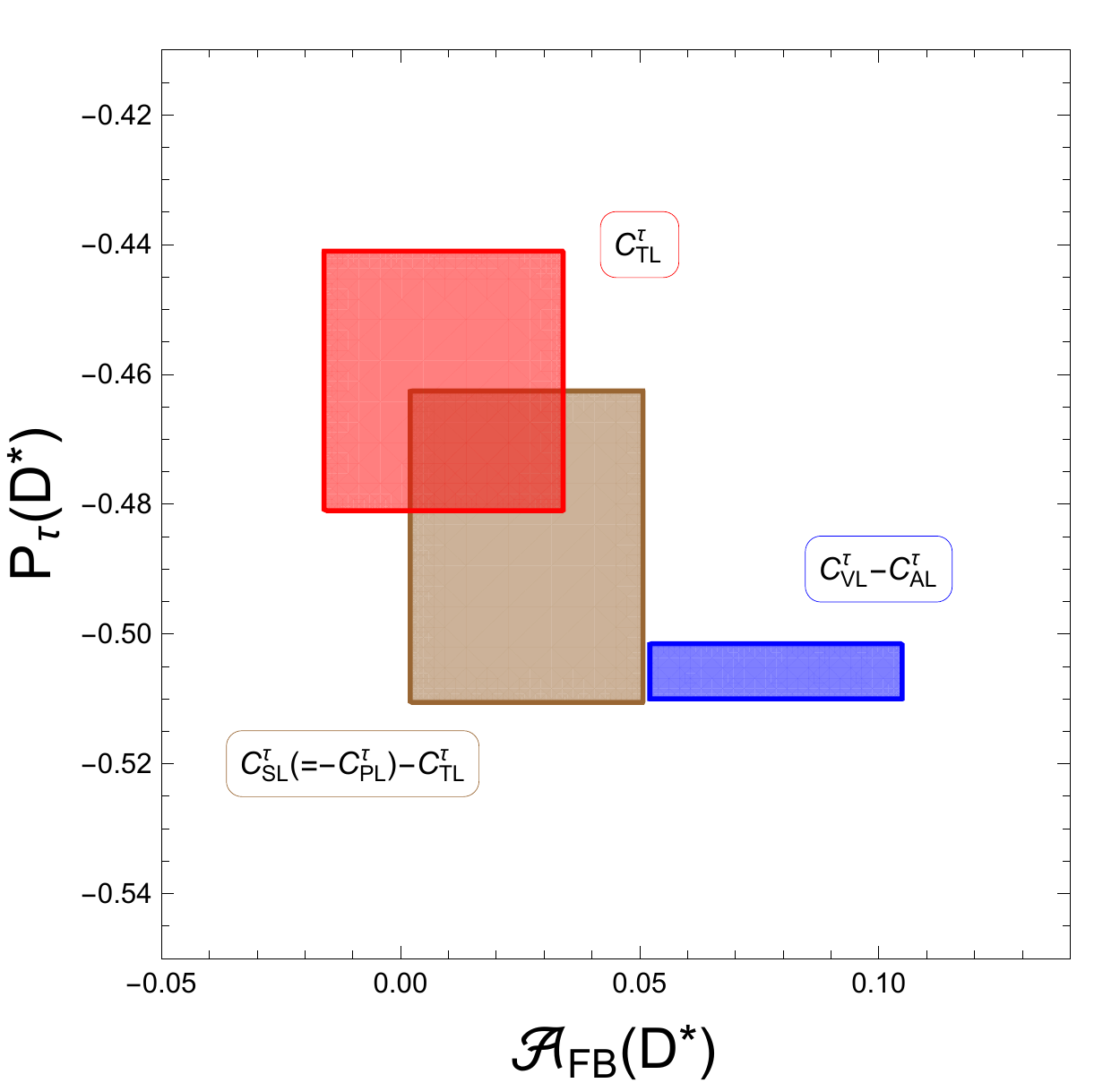} &  
\hspace*{-8mm} \includegraphics[width=7cm, height=7cm]{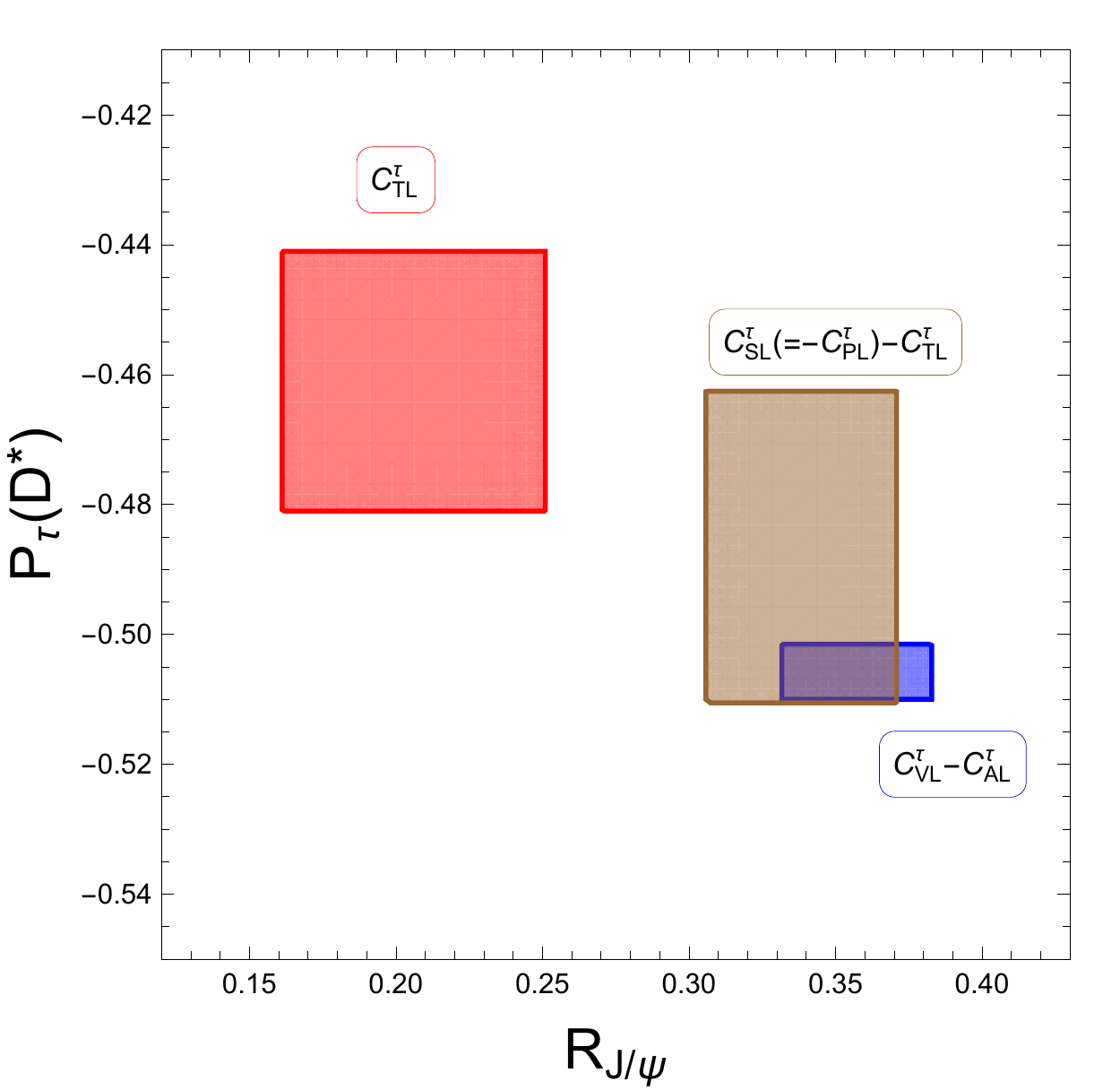}
\end{tabular}
\caption{ \sf Predictions for $P_\tau(D^*)$, ${\cal A}_{\rm FB}(D^*)$ and $R_{J/\psi}$ for values of the WCs that correspond to various 
simultaneous solutions of $R_D$ and $R_{D^*}$ anomalies. See text for more details. \label{fig:diff}}
\end{figure}

The second solution is quite strongly disfavoured by the existing indirect upper bound on the branching ratio of $B_c \to \tau \nu$.
So we ignore it here. We also ignore the scenario with a combination of vector and scalar operators because
of the reason mentioned in the last paragraph of section \ref{sec:scalar-vector}. 

We now very briefly comment on the possibility of distinguishing the three possible solutions 1), 3) and 4) by measuring
the $\tau$-polarisation ($P_\tau(D^*)$), forward-backward asymmetry (${\cal A}_{\rm FB}(D^*)$) and more interestingly,
$R_{J/\psi}$. In Fig.~\ref{fig:diff}, we plot the predictions for $P_\tau(D^*)$, ${\cal A}_{\rm FB}(D^*)$ and $R_{J/\psi}$ for
values of the WCs that correspond to various simultaneous solutions of $R_D$ and $R_{D^*}$ anomalies.

It can be seen that it is indeed possible to discriminate the three solutions by measuring $P_\tau(D^*)$, ${\cal A}_{\rm FB}(D^*)$ and
$R_{J/\psi}$. In fact,  as can be seen from the right panel of Fig.~\ref{fig:diff}, $R_{J/\psi}$ can be a very good discriminating
observable between the solutions 1) and 3). Of course, with more data, various kinematical distributions can also be used to
discriminate the different Lorentz structures \cite{Nierste:2008qe,Datta:2012qk,Duraisamy:2013kcw,Ligeti:2016npd,Alonso:2016gym,
Alok:2016qyh,Celis:2016azn,Colangelo:2018cnj}.

\section{Linearly realised SU(2)$_{\rm L}$ $\times$ U(1)$_{\rm Y}$ gauge invariance}
\label{gauge-invariance}

In the previous sections, we considered operators which were manifestly SU(3) $\times$ U(1)$_{\rm em}$ invariant, but invariance
under the full electroweak group was not demanded. We investigate the consequences of SU(2)$_{\rm L}$ $\times$ U(1)$_{\rm Y}$
invariance in this section. 

\subsection{List of operators}
\label{sec:dim6-op}
The SU(3) $\times$ SU(2)$_{\rm L}$ $\times$ U(1)$_{\rm Y}$ invariant dimension-6 operators that lead to $b \to c \, \tau \, \nu$ 
decay are given by (using the same notation as in \cite{Grzadkowski:2010es}; the primes represent the fact that the operators 
and couplings are written in the gauge basis)
\bal
{\cal L}^{\rm dim6} =-\frac{1}{\Lambda^2} \sum_{p'r's't'} \bigg \{ & [C_{lq}^{(3)}]^\prime_{p'r's't'}  \left(\bar{l^\prime}_{p'} \gamma_\mu \sigma^I l^\prime_{r'} \right) 
\left( \bar{q^\prime}_{s'} \gamma^\mu \sigma^I q^\prime_{t'} \right)  + \rm h.c.  \label{dim6-1}\\
+  & [C_{ledq}]^\prime_{p'r's't'}  \left(\bar{l^\prime}_{p'}^j e^\prime_{r'} \right) \left( \bar{d^\prime}_{s'} q_{t'}^{\prime j} \right) + \rm h.c.  \\
+  & [C_{lequ}^{(1)}]^\prime_{p'r's't'}  \left(\bar{l^\prime}_{p'}^j e^\prime_{r'} \right) \epsilon_{jk} \left( \bar{q^\prime}_{s'}^k u^\prime_{t'} \right) + \rm h.c. \\
+  & [C_{lequ}^{(3)}]^\prime_{p'r's't'} \left(\bar{l^\prime}_{p'}^j \sigma_{\mu\nu} e^\prime_{r'} \right) \epsilon_{jk}  
\left( \bar{q^\prime}_{s'}^k \sigma^{\mu\nu} u^\prime_{t'} \right) + \rm h.c. \\
------- & ----------------------- \nn \\
+ & [C_{\phi l}^{(3)}]^\prime_{p'r'} \left(\phi^\dagger i \overleftrightarrow{D}_\mu^I \phi \right) 
\left( \bar{l^\prime}_{p'} \, {\sigma^I} \gamma^\mu \, l^\prime_{r'} \right) + \rm h.c. \label{dim6-5}\\
------- & ----------------------- \nn \\
+ & [C_{\phi q}^{(3)}]^\prime_{p'r'} \left(\phi^\dagger i \overleftrightarrow{D}_\mu^I \phi \right) 
\left( \bar{q^\prime}_{p'} \, {\sigma^I} \gamma^\mu \, q^\prime_{r'} \right) + \rm h.c. \\
+ & [C_{\phi ud}]^\prime_{p'r'} \left(\phi^j \epsilon_{jk} \, i(D_\mu \phi)^k \right) 
\left( \bar{u^\prime}_{p'} \gamma^\mu d^\prime_{r'}\right) + \rm h.c. \bigg \}   \label{dim6-7}
\eal
where, $\epsilon_{ij}$ is antisymmetric with $\epsilon_{12} = +1$, and 
\bal
\phi^\dagger i \overleftrightarrow{D}_\mu^I \phi &= 
i \left( \phi^\dagger \sigma^I D_\mu \phi - (D_\mu \phi)^\dagger \sigma^I \phi \right) \\
D_\mu \phi &= (\del_\mu + i g_2 \frac{\sigma^I}{2} W^I_\mu  + i g_1 Y_\phi B_\mu) \phi \, \,  , Y_\phi = \frac{1}{2} \, . \\
\text{(Note that, the operator } &\text{structure $\left(\bar{l^\prime}_{p'}^j \sigma_{\mu\nu} e^\prime_{r'} \right) 
\left( \bar{d^\prime}_{s'} \sigma^{\mu\nu}q_{t'}^{\prime j} \right)$ vanishes algebraically.)} \nn 
\eal

The operators $[{\cal O}_{\phi q}^{(3)}]_{p'r'}$ and $[{\cal O}_{\phi ud}]_{p'r'}$ modify the charged 
current vertex of the quarks, in particular, the one of our interest $\bar{c} b W$. However, this affects both the 
$b \to c \, \tau \, \nu$ and $b \to c \, \llx \, \nu$ decays in the same way. Consequently, the operators 
$[{\cal O}_{\phi q}^{(3)}]_{p'r'}$ and $[{\cal O}_{\phi ud}]_{p'r'}$ are not relevant for the explanation of 
the $R_D$ and $R_{D^*}$ anomalies, and we will ignore them in the rest of the paper.

It can be seen from Eq.~\eqref{dim6-1} and Eq.~\eqref{dim6-5} that the operators only involve left chiral fields. 
Consequently, these operators only lead to V-A interactions. We stress that it is not true in general that linearly realised 
SU(2)$_{\rm L}$ $\times$ U(1)$_{\rm Y}$ gauge invariance forbids V+A operators at the dimension-6 level. For example, 
the operator in Eq.~\eqref{dim6-7} generates V+A  operator, but, as mentioned before, it does not lead to 
lepton non-universality at the dimension-6 level. This is an important consequence of linearly realised 
SU(2)$_{\rm L}$ $\times$ U(1)$_{\rm Y}$ gauge invariance.

Note however that at the dimension-8 level, the operator (${\cal O}_{\rm VL}^\tau + {\cal O}_{\rm AL}^\tau$) with the possibility 
of lepton non-universality can be generated. For example, consider the operator 
\bal 
{\cal O}_{\rm RL}^8 = \frac{1}{\Lambda^4}\left(\bar{l^\prime}_{p'} \phi \right) \gamma_\mu \left( l^\prime_{r'}  \phi \right)  \left( \bar{u^\prime}_{s'} 
\gamma^\mu d^\prime_{t'}\right) \label{dim8}
\eal
where the objects inside each of the parenthesis are constructed as SU(2) singlets. After electroweak symmetry 
breaking, this operator generates the following interaction term
\bea
\frac{v^2}{\Lambda^2} \, \frac{1}{\Lambda^2} \,   [\bar \ell \, \gamma_\mu \, P_L \, \nu] [\bar{c} \, \gamma^\mu \, P_R \, b]
 \eea
with right handed current in the quark sector. We will however ignore dimension-8 operators in the rest of this paper. 


\subsection{Correspondence with section \ref{section2}}

We now expand the various SU(2) structures in order to relate the WCs of the  SU(2)$_{\rm L}$ $\times$ U(1)$_{\rm Y}$ 
invariant operators to those in section \ref{section2}:
\bal
\left(\bar{l^\prime}_{p'} \gamma_\mu \sigma^I l^\prime_{r'} \right) 
\left( \bar{q^\prime}_{s'} \gamma^\mu\sigma^I q^\prime_{t'} \right) &= 
\left(\bar{\nu}^\prime_{p'} \gamma ^{\mu } P_L \nu '_{r'}\right) \left(\bar{u}^\prime_{s'} \gamma_{\mu  } P_L u'_{t'}\right) + 
\left(\bar{e}^\prime_{p'} \gamma ^{\mu} P_L e^\prime_{r'}\right) \left(\bar{d}^\prime_{s'} \gamma _{\mu} P_L d^\prime_{t'}\right)
 \nn \\ 
&-
\left(\bar{e}^\prime_{p'} \gamma ^{\mu } P_L e^\prime_{r'}\right) \left(\bar{u}^\prime_{s'} \gamma _{\mu } P_L u^\prime_{t'} \right)
- \left(\bar{\nu}^\prime_{p'} \gamma ^{\mu } P_L \nu^\prime_{r'} \right) \left(\bar{d}^\prime_{s'} \gamma_{\mu } P_L d^\prime_{t'}\right) \nn \\
&+ 
2 \left(\bar{\nu}^\prime_{p'} \gamma ^{\mu } P_L e^\prime_{r'}\right) \left(\bar{d}^\prime_{s'} \gamma_{\mu} P_L u^\prime_{t'} \right) 
+ 2 \left(\bar{e}^\prime_{p'} \gamma ^{\mu } P_L \nu^\prime_{r'}\right) \left(\bar{u}^\prime_{s'} \gamma _{\mu} P_L d^\prime_{t'}\right)
\label{break-1}\\ 
\left(\phi^\dagger i \overleftrightarrow{D}_\mu^I \phi \right) 
\left( \bar{l^\prime}_{p'} \, \sigma^I \gamma^\mu \, l^\prime_{r'} \right)  
&=  \bigg[  - \frac12 \frac{g_2}{{\rm cos} \theta_W}  Z_\mu \left(\overline{\nu'}_{p'} \gamma ^{\mu } 
P_L \nu '_{r'}\right) + \frac12 \frac{g_2}{{\rm cos} \theta_W}  Z_\mu \left(\overline{e'}_{p'} 
\gamma ^{\mu } P_L e '_{r'}\right) \nn \\
& \hspace{-3cm} - \frac{g_2}{ \sqrt{2}} W^+_\mu \left(\overline{\nu'}_{p'} \gamma ^{\mu } P_L e '_{r'}\right) 
- \frac{g_2}{ \sqrt{2}}  W^-_\mu \left(\overline{e'}_{p'} \gamma ^{\mu } P_L \nu '_{r'}\right)  \bigg] \left( v^2+  2v h +h^2 \right) \label{break-5} \, . 
\eal
The scalar and tensor operators can be decomposed similarly.
It is clear that, as a consequence of the manifest SU(2)$_{\rm L}$ $\times$ U(1)$_{\rm Y}$ gauge invariance, the operators 
relevant for the explanation of the $R_D$ and $R_{D^*}$ anomalies get related to other operators, in particular, to operators 
that give rise to neutral current decays.  However, in order to understand these correlations more concretely, we have to first 
rotate the fields from the gauge to the mass eigenstates.

\subsubsection*{From the gauge to the mass eigenstates:}

We introduce the following mixing matrices which relate the gauge and mass eigenstates 
\bal
(e_{L,R})_{r'} &= (V_{L,R}^e)_{r'r} (e_{L,R})_{r}, ~~~
(\nu_{L,R})_{r'} = (V_{L,R}^\nu)_{r'r} (\nu_{L,R})_{r} \, , \nn \\
(u_{L,R})_{r'} &= (V_{L,R}^u)_{r'r} (u_{L,R})_{r}, ~~~
(d_{L,R})_{r'} = (V_{L,R}^d)_{r'r} (d_{L,R})_{r}  \label{rot-matrix}
\eal

The CKM and PMNS matrices are defined as 
\bal
\label{ckmdef}
V_{\rm CKM} &= (V_{L}^u)^\dagger V_{L}^d, ~~~V_{\rm PMNS} = (V_{L}^\nu)^\dagger V_{L}^e \, .
\eal

Using the above definition of the mixing matrices, we get (see appendix~\ref{app:gauge2mass})
%
\begin{eqnarray}
&& \Delta C_{\rm VL}^{cb\tau\nu_3} = \frac{\Lambda_{\rm SM}^2}{\Lambda^2}
\bigg[[\tilde C_{l q}^{ (3) e \nu u d }]_{3323} + 
\big( [\tilde C_{l q}^{ (3) \nu e d u}]_{3332}  \big)^* \bigg]  \nn \\
&& \hspace{1.6cm} - 
\frac{\Lambda_{\rm SM}^2}{\Lambda^2}  \bigg[[\tilde C_{\phi l}^{ (3) e \nu }]_{33} + 
\big( [\tilde C_{\phi l}^{ (3) \nu e}]_{33}  \big)^*  \bigg] V_{cb}  \,  \label{Vcb-suppression}\\
&& \Delta C_{\rm AL}^{cb\tau\nu_3} =  - \Delta C_{\rm VL}^{cb\tau\nu_\tau} 
%
%
%
\end{eqnarray}
Similar relations can also be found for the scalar and tensor operators, see appendix~\ref{app:gauge2mass}. 
The $[{\tilde C}]$ couplings are related to the $[C]^\prime$ couplings of section \ref{sec:dim6-op} by appropriate mixing 
matrices. For example, 
\begin{eqnarray}
\sum_{p',r',s',t'} [C_{lq}^{(3)}]^\prime_{p'r's't'} \vnd{L}{p} \vn{L}{r} 
\vud{L}{s} \vu{L}{t}  &\equiv& \clqT{\nu}{\nu}{u}{u} \, \, 
\end{eqnarray}
see appendix~\ref{app:gauge2mass} for more details.

Note that the operator $\left(\phi^\dagger i \overleftrightarrow{D}_\mu^I \phi \right) \left( \bar{l^\prime}_{p'} \, 
\sigma^I \gamma^\mu \, l^\prime_{r'} \right) $ modifies the leptonic charged and neutral current vertices 
of $W$ and $Z$ bosons respectively (see Eq.~\ref{break-5}). So in order to explain the $R_{D^{(*)}}$ data by this operator, lepton non-universality 
has to be introduced at these vertices. However, a strong bound on such lepton non-universality exists from LEP \cite{ALEPH:2005aa}:
\bal
\label{eq:tau}
\dfrac{\text{Br}(W^+ \to \tau^+ \nu)}{[\text{Br}(W^+ \to \mu^+ \nu) + \text{Br}(W^+ \to e^+ \nu)]/2} = 1.077 \pm 0.026 \, .
\eal
This means that the branching ratio of $W^+ \to \tau^+ \nu$ can exceed that of $W^+ \to \mu^+ \nu$ or $W^+ \to e^+ \nu$ 
at most by 10.3\% at $1\sigma$. Thus the correction to the $W\tau \nu$ vertex can at most be 5\% of the SM, assuming that 
the $W \bar{\mu} \nu$ and $W \bar{e} \nu$ vertices have no NP.  This gives 
(using Eq.~\ref{break-5} and appendix \ref{app:gauge2mass})\footnote{The SM couplings are defined by
\bea
{\cal L}_{W\tau\nu}^{\rm SM} =   
- \frac{g_2}{\sqrt{2}} g_W^\tau \left( W_\mu^- \bar{\tau} \gamma^\mu P_L \nu_\tau + W_\mu^+ \bar{\nu_\tau} \gamma^\mu P_L \tau \right) \, ,
\eea
where $g_W^\tau = 1$. Any deviation from the SM will be denoted by $\Delta g_W^\tau$.
 }
\bea
- \bigg[ \left([\tilde C_{\phi l}^{ (3) e \nu }] + [\tilde C_{\phi l}^{ (3) \nu e}]^\dagger\right)_{33}  \bigg] \frac{v^2}{\Lambda^2} \lesssim 0.05 \,, \label{Wtaunu-bound}
\eea
which, from the second line of Eq.~\ref{Vcb-suppression}, implies 
\bea
\Delta C_{\rm VL}^\tau = - \Delta C_{\rm AL}^\tau < 0.05\, ,
\eea
where we have used $v^2 = 1/(\sqrt{2} G_F) = \Lambda_{\rm SM}^2 V_{cb} \approx (246 ~ \rm GeV)^2 $.

As we saw in section \ref{res:VA}, $\Delta C_{\rm VL}^\tau = - \Delta C_{\rm AL}^\tau < 0.05$ is not enough to explain the 
$R_{D^{(*)}}$ data within their $1 \, \sigma$ experimental ranges.  
Moreover, as can be seen from Eq.~\eqref{Vcb-suppression}, 
contribution of this operator to the WCs $C_{\rm VL}^\tau = -C_{\rm AL}^\tau$ is suppressed by $V_{cb}$ compared to the other 
contribution. 
This operator also modifies the $Z_\mu \bar{\tau} \gamma^\mu P_L \tau$ coupling, and the modification is given 
by\footnote{We define the SM couplings to be 
\bea
{\cal L}_{Z\tau\tau}^{\rm SM} =   
- \frac{g_2}{\cos\theta_W} Z_\mu \left( g_L^\tau \bar{\tau} \gamma^\mu P_L \tau + g_R^\tau \bar{\tau} \gamma^\mu P_R \tau \right) \, ,
\eea
where $g_L^\tau = -1/2 + \sin^2\theta_W \approx -0.27$ and $g_R^\tau = \sin^2\theta_W \approx 0.23$.  The vector and axial-vector couplings are defined by 
$g_{V, A}^\tau = g_L^\tau \pm g_R^\tau$. }
\bea
\label{delta-gL}
\Delta g_L^\tau =  \frac{1}{2} \,  
\bigg[ \left([\tilde C_{\phi l}^{ (3) e \nu }] + [\tilde C_{\phi l}^{ (3) \nu e}]^\dagger\right)V_{\rm PMNS}  \bigg]_{33}  \, \frac{v^2}{\Lambda^2}\, \, , \, \, \Delta g_R^\tau = 0. 
\eea
Using the experimental constraint from LEP \cite{ALEPH:2005ab} , and assuming that there is no NP in the decays to light leptons, we 
get\footnote{Here we have assumed $(V_{\rm PMNS})_{33} = 1$. However, given the strong experimental constraints,
our results will not
change if correct values of $(V_{\rm PMNS})_{13}$ and $(V_{\rm PMNS})_{23}$ are used.}
\bea
\label{Ztautau-bound}
|\Delta g_L^\tau| \lesssim 6\times 10^{-4}  , ~~&\Rightarrow& \left| \left([\tilde C_{\phi l}^{ (3) e \nu }] + [\tilde C_{\phi l}^{ (3) \nu e}]^\dagger \right)_{33}\right| 
\lesssim 0.02 \left(\frac{\Lambda^2}{\hbox{TeV}^2}\right) \\
~~&\Rightarrow& \Delta C_{\rm VL}^\tau \lesssim 0.001 \, .
\eea
Similarly, 
\bea
\label{delta-gLnu}
\Delta g_L^\nu =  -\frac{1}{2} \,
\bigg[ V_{\rm PMNS} \left( [\tilde C_{\phi l}^{ (3) e \nu }] + [\tilde C_{\phi l}^{ (3) \nu e}]^\dagger\right)V_{\rm PMNS}^\dagger  \bigg]_{33}  \, \frac{v^2}{\Lambda^2}\, 
\eea
which should be compared with the experimental constraint \cite{ALEPH:2005ab}
\bea
\label{Znunu-bound}
|\Delta g_L^\nu| \lesssim 1.2 \times 10^{-3} \, .
\eea
Given these constraints, it is clear that the operator
${\cal O}_{\phi l}^{(3)}$ alone
is unable to explain the $R_{D,D^{*}}$ data. We will thus not consider this operator anymore in this work. 
 Before closing this section, we would like to mention that a much stronger indirect constraint (compared to Eq.~\ref{Wtaunu-bound})
on the $W \tau \nu$ coupling 
can be obtained from measurements of leptonic tau decays assuming that no other four-fermion operator that can either contribute to
$\tau \to e \, \nu \, \bar \nu$ or $\mu \to e \, \nu \, \bar \nu$ exists\cite{Pich:2013lsa}. Assuming no NP in the $W \mu \, \nu$ vertex, this gives, 
\bea
\label{eq:pich}
-0.4 \times 10^{-3} \lesssim -\bigg[ \left([\tilde C_{\phi l}^{ (3) e \nu }] + [\tilde C_{\phi l}^{ (3) \nu e}]^\dagger\right)_{33}  \bigg] \frac{v^2}{\Lambda^2}
\lesssim 2.6 \times 10^{-3} \,
\label{Wtaunu-bound-2} \, . 
\eea

\subsection{Correlations}
\label{sec:corr}
The linearly realised SU(3) $\times$ SU(2)$_{\rm L}$ $\times$ U(1)$_{\rm Y}$ symmetry leads to various correlations
among the flavour violating neutral and charged current observables. To start with, we assume that only the operator(s) that is (are)
needed for the explanation of the $R_{D,D^*}$ anomalies is (are) present. In particular, we consider the operator of Eq.~\ref{dim6-1}
and investigate the correlations arising from it.
We will not consider the scalar and tensor operators anymore because, as discussed in the previous sections, an explanation
of $R_{D,D^*}$ anomalies by scalar (or a combination of scalar and tensor) operators is strongly disfavoured by the upper bound
on Br($B_c \to \tau \, \nu$), and it is rather difficult to generate only the tensor operator.

Without loss of generality, we now go to a basis where the left-chiral down quarks and left-chiral charged leptons are in the 
mass basis. 
This amounts to setting
\begin{equation}
V_L^e = \dblone_{3 \times 3} \quad , \quad \quad V_L^d = \dblone_{3 \times 3} \, .  \label{VL-basis}
\end{equation}
It should be emphasised that we are not making any assumption here by going to a particular basis. 
This just means that our primed WCs of section \ref{sec:dim6-op} are defined in this basis.
In this basis, we have
\begin{equation}
V_{\rm CKM} = V_L^{u \dagger} \quad ,  \quad \quad V_{\rm PMNS} = V_L^{\nu \dagger} \, .
\end{equation}

Let us first consider the contribution to the operator $\left(\bar{\tau} \gamma ^{\mu} P_L \nu \right) \left(\bar{c} \gamma _{\mu} P_L b\right)$. 
From Eq.~\eqref{op1-massbasis} and \eqref{op1-massbasis-def}, one can read off the coefficient of this operator. 
For simplicity, we assume that the NP Wilson coefficients, $[C_{lq}^{(3)}]'_{p'r's't'}$, are diagonal in the Lepton flavours. We get 
\bal
& -2\left( [\tilde{C}_{lq}^{(3) e \nu u d}]_{3r23} + ([\tilde{C}_{lq}^{(3) \nu e d u}]_{r332})^* \right)   \left(\bar{\tau} \gamma ^{\mu} P_L \nu_r \right) \left(\bar{c} \gamma _{\mu} P_L b\right) \\ 
&= -2 \left( ([C_{lq}^{(3)}]'_{3313} + ([C_{lq}^{(3)}]'_{3331})^*)  V_{cd}  +  ([C_{lq}^{(3)}]'_{3323} + ([C_{lq}^{(3)}]'_{3332})^*) V_{cs} 
 \right.
  \nn \\
&\left.+ ( [C_{lq}^{(3)}]'_{3333} + ([C_{lq}^{(3)}]'_{3333})^* ) V_{cb} \right) \times [V_{\rm PMNS}^\dagger]_{3 r} \left(\bar{\tau} \gamma ^{\mu} P_L \nu_r \right) \left(\bar{c} \gamma _{\mu} P_L b\right)  
\label{rdrds-dim6-coup}
\eal

Note that $\nu_{L\tau} = [V_{\rm PMNS}^\dagger]_{3 r} \, \nu_{L \, r}$, $\nu_{L\tau}$ being the $\tau$-flavour neutrino. As we discussed in the previous
sections, in order to explain the anomalies at the $1 \, \sigma$ level, the coefficient of the operator 
$\left(\bar{\tau} \gamma ^{\mu} P_L \nu_\tau \right) \left(\bar{c} \gamma _{\mu} P_L b\right)$ in Eq.~\ref{rdrds-dim6-coup} should at least be
$\sim 0.16$ for a NP scale $\Lambda = \Lambda_{\rm SM}$. This gives, 
\bea
\label{eq:rdlinear}
&& \hspace*{-1cm} ([C_{lq}^{(3)}]'_{3313} + ([C_{lq}^{(3)}]'_{3331})^*)  V_{cd}  +  ([C_{lq}^{(3)}]'_{3323} + ([C_{lq}^{(3)}]'_{3332})^*) V_{cs} 
+ ( [C_{lq}^{(3)}]'_{3333} + ([C_{lq}^{(3)}]'_{3333})^* ) V_{cb} \nonumber\\ 
&&\hspace*{10cm} \gtrsim 0.06 \left(\frac{\Lambda^2}{\hbox{TeV}^2}\right) \, . 
\eea

We would now like to understand whether this condition is consistent with the other measurements of $B$ meson decays.
We first consider the decay $B \to K^* \bar{\nu} \nu$, or in other words, the operator
$\left(\bar{\nu} \gamma^{\mu} P_L \nu \right) \left(\bar{s} \gamma _{\mu} P_L b\right)$. 
Contribution to this operator is given by
\bal
&  \left( [C_{lq}^{(3)}]'_{p'r'23} + [C_{lq}^{(3)}]^{'\, ^\ast}_{r'p'32}\right) 
\left(\bar{\nu}_{p'} \gamma^\mu P_L \nu_{r'} \right) \left(\bar{s}\gamma_\mu P_L b\right) \, .\label{bsnunu-dim6-coup}
\eal

Experimental bound on Br($B^0 \to K^{*\,0}  \bar \nu \nu$) \cite{Grygier:2017tzo} then requires the Wilson coefficients to satisfy (the
SM prediction is taken from \cite{Buras:2014fpa}),
\bea
\label{bknunu-bound}
- 0.005 \left(\frac{\Lambda^2}{\hbox{TeV}^2}\right)  \lesssim [C_{lq}^{(3)}]'_{3323} +[C_{lq}^{(3)}]^{'\,*}_{3332} \leq 0.025 \left(\frac{\Lambda^2}{\hbox{TeV}^2}\right) \, . 
\eea
The first term on the left hand side of Eq.~\ref{eq:rdlinear} contributes to $b \to d \, \bar{\nu} \, \nu$ processes. Using the experimental
bound on the Br($B^0 \to  \pi^0  \bar\nu \, \nu$) \cite{Grygier:2017tzo} and the corresponding SM prediction from \cite{Hambrock:2015wka}, 
we obtain 
\bea
\label{bdnunu-bound}
-0.018 \left(\frac{\Lambda^2}{\hbox{TeV}^2}\right) \lesssim 
[C_{lq}^{(3)}]'_{3313} +[C_{lq}^{(3)}]^{' \,*}_{3331}  \lesssim 0.023 \left(\frac{\Lambda^2}{\hbox{TeV}^2}\right) \, . 
\eea
{ 
The same coupling can also be constrained by measurement of Br($B_u \to \tau \, \nu_\tau$). Assuming the maximum allowed value of
Br($B_u \to \tau \, \nu_\tau$) to be twice that of the SM \cite{Manoni:2017lxj} and, in the absence of cancellations (see
Eq.~\ref{butaunu-dim6-coup} below), we get 
\bea
-0.15 \left(\frac{\Lambda^2}{\hbox{TeV}^2}\right) \lesssim 
[C_{lq}^{(3)}]'_{3313} +[C_{lq}^{(3)}]^{' \,*}_{3331}  \lesssim 0.025 \left(\frac{\Lambda^2}{\hbox{TeV}^2}\right) \, . 
\eea
}
Thus, the maximum contribution from the first two terms of Eq.~\ref{eq:rdlinear} subject to the constraints in
Eqs.~\ref{bknunu-bound} and \ref{bdnunu-bound} is $\approx 0.03 \left(\Lambda^2/\hbox{TeV}^2\right)$. 
This requires 
\bea 
\label{eq:tttt}
( [C_{lq}^{(3)}]'_{3333} + [C_{lq}^{(3)}]^{'\, *}_{3333}) V_{cb}  \gtrsim 0.03 \left(\frac{\Lambda^2}{\hbox{TeV}^2}\right) \, .
\eea
See also \cite{Capdevila:2017iqn} for a related discussion. We now investigate whether the coupling 
$( [C_{lq}^{(3)}]'_{3333} + [C_{lq}^{(3)}]^{'\, *}_{3333})$ is constrained by other measurements. First of all, notice that the coefficient of the 
operator $\left(\bar{\tau} \gamma ^{\mu} P_L \tau\right) \left(\bar{b} \gamma _{\mu} P_L b\right)$ is given by 
\bal
[\tilde{C}_{lq}^{(3) e e d d}]_{3333} +([\tilde{C}_{lq}^{(3) e e d d}]_{3333})^* 
& =  [C_{lq}^{(3)}]'_{3333} +  ([C_{lq}^{(3)}]'_{3333})^*  \, . 
\label{eqn:clq}
\eal
Direct searches of processes involving two $\tau$ leptons in the final state constrain this coupling weakly\cite{Faroughy:2016osc}:
\bea
\label{eq:bbtt}
 \left| [C_{lq}^{(3)}]'_{3333} + ([C_{lq}^{(3)}]'_{3333})^* \right| < 2.6 \left(\frac{\Lambda^2}{\hbox{TeV}^2}\right) \,. 
\eea
Moreover, as we show below, the same coupling also appears in the coefficient of the operator 
$\left(\bar{\tau} \gamma ^{\mu} P_L \tau\right) \left(\bar{t} \gamma _{\mu} P_L t\right)$. From Eq.~\eqref{op1-massbasis} and 
\eqref{op1-massbasis-def} we get, for the coefficient of this operator,  \bea
 \label{t-t-tau-tau-0}
 \left[\tilde{C}_{lq}^{(3)eeuu}\right]_{3333}  + \left[\tilde{C}_{lq}^{(3)eeuu}\right]^*_{3333}
\eea
where, 
\bea
& &\left[\tilde{C}_{lq}^{(3)eeuu}\right]_{3333} = 
 \left[C_{lq}^{(3)}\right]'_{p'r's't'} \left(V_L^e\right)^\dagger_{3p'} \left(V_L^e\right)_{r'3} \left(V_L^u\right)^\dagger_{3s'} \left(V_L^u\right)_{t'3}  \nn \\
& = &  \left[C_{lq}^{(3)}\right]'_{3333} \left| V_{tb}\right|^2 + 
          \left[C_{lq}^{(3)}\right]'_{3323}  V_{ts} V_{tb}^*  + 
           \left[C_{lq}^{(3)}\right]'_{3332}  V_{tb} V_{ts}^*  + \nn \\
  &&   \left[C_{lq}^{(3)}\right]'_{3311} \left| V_{td}\right|^2 + 
           \left[C_{lq}^{(3)}\right]'_{3313}  V_{td} V_{tb}^*  + 
         \left[C_{lq}^{(3)}\right]'_{3331}  V_{tb} V_{td}^*  + \nn \\
  &&   \left[C_{lq}^{(3)}\right]'_{3322} \left| V_{ts}\right|^2 + 
        \left[C_{lq}^{(3)}\right]'_{3312}  V_{td} V_{cb}^*  + 
            \left[C_{lq}^{(3)}\right]'_{3321}  V_{cb} V_{td}^* \, .  \label{t-t-tau-tau-1}
\eea
The 2nd, 3rd, 5th and 6th terms in Eq.~\ref{t-t-tau-tau-1} are small because of Eqs.~\ref{bknunu-bound} and \ref{bdnunu-bound}. 
{ All the other terms which are of the form $ [C_{lq}^{(3)}]'_{33ij}$, $i,j=1,2$ are constrained by direct searches of $\tau \, \tau$ final state
(taking into account the enhancement of the di-jet $ \to \tau \tau$ cross-section compared to that of $\bar b \, b \to \tau^+ \, \tau^-$ due
to larger parton distribution functions, we get
bounds which are stronger than Eq.~\ref{eq:bbtt} by a factor of $\sim 2$ for  ($ [C_{lq}^{(3)}]'_{3322} + \rm c.c$) to a factor of $\sim 8$ for
($[C_{lq}^{(3)}]'_{3311} + \rm c.c$)).}
Thus, the only term which remains is of the form 
\bea
 \left( \left[C_{lq}^{(3)}\right]'_{3333} +\left[C_{lq}^{(3)}\right]^{\prime \, *}_{3333}  \right) \left| V_{tb}\right|^2 .
\label{Ztautau-coupling}
\eea
This term, once the top quarks in the operator $\left(\bar{\tau} \gamma ^{\mu} P_L \tau\right) \left(\bar{t} \gamma _{\mu} P_L t\right)$ form a loop and 
are attached to $Z$, contributes to $\Delta g_L^\tau$ \cite{Feruglio:2016gvd}. As $\Delta g_L^\tau$ is very strongly constrained, see Eq.~\ref{Ztautau-bound}, 
this provides a stringent constraint on the coupling of Eq.~\ref{Ztautau-coupling}. Indeed, from Eq.~\ref{eq:running} in appendix \ref{Ztautau-running}, we find 
{
\bea
\left| [C_{lq}^{(3)}]'_{3333} + [C_{lq}^{(3)}]^{' \, *}_{3333} \right| \lesssim \frac{0.017}{V_{cb}}\left(\frac{\Lambda}{\text{TeV}}\right)^2
\frac{1}{1+0.6\log\frac{\Lambda}{\text{TeV}}}\, , 
\eea
}which clearly rules out the possibility of explaining $R_{D, D^*}$ anomalies by this term, unless there are 
other contributions to the modifications of the $Z$ couplings making it compatible with the experimental observations.

It is worth mentioning that in our discussion so far we have made the assumption that no other operator is present except the one required 
for the $R_{D,D^*}$ anomaly. The presence of other operator(s) can however help evade some of these constraints\cite{Buttazzo:2017ixm}.
For example, one possibility is to assume the presence of the singlet operator
\bea
\label{eq:singop}
{\cal L}^{\rm dim6} \supset -\frac{1}{\Lambda^2} \sum_{p'r's't'} & [C_{lq}^{(1)}]^\prime_{p'r's't'} 
 \left(\bar{l^\prime}_{p'} \gamma_\mu l^\prime_{r'} \right) 
\left( \bar{q^\prime}_{s'} \gamma^\mu  q^\prime_{t'} \right)  + \rm h.c.
\eea 
which, for appropriate values of the WC, can cancel the large contribution from the triplet operator both in $b \to s \, \bar \nu \, \nu$ 
\cite{Buttazzo:2017ixm}  and $\Delta g_L^\tau$\cite{Feruglio:2016gvd,Feruglio:2017rjo}.  However,  NP contributions to $\Delta g_L^\tau$,
$\Delta g_L^\nu$ and $\Delta g_W^\tau$ cannot be cancelled simultaneously. This can be understood in the following way. Note that, while the operator
$[{\cal O}_{lq}^{(3)}]^\prime$ generates the operator $[{\cal O}_{\phi l}^{(3)}]^\prime$ (and not $[{\cal O}_{\phi l}^{(1)}]^\prime$) through RG running, the operator
$[{\cal O}_{lq}^{(1)}]^\prime$ generates the operator $[{\cal O}_{\phi l}^{(1)}]^\prime$ (and not $[{\cal O}_{\phi l}^{(3)}]^\prime$).
The operator $[{\cal O}_{\phi l}^{(3)}]^\prime$ 
contributes to $\Delta g_L^\tau$ and $\Delta g_L^\nu$ with opposite signs (see Eq.~\ref{break-5})while $[{\cal O}_{\phi l}^{(1)}]^\prime$ 
contributes to them with the same sign. { Thus, taking into account constraints from  $\Delta g_L^\nu$ , $\Delta g_L^\tau$ and $\Delta g_W^\tau$, we get
(see appendix \ref{Ztautau-running} for details)
\bea
\left| [C_{lq}^{(3,1)}]'_{3333} + [C_{lq}^{(3,1)}]^{' \, *}_{3333} \right| \lesssim \frac{0.025}{V_{cb}} \left(\frac{\Lambda}{\text{TeV}}\right)^2
\frac{1}{1+0.6\log\frac{\Lambda}{\text{TeV}}}\, .
\label{eq:boundgZ}
\eea
}
This makes  the explanation
of the $R_{D,D^*}$ anomalies by the third term of Eq.~\ref{eq:rdlinear} impossible even in the presence of the operator of
Eq.~\ref{eq:singop}.
This leaves us with two possibilities :
\begin{enumerate}
\item[I.] The anomaly is explained by the second term in Eq.~\ref{eq:rdlinear}. The tension with the Br($B^0 \to K^{*\,0}  \bar \nu \nu$) in
Eq.~\ref{bknunu-bound} is assumed to be cured by cancellation against the contribution of the operator in Eq.~\ref{eq:singop}. 
 However, in this case, the flavour structure of the BSM sector must be such that the last term of Eq.~\ref{eq:rdlinear} is smaller
than the second by at least factor of $\sim 3$. 
\item[II.]  The other possibility is to assume the presence of appropriate UV contribution at the matching scale that takes care of
the $\Delta g_{L}^{\tau,\nu}$ constraints. In this case,  one can explain the anomalies by the third term of Eq.~\ref{eq:rdlinear} alone.
\end{enumerate}
{ As we will see in section \ref{NoLeptoquark} and \ref{Leptoquark}, where we study the explanation of the $R_{D,D*}$ anomalies within
the partial compositeness framework, elements of both the above mechanisms can in principle be present there. 

Before concluding this section, we would like to comment on a few observables which do not provide relevant constraints at this moment, but may
become important in the future.}
\begin{itemize}
\item $b\to s \tau \tau$ transition:
The coefficient of the operator 
$\left(\bar{\tau} \gamma ^{\mu} P_L \tau\right) \left(\bar{s} \gamma _{\mu} P_L b\right)$ is given by
\bal
& - ([C_{lq}^{(3)}]'_{3323} + ([C_{lq}^{(3)}]^{'\, *}_{3332})   \label{bstautau-dim6-coup}  \, . 
\eal
Now we assume that we are in scenario (I) where $B \to K^*  \bar \nu \nu$ transition is cancelled by the singlet operator, and 
Eq.~\ref{eq:rdlinear} is saturated by the second term.
{ Then, in the standard notation, we get for the WCs, $\Delta C_9^\tau = - \Delta C_{10}^\tau = -35$, with the corresponding Lagrangian:
\bea
{\cal L}_{b\to s \tau \tau} = - 35 \, \frac{4 G_F}{\sqrt{2}} \, V_{tb} V_{ts}^* \, \frac{\alpha_{em}}{4\pi} \, 
[\bar{\tau} \gamma^\mu (1-\gamma_5) \tau ] [\bar{s}\gamma_\mu P_L b] \, . 
\eea
giving rise to large enhancement in 
$B_s \to \tau^+ \tau^-$ (by a factor of $\sim 50$ compared to the SM in the branching fraction) and $B \to K/K^*  \, \tau^+ \tau^-$ decays
(by a factor of $\sim 60 \, (\text{for} \, K), 75 \, (\text{for} \, K^*)$ compared to the SM in the branching fraction).} It is interesting to note 
that large enhancement in Br($B_s \to \tau^+ \tau^-$) was also proposed as a possible solution to the like-sign 
di-muon charge asymmetry observed in one of the experiments in Tevatron\cite{Bobeth:2011st,Dighe:2012df}.

\item $b\to u \tau \nu$ transition: in this case, the operator 
$\left(\bar{\tau} \gamma ^{\mu} P_L \nu \right) \left(\bar{u} \gamma _{\mu} P_L b\right)$ is generated with the Wilson coefficient:
\bal
& -\left( 2[\tilde{C}_{lq}^{(3) e \nu u d}]_{3r13} +2 ([\tilde{C}_{lq}^{(3) \nu e d u}]_{r331})^* \right)   \left(\bar{\tau} \gamma ^{\mu}
P_L \nu_r \right) \left(\bar{u} \gamma _{\mu} P_L b\right) \nonumber \\ 
&= -2 \left( ([C_{lq}^{(3)}]'_{3313} + ([C_{lq}^{(3)}]'_{3331})^*)  V_{ud}  +  ([C_{lq}^{(3)}]'_{3323} + ([C_{lq}^{(3)}]'_{3332})^*) V_{us} \right. \nn \\
& \left.  \hspace{1cm} + ( [C_{lq}^{(3)}]'_{3333} + ([C_{lq}^{(3)}]'_{3333})^* ) V_{ub} \right) 
 \times \left(\bar{\tau} \gamma ^{\mu} P_L \nu_\tau \right) \left(\bar{u} \gamma _{\mu} P_L b\right)  \label{butaunu-dim6-coup}
\eal
where, we have again used  $\nu_{L\tau} = [V_L^\nu]_{3 r} \nu_r$, and assumed that the NP Wilson coefficients, $[C_{lq}^{(3)}]'_{p'r's't'}$, are 
diagonal in the Lepton flavours. 
Assuming that we have both singlet and triplet operators, the $b\to s(d )\nu\nu$ transitions do not lead to any constraints. Then,
if the second (third) term in Eq.~\ref{eq:rdlinear} is responsible for $R_{D,D^*}$ anomalies (i.e., saturates the inequality), we get
\bea
{\cal L}_{b\to u \tau \nu} &\approx& 
 -0.2 (0.1) \, \frac{4 G_F}{\sqrt{2}} \, V_{ub} \, 
\left(\bar{\tau} \gamma^\mu P_L \nu_{\tau} \right) \left(\bar{u}\gamma_\mu P_L b\right) \, ,
\eea
which leads to approximately 45\% (20\%) increase in Br($B_u \to \tau \, \nu_\tau$). 
{ Instead, if one assumes that the first term in
Eq.~\ref{eq:rdlinear} is responsible for $R_{D,D^*}$ anomalies and saturates the inequality, the corresponding NP coupling 
for  $b \to u \tau \nu$ becomes, 
\bea
{\cal L}_{b\to u \tau \nu} &\approx& 
 -4 \, \frac{4 G_F}{\sqrt{2}} \, V_{ub} \, 
\left(\bar{\tau} \gamma^\mu P_L \nu_{\tau} \right) \left(\bar{u}\gamma_\mu P_L b\right) \, ,
\eea
which is obviously ruled out by experiment. 
Thus, even in the presence of cancellation in $b \to d \, \bar{\nu} \, \nu$, an explanation of 
$R_{D,D^*}$ by the first term in Eq.~\ref{eq:rdlinear} seems very unlikely. }
\end{itemize}

Similar analysis can also be done for the scalar, pseudo-scalar and tensor operators. Since the scalar and pseudo-scalar 
operators alone cannot explain the anomalies because of the strong constraint from $B_c \to \tau \nu$ (see section \ref{res:SP}), 
we do not discuss them anymore. 
The tensor operator, $[C_{lequ}^{(3)}]^{\prime}_{p'r's't'} \left(\bar{l^\prime}_{p'}^j \sigma^{\mu\nu} e^\prime_{r'} \right) \epsilon_{jk} 
\left( \bar{q^\prime}_{s'}^k \sigma_{\mu\nu} u^\prime_{t'} \right)$, on the other hand, is not affected by the process $B_c \to \tau \nu$, 
and generates, along with the charged current operator which is relevant for $R_{D^{(*)}}$, also neutral current operators involving 
up-type quarks.

\section{Going beyond the dimension-6 analysis: partial compositeness}
\label{comp-Higgs}
In the previous section, we illustrated that some other processes e.g., $B \to K^* \nu \, \nu$, $Z\, \tau\, \tau$ and $Z\, \nu\, \nu$ couplings
can provide stringent restrictions on the possible explanations of $R_D$ and $R_{D^*}$ anomalies.  
It would be interesting also to study the correlations with the various $\Delta F =2$ observables where the constraints on NP are
particularly strong. Such an analysis requires specific assumptions on the underlying UV theory, or some power-counting rules. 
As we discussed before, explanations for the $R_D$ and $R_{D^*}$ anomalies call for NP close to the TeV scale, which is 
also expected for the naturalness of the Higgs mass.  
This coincidence of scales advocates for the speculation of a common origin of these two seemingly unrelated phenomena. This motivates us
to consider the Composite Higgs paradigm \cite{Kaplan:1983fs}, and, in particular, the models where fermion masses are generated via 
the Partial Compositeness (PC) mechanism \cite{Kaplan:1991dc}. In fact, recently there has been a lot of effort invested in analysing the $B$-meson
anomalies within this framework \cite{Barbieri:2017tuq,DAmbrosio:2017wis,Barbieri:2016las,Sannino:2017utc,Niehoff:2016zso,
Niehoff:2015iaa,Carmona:2017fsn}, all of which, however, focusses on specific models. 
A novel feature of our study would instead be to carry out the analysis in the EFT language, emphasising the correlations among the 
various observables. In particular, our aim would be to identify the key features that these models should possess in order to satisfy 
the experimental data. Our main results will be independent of the concrete realisation of PC, and are thus expected to be quite generic.
%


\subsection{Two-site Lagrangian }
\label{App_NoLeptoquark}
In this subsection we will briefly sketch the minimal Composite Higgs construction (for the details see the orginal paper \cite{Contino:2003ve}  and reviews \cite{Contino:2010rs,Panico:2015jxa}) and the familiar reader can directly proceed to the subsection \ref{NoLeptoquark}. The global symmetry breaking pattern is taken as follows:
\begin{equation}
\label{MCH}
\text{MCHM}\,:\, \rm{U(1)}_X \times {\rm SU(3)} \times \rm{SO(5) }\rightarrow U(1)_X \times \rm{SU(3)} \times SU(2)_L \times SU(2)_R.
\end{equation}
We will study the phenomenology  within effective field theory approach, using  so called two site model \cite{Contino:2006nn}. 
The model consists of  two sectors: the composite sector invariant under $ \rm SO(5)\times SU(3)\times U(1)_X$ and the elementary sector invariant under $\rm SU(2)\times SU(3)\times U(1)_Y$. The SM gauge symmetry is identified with the diagonal subgroup, where the ``composite hypercharge"  generator is defined as follows
\bea
T_Y=T_{X}+T^{3}_R.
\eea
The Higgs boson appears as the Goldstone boson of the spontaneous symmetry breaking $\rm SO(5)/SO(4)$. 
We will use the CCWZ formalism \cite{Coleman:1969sm,Callan:1969sn} to 
parametrise the nonlinearly realised symmetry  $\rm SO(5)/SO(4)$ for the composite 
sector (in our discussion we will follow closely the notations of \cite{Contino:2011np}). 
Then 
the Higgs boson appears inside the usual Goldstone boson matrix $U$ which in the unitary gauge is equal to: 
\bea
U=e^{i\Pi(x)}=\left(\begin{array}{ccc}\dblone_{3\times 3}&&\\
&\cos \frac{h}{f}&\sin\frac{h}{f}\\
&-\sin\frac{h}{f} & \cos \frac{h}{f}
\end{array}\right),
\eea
where $h$ is the Higgs boson and the $f$ is the scale of the global symmetry breaking.
It is customary to define two covariant derivatives (Maurer-Cartan 1-form)
\bea
-i U^\dagger D_\mu U=d^{\hat a}_\mu T^{\hat{a}}+ E_\mu^a T^a=d_\mu + E_\mu
\eea
decomposing it along the broken $T^{\hat a}$ and unbroken $T^a$ generators.  The Higgs kinetic term and the mass of the gauge terms come from the two derivative term of the chiral Lagrangian
\bea
\frac{f^2}{4}Tr (d_\mu d^\mu)=\frac{1}{2}(\partial_\mu h)^2+\frac{1}{2}\left(2 m_W^2W_\mu^+ W_\mu^-+m_Z^2 Z_\mu Z^\mu\right)\sin^2\frac{h}{f}.
\eea
\subsubsection{Fermion masses}
Let us proceed to  the fermion mass generation.  For concreteness we  consider  the model where the composite fields appear as a fiveplets of SO(5), i.e. MCHM5 model \cite{Contino:2003ve}.  However we will show explicitly that our results depend only mildly on this assumption and practically do not change for the other  fermion embeddings. 
The fivepletes after the 
$\rm SO(5)/SO(4)$ breaking can be decomposed as a fourplet of SO(4) and a singlet.
The fourplet of SO(4) has in its turn two $\rm SU(2)_L$ doublets: one with the standard model quantum numbers denoted as ${\cal O}_\textbf{SM}$\footnote{We intend that it has the same quantum numbers as the elementary doublet under the $\rm SU(3)\times SU(2)_L \times U(1)_Y$ subgroup. } 
and another one ${\cal O}_\textbf{EX}$, where the doublets are related by $\rm SU(2)_R$ transformations.
The singlet operators are denoted as $\tilde {\cal O}_{u,d,e}$ and the full spectrum is
\bea
\label{spectrum1}
\tilde{\mathcal{O}}_{q_1} =\left( \tilde{\mathcal{O}}^{q_1}_\textbf{EX}\,\tilde{\mathcal{O}}^{q_1}_\textbf{SM}\right)\,,\qquad  \tilde{\mathcal{O}}^{q_1}_\textbf{SM} = \left( \begin{array}{c}
U\\
D
\end{array}\right)\,,\qquad  \tilde{\mathcal{O}}^{q_1}_\textbf{EX} = \left( \begin{array}{c}
\chi_{5/3}\\
\chi_{2/3}
\end{array}\right)\nonumber\\
\hbox{5-plet   ~~}\Psi_{q_1}=\left(\tilde{\mathcal{O}}_{q_1}, \tilde {\cal O}_u\right)\\
\tilde{\mathcal{O}}_{q_2} =\left( \tilde{\mathcal{O}}^{q_2}_\textbf{SM}\,\tilde{\mathcal{O}}^{q_2}_\textbf{EX}\right)\,,\qquad  \tilde{\mathcal{O}}^{q_2}_\textbf{SM} = \left( \begin{array}{c}
U'\\
D'
\end{array}\right)\,,\qquad  \tilde{\mathcal{O}}^{q_2}_\textbf{EX} = \left( \begin{array}{c}
\chi_{-1/3}\\
\chi_{-4/3}
\end{array}\right) \nonumber \\
 \hbox{5-plet   ~~}\Psi_{q_2}=\left(\tilde{\mathcal{O}}_{q_2}, \tilde {\cal O}_d\right)\\
\tilde{\mathcal{O}}_{\ell_1} = \left( \tilde{\mathcal{O}}^{\ell_1}_\textbf{EX}\,\tilde{\mathcal{O}}^{\ell_1}_\textbf{SM}\right)\,,\qquad  \tilde{\mathcal{O}}^{\ell_1}_\textbf{SM} = \left( \begin{array}{c}
N\\
E
\end{array}\right)\,,\qquad  \tilde{\mathcal{O}}^{\ell_1}_\textbf{EX} = \left( \begin{array}{c}
\chi_{+1}\\
\chi_{0}
\end{array}\right)\nonumber\\
\hbox{5-plet   ~~}\Psi_{l_1}=\left(\tilde{\mathcal{O}}_{l_1}, \tilde {\cal O}_N\right) \\
\tilde{\mathcal{O}}_{\ell_2} =\left( \tilde{\mathcal{O}}^{\ell_2}_\textbf{SM}\,\tilde{\mathcal{O}}^{\ell_2}_\textbf{EX}\right)\,,\qquad  \tilde{\mathcal{O}}^{\ell_2}_\textbf{SM} = \left( \begin{array}{c}
N'\\
E'
\end{array}\right)\,,\qquad  \tilde{\mathcal{O}}^{\ell_2}_\textbf{EX} = \left( \begin{array}{c}
\chi_{-1}\\
\chi_{-2}
\end{array}\right)\nonumber\\
\hbox{5-plet   ~~}\Psi_{l_2}=\left(\tilde{\mathcal{O}}_{l_2}, \tilde {\cal O}_e\right)
\eea
where the charges of the components of a $\rm SU(2)_R$ doublet $\tilde{\mathcal{O}} = \left(\tilde{\mathcal{O}}_1,\tilde{\mathcal{O}}_2 \right) $ under $T_R^3$ are equal to $+\frac{1}{2}$ and $-\frac{1}{2}$ respectively.
The elementary fields are denoted as 
 $\tilde{q}_L$, $\tilde{l}_L$, $\tilde{u}_R$, $\tilde{d}_R$, $\tilde{e}_R$ .
Each field is a 3-vector in the flavour generation space and the subscript of $\chi$ field indicates its electric charge.

The elementary $\rm SU(2)_L$ doublet $\tilde{q}_L$ is embedded in the incomplete fiveplet of  SO(5). Thus, the group representations and charges of the fermion states are depicted in table~\ref{Sp}.
Note that we have two composite doublets $\tilde{\mathcal{O}}^{q_1}_\textbf{SM}$ and $\tilde{\mathcal{O}}^{q_2}_\textbf{SM}$ which have the same quantum numbers under the SM gauge group; similarly for the leptons.\\
\begin{table}
\begin{center}
\begin{tabular}{c c c c c}
\hline
 & $SU(3)^\text{co}$ & $SU(2)_L^\text{co}$&$SU(2)_R^\text{co}$ & $U(1)_X^\text{co}$\\
\hline 
 $\tilde{\mathcal{O}}_{q_1}$ & $\textbf{3}$ & $\textbf{2}$& $\textbf{2}$ & $2/3$\\
 \hline 
 $
\tilde{\cal O}_{q_2}$ & $\textbf{3}$ & $\textbf{2}$& $\textbf{2}$ & $-1/3$\\
  \hline 
 $\tilde{\mathcal{O}}_{u}$ & $\textbf{3}$ & $\textbf{1}$& $\textbf{1}$ & $2/3$\\
   \hline 
 $\tilde{\mathcal{O}}_{d}$ & $\textbf{3}$ & $\textbf{1}$& $\textbf{1}$ & $-1/3$\\
 \hline 
 $\tilde{\mathcal{O}}_{\ell_1}$ & $\textbf{1}$ & $\textbf{2}$& $\textbf{2}$ & $0$\\
   \hline 
 $\tilde{\mathcal{O}}_{\ell_2}$ & $\textbf{1}$ & $\textbf{2}$& $\textbf{2}$ & $-1$\\
  \hline 
$\tilde{\mathcal{O}}_{e}$ & $\textbf{1}$ & $\textbf{1}$& $\textbf{1}$ & $-1$\\
\end{tabular}\quad \begin{tabular}{c c c c c}
\hline
 & $SU(3)^\text{el}$ & $SU(2)_L^\text{el}$&${U(1)_Y}^\text{el}$ &\\
\hline 
$\tilde{q}_L$ & $\textbf{3}$ & $\textbf{2}$& $1/6$ &\\
 \hline 
$\tilde{u}_R$ & $\textbf{3}$ & $\textbf{1}$& $2/3$ & \\
  \hline 
$\tilde{d}_R$ & $\textbf{3}$ & $\textbf{1}$& $-1/3$ &\\
  \hline 
$\tilde{\ell}_L$ & $\textbf{1}$ & $\textbf{2}$& $-1/2$ & \\
  \hline 
$\tilde{e}_R$ & $\textbf{1}$ & $\textbf{1}$& $-1$ & \\
  \hline 
$\tilde{\nu}_R$ & $\textbf{1}$ & $\textbf{1}$& $0$ & \\
\end{tabular}
\caption{\sf Group representations and charges of the fermion composite resonances and elementary fields {}}
\label{Sp}
\end{center}
\end{table}
The symmetries of the composite sector are broken explicitly to the diagonal subgroup
 by the mixing with the elementary sector, which is  given by:
\bea
\label{Fmixing}
\mathcal{L}_{flavour} =  \lambda_q M_* \overline{\tilde{q}}_L  U(h)\Psi_{q_1} +\tilde \lambda_q M_* \overline{\tilde{q}}_L U(h) \Psi_{q_2} + \lambda_u M_* \overline{\tilde{u}}_R U(h)\Psi_{q_1} + \lambda_d M_* \overline{\tilde{d}}_R \Psi_{q_2}\nonumber\\
 +\lambda_l M_* \overline{\tilde{l}}_L  U(h)\Psi_{l_1} + \tilde\lambda_l M_* \overline{\tilde{l}}_L U(h) \Psi_{l_2} +
 \lambda_e M_* \overline{\tilde{e}}_R U(h)\Psi_{l_2} 
\eea
where the SM doublets where uplifted to incomplete 5-plets as follows:
\bea
\label{eq:uplift}
\lambda_q \tilde q_L\equiv \lambda_q\left[ (0,q_L),0\right]\nonumber\\
\tilde\lambda_q \tilde  q_L\equiv \tilde\lambda_q\left[ (q_L,0),0\right],
\eea
where we put zeros in all the missing components and $(q_L,0)$ singles out the SO(4) multiplet. Note also that symmetries of the model allow us to further split $\lambda_q$ mixing into two independent parameters
\bea
\label{splitting}
\lambda_q \overline {\tilde{q}}_L U(h)\Psi_{q_1}\to \left \{ \begin{array}{c}
\left[
\lambda^{(4)}_q \overline {\tilde {q}}_L\right]_I U(h)_{I i} \left[{\cal O}_{q_1}\right]_i, \hbox{ where }  I=1,...5,~~i=1,...4\\
\left[\lambda_q^{(1)} \overline {\tilde {q}}_L\right]_I U(h)_{I 5}{\cal  O}_{u}
 \end{array}\right. ,
\eea
where the sum over repeating  indices is understood.
Let us look at the fermion spectrum before EWSSB. Due to the mixing $\lambda$ we will have one massless SM state and one heavy field with the mass 
 $M_*(1+\lambda)/\sqrt{1+\lambda^2}$ , which becomes $M_*$ in the limit $\lambda\ll 1$ .
This leads to the mixing between the elementary and composite states which can be described by the mixing angles defined as follows:
\begin{equation}
\left(\begin{array}{c}
\tilde{\psi}\\
\tilde{\mathcal{O}}
\end{array}\right) = \left(\begin{array}{c c}
\cos \theta_\psi & -\sin\theta_\psi \\
 \sin \theta_\psi & \cos \theta_\psi
\end{array} \right)\left(\begin{array}{c}
\psi'\\
\mathcal{O}
\end{array}\right)
\label{eq:fermmix} 
\end{equation}
with 
\begin{equation}
\sin \theta_\psi \equiv \hat{s}=\dfrac{\lambda}{\sqrt{1+\lambda^2}}, \qquad \cos \theta_\psi\equiv\hat{c}=\dfrac{1}{\sqrt{1+\lambda^2}},
\end{equation}
where $\hat s, \hat c$ are the sine and cosine of the corresponding mixing angles. Then the SM Yukawa coupling will scale as 
\bea
y_{u,d}\sim\frac{ s_{q_{1,2}} s_{u,d}  M_*}{f}.
\eea

\subsubsection{Vector fields}
\label{vector}
We are interested in the interactions between the SM fermions and the composite vector fields.  We will follow the vector formalism \cite{Contino:2011np}  (see for example the Ref.\cite{Ecker:1989yg} for the comparison of various formalisms) for a spin-1 fields  where it is assumed   that the vector fields transform non-homogeneously 
\bea
\tilde \rho_\mu\to  {\cal H} \tilde  \rho_\mu {\cal H}^\dagger - i {\cal H}\partial_\mu {\cal H }^\dagger,
\eea
where $\mathcal{H}$ is unbroken subgroup ($SO(4)$) transformation. Then the following interactions are 
allowed by the CCWZ symmetries:
\bea
\label{Lvec}
{\cal L}^{vec}=-\frac{1}{4 g_*^2}\tilde \rho^a_{\mu\nu}\tilde \rho^{\mu\nu}_a+
\frac{M_*^2}{2 g_*^2}(\tilde \rho_\mu^a-E_\mu^a)^2+...
\eea
where $g_*$ is a strength of interactions between the composite fields
and we have ignored the higher derivative terms. The Lagrangian Eq.(\ref{Lvec}) in the limit of vanishing Higgs vev reduces to
\begin{equation}
\mathcal{L}_{vec} =  -\frac{1}{4}\tilde{\rho}^a_{\mu\nu}\tilde{\rho}^{\mu\nu}_a+\frac{M_*^2}{2}\tilde{\rho}_\mu^a\tilde{\rho}^\mu_a - M_*^2 \frac{g_{el}}{g_*} \tilde{\rho}_\mu^a A^\mu_a + \frac{M_*^2}{2}\frac{g^2_{el}}{g_*^2} A_\mu^a A^\mu_a,
\end{equation}
where $A^\mu_a$ are the elementary vector fields.
The interaction between $\tilde\rho$ and the composite fermions can be deduced from the symmetries 
\bea
\label{L_fermiongauge}
{\cal L}_{ferm}=\bar \Psi \gamma^\mu\left( i \partial_\mu + g_* \tilde  \rho_\mu\right)\Psi,
\eea
where $\tilde{\rho}_\mu =\tilde{\rho}_\mu^a T_a^\text{co} $and $T_a^\text{co} $ are the generators of the global symmetry group of the composite sector\footnote{Of course, we can have different values
of $g_*$ for $SU(3)$, $SO(5)$ and $U(1)_X$ parts of the global group.}.
In order to get mass eigenstates vectors, a diagonalisation of the matrix of masses and mixing is needed
\begin{equation}
\label{eq:vecmix}
\left( \begin{array}{c}
A_\mu\\
\tilde\rho_\mu 
\end{array} \right) \rightarrow \left( \begin{array}{cc}
\cos \theta & -\sin \theta\\
\sin \theta & \cos\theta
\end{array}\right)\left( \begin{array}{c}
A^{SM}_\mu\\
\rho_\mu 
\end{array} \right) \,,\qquad \cos\theta = \frac{g_*}{\sqrt{g_*^2+g_{el}^2}}
\end{equation}
where $\tilde\rho_\mu$ is an eigenstate with mass of $M_*\sqrt{1+g_{el}^2/g_*^2}$ and the orthogonal $A^{SM}_\mu$ is the massless state, that is identified with the SM gauge boson. Rotating to the mass eigenstate basis we get
\begin{equation}
\label{RhoInt}
\bar{\psi'}_i\left[\sqrt{g_*^2-g^2}\left[\hat{s}^\dagger T_a^\text{co} \hat{s}\right]^i_j - 
\frac{g^2}{ \sqrt{g_*^2 - g^2}}\left[\hat{c}^\dagger T_a^\text{el} \hat{c}\right]^i_j \right]\gamma^\mu \psi'^{j}\rho_\mu^a \, , 
\end{equation}
where the first term comes from the mixing of the elementary and composite fermions and the second term corresponds to the mixing between composite and elementary vector bosons 
(Eq.\ref{eq:vecmix}). In this paper we are mainly interested in the flavour non universal and flavour violating effects, so the contribution of the last term will be subleading since $g_*\gg g$ and the non-universalities in $\hat c\sim 1- \hat s^2/2$ have an extra $\hat s$ supression. Note that the Eq. \ref{RhoInt}  is a generic prediction of the partial compositness and the various fermion embeddings lead  only to the  generators 
$T^a_\text{co}$ for the different group representations.

\vspace{-4mm}
\subsection{$R_{D,D^*}$ from the composite electroweak resonances}
\label{NoLeptoquark}

We are interested in the dimension-6 four-fermion operators. These operators are generated by the exchange of the composite 
vector resonances.  Using the Eq.~\ref{RhoInt} and assuming $g_*\gg g$  we can see that
 the these operators at the dimension-6 level  schematically take the form
\begin{equation}
\label{6dimOp}
\frac{g_*^2}{M_*^2} \left[\bar{\psi'}\, \hat{s}^\dagger T_a^\text{co}\hat{s}\, \gamma^\mu  \psi' \right]\left[\bar{\psi'}\,  \hat{s}^\dagger T_a^\text{co} \hat{s}\, \gamma_\mu \psi'\right] \, .
\end{equation}
Our aim would now be to understand the correlations among the flavour-changing $\Delta F =2$ operators and those that contribute to the $R_{D^{(*)}}$
anomalies. 
The effective Lagrangian for the $\Delta F = 2$ transitions can be written as 
\bea
\label{df2}
{\cal L}_{\Delta F=2} = - \text{const} \times \frac{g_*^2}{M_*^2} \left(\bar\psi_{i \, L}\, \left[V_L^{d\dagger}\hat{s}_q^\dagger \hat{s}_qV_L^d\right]^i_{~j}\, 
\gamma^\mu \,\psi_{j \, L} \right)^2 \, ,
\eea
where $V_L^d$ is the rotation matrix for the left-handed  quarks defined in Eq.\ref{ckmdef} and the constant in front for the case of MCHM5 is equal to 
\bea
\text{const} = \frac{M_*^2}{2g_*^2}\left( \frac{1}{3}\frac{g_{*3}^2}{M_{*3}^2} +\frac{1}{2}\frac{g_{*2}^2 }{ M_{*2}^2}+\frac{4}{9} \frac{g_{* X}^2}{M_{* X}^2}\right) \, .
\eea
The first term inside the parenthesis corresponds to the contribution of the composite gluon,
the second to the $\rm SU(2)_{L,R}$ triplets and the third to $\rm U(1)_X$ vector bosons
(the number 4/9 is fixed by the $U(1)_X$ charge assignment of the up-like multiplet $\tilde{O}_{q_1}$, see Eq.\ref{spectrum1}).
Experimental data on $\bar{K}$-$K$, $\bar{B}_d$-$B_d$ and $\bar{B}_s$-$B_s$ mixings give the following constraints\footnote{ Here, we 
have assumed that only one $\Delta F = 2$ operator (the operator $Q_1$ in the basis of \cite{Bona:2007vi}) is generated. In principle, other 
operator(s) may also be generated at the matching scale, and cancel part of the contribution from $Q_1$. However, barring large accidental 
cancellations, our results should always hold.},
\bea
\label{eq:utfit}
\left| \left[V_L^{d\dagger}\hat{s}_q^\dagger \hat{s}_qV_L^d\right]^i_{~j} \right| \lesssim \frac{(M_*/\text{TeV})}{g_*\sqrt{\text{const}}}
\left \{
\begin{array}{c}
{ 10^{-3} } \hbox{~,~from $\bar{K}$-$K$ mixing, i.e., $i=1, j=2$ \cite{Bona:2007vi}}\\
{ 1.1\times 10^{-3} }\hbox{~,~from $\bar{B}_d$-$B_d$ mixing, i.e., $i=1, j=3$ \cite{Carrasco:2013zta}}\\
{ 4 \times 10^{-3} } \hbox{~,~from $\bar{B}_s$-$B_s$ mixing, i.e., $i=2, j=3$ \cite{Carrasco:2013zta} \, , }\\
\end{array}\right. 
\eea
where the numerical values are obtained by running the couplings to the scale $M_*$.
Keeping the above constraints from $\Delta F = 2$ processes in mind, we now look at the $b\to c \tau \nu$ transitions. We assume that the NP contribution
arises from the exchange of a composite vector field which is a triplet of $\rm SU(2)_L$. This generates the interaction Lagrangian
\begin{align} 
\label{Rdmass}
{\cal L}_{b \to c \, \tau \, \nu}  & = 
 {-}\frac{g_{*2}^2}{2M_{*2}^2} \left(\bar{\tau}_L\, \left[V_L^{e\dagger}\hat{s_l}^\dagger \hat{s_l}V_L^{\nu\dagger}\right]_{~3}^3\, \gamma^\mu  \, \nu_{\tau L} \right) 
 \left(\bar{c}_L\, \left[V_L^{u\dagger}\hat{s_q}^\dagger \hat{s_q}V_L^d\right]_{~3}^2\, \gamma^\mu  \, b_L \right) \\
&= {-} \frac{g_{*}^2}{2M_{*}^2}  \left(\bar{\tau}_L\, \left[ V_L^{e\dagger}\hat{s_l}^\dagger \hat{s_l}V_L^{\nu}\right]_{~3}^3\, \gamma^\mu  \, \nu_{\tau L} \right) 
\left(\bar{c}_L\, \left[V_{\text{CKM}}V_L^{d\dagger}\hat{s_q}^\dagger \hat{s_q}V_L^d\right]_{~3}^2\, \gamma^\mu  \, b_L \right) \, , 
\end{align}
where we have assumed $g_{*2}=g_{*}, M_{*2}=M_{*}$ and the roational matrices are defined in Eq.\ref{ckmdef}.
If we decide to remain agnostic about the leptonic sector, we can still use the loose upper bound 
\bea
\left| \left[ V_L^{e\dagger}\hat{s_l}^\dagger \hat{s_l}V_L^{\nu}\right]_{~3}^3 \right| <1
\eea
which is satisfied even for maximal possible $\tau$ compositeness. Thus, for the explanation of $R_D$ and $R_{D^\ast}$ anomalies at the $1\, \sigma$ level , we 
need 
\bea
\label{constraint}
\left[V_{\text{CKM}}V_L^{d\dagger}\hat{s_q}^\dagger \hat{s_q}V_L^d\right]_{~3}^2\gtrsim  0.2 \left(\frac{M_*/g_*}{\text{TeV}} \right)^2  \, , 
\eea
where the numerical factor 0.2 corresponds to $\Delta C_{\rm VL}^\tau = - \Delta C_{\rm AL}^\tau = 0.08$ (see Fig.~\ref{fig:VL-AL}).
Expanding Eq.~\ref{constraint}, we get 
\bea
 V_{cd} \left[V_L^{d\dagger}\hat{s_q}^\dagger \hat{s_q}V_L^d\right]_{~3}^1  + V_{cs} \left[V_L^{d\dagger}\hat{s_q}^\dagger \hat{s_q}V_L^d\right]_{~3}^2
+ V_{cb} \left[V_L^{d\dagger}\hat{s_q}^\dagger \hat{s_q}V_L^d\right]_{~3}^3    &\gtrsim&  0.2 \left(\frac{M_*/g_*}{\text{TeV}} \right)^2   \nn \\
\implies  \left| V_{cd} \right| \left| \left[V_L^{d\dagger}\hat{s_q}^\dagger \hat{s_q}V_L^d\right]_{~3}^1 \right|  + 
\left| V_{cs} \right|  \left| \left[V_L^{d\dagger}\hat{s_q}^\dagger \hat{s_q}V_L^d\right]_{~3}^2 \right|
+ \left| V_{cb} \right|  \left| \left[V_L^{d\dagger}\hat{s_q}^\dagger \hat{s_q}V_L^d\right]_{~3}^3 \right|   &\gtrsim&  0.2 \left(\frac{M_*/g_*}{\text{TeV}} \right)^2  \nn 
\eea
Using the upper bounds on $\left| \left[V_L^{d\dagger}\hat{s_q}^\dagger \hat{s_q}V_L^d\right]_{~3}^1 \right| $ and 
$\left| \left[V_L^{d\dagger}\hat{s_q}^\dagger \hat{s_q}V_L^d\right]_{~3}^2 \right|$ from Eq.~\ref{eq:utfit} and the trivial inequality 
$\left| \left[V^{d\dagger}\hat{s_q}^\dagger \hat{s_q}V^d\right]_{~3}^3 \right| \leq 1$, we now get 
\bea
{ 1.1 \times 10^{-3} } \left|V_{cd} \right|   \, \frac{(M_*/\text{TeV})}{g_*\sqrt{\text{const}}} 
+{  4 \times 10^{-3}}  \left| V_{cs} \right| \,   \frac{(M_*/\text{TeV})}{g_*\sqrt{\text{const}}} 
+ \left| V_{cb}  \right|  \gtrsim  0.2 \left(\frac{M_*/\text{TeV}}{g_*} \right)^2  \label{no-leptoquark}
\eea
As the first two terms are negligibly small compared to the third term (for small $(M_*/\text{TeV})/g_*$) on the left hand side, we finally get 
\bea
M_*/g_* \lesssim ~ 0.45 \, \hbox{TeV} \, \label{eqn:CHMmin}
\eea
Note that partial compositeness automatically selects the scenario (II) (see discussion after Eq.~\ref{eq:boundgZ}) for fitting the $R_{D, D^*}$
anomalies.
This solution, as mentioned in section \ref{sec:corr}, requires the presence of additional UV contributions to protect
$g_{L}^{\tau,\nu}$ couplings of the $Z$ boson. In the appendix \ref{sec:Zint}, we explicitly show how this can be achieved.
Interestingly, the generated operators automatically satisfy the condition of scenario (I) due to the SO(4) structure of the model.

We would like to make a few comments here regarding the robustness of this result and its applicability to the various models employing 
partial compositeness. The only assumption that we have made in deriving the Eq. \ref{eqn:CHMmin} is that the charged current operator 
(see Eq.~\ref{Rdmass}) is generated by a vector field, which is a triplet of electroweak $\rm SU(2)_L$.
 The rest of the discussion is completely
model independent and applies to various embeddings of the SM fermions into the composite multiplets,  choices of the
off-diagonal elementary-composite mixing parameters $\hat s$, and is practically independent of the mass of the composite gluon  and the mass of the $\rm U(1)_X$ vector. 
It should also be emphasised that we have been completely agnostic of the dynamics that allows the model under consideration to satisfy the
constraints from $\Delta F = 2$ processes namely, those given in Eq.~$\ref{eq:utfit}$.
For example, in anarchic partial compositness,  where the left-handed quark mixing parameters scale as the CKM matrix elements
\bea
\left[ \hat{s}_q \right]_{i} \sim  V_{t i}
\eea
these bounds are roughly $M_* \gtrsim 10-20$ TeV \cite{Csaki:2008zd,Agashe:2008uz} \footnote{The strongest constraint in this case comes from the $\epsilon_K$ bound.}
, which is a  too high  
scale to explain the $R_D$ and 
$R_{D^*}$ anomalies.
However the scale of the compositeness can  be lowered and made consistent with the $R_{D^{(*)}}$ anomalies by 
invoking additional flavour symmetries, for example $\rm U(2)$\cite{Barbieri:2011ci,Barbieri:2012tu,Barbieri:2015yvd}.  Interestingly the bounds from the direct searches at the LHC \cite{Sirunyan:2018omb, Aaboud:2018pii} on the composite partners of the top quarks are still in the range of $M_*\gtrsim 1.2$ TeV,  making them consistent with the requirement of Eq.\ref{eqn:CHMmin}

The constraint in Eq.~\ref{eqn:CHMmin}, in general, can pose serious difficulties with the electroweak precision observables and  
measurement of Higgs's couplings to electroweak vector bosons. Indeed the constraints from electroweak precision tests  \cite{Ciuchini:2013pca,Baak:2014ora,deBlas:2016ojx}
require  the scale of compositeness to be $\gtrsim1.2$ TeV in order to satisfy the data at $2\sigma $ level. At the same time 
 the mass of the vector resonance is related to the scale of compositeness, $f$, as 
\bea
M_*^2= a_\rho g_*^2 f^2,
\eea
where $a_\rho$ is a number of ${\cal O}(1)$. In an explicit two-site construction, $a_\rho=1/\sqrt{2}$
(see for example \cite{Panico:2015jxa}) so that  the  compositeness scale is constrained to $f\lesssim 0.64 \, \text{TeV}$. 
This is incompatible with the bound from electroweak precision measurements mentioned above. 
It may however be possible to accommodate the electroweak precision observables by additional UV contributions, see for example, 
\cite{Azatov:2013ura,Grojean:2013qca,Ghosh:2015wiz}. 

The tension with meson mixing data makes it interesting to think of other possibilities of enhancing the contributions to $R_D$ and 
$R_{D^*}$ without modifying the $\Delta F=2$ observables considerably. This can be partially achieved in scenarios with composite vector 
leptoquarks which we discuss in the next section.


\subsection{Leptoquark  contribution}
\label{Leptoquark}
The composite vector leptoquark scenario in connection to the $B$ meson anomalies was first proposed in
\cite{Barbieri:2015yvd,Barbieri:2016las,Barbieri:2017tuq}. In this 
construction, the global symmetry of the composite sector is extended from $\rm SO(5) \times SU(3)$ (where SU(3) is weakly gauged later
and becomes the SU(3) of QCD)  
to $\rm SO(5) \times SU(4)$. The composite gluon, which is an octet of SU(3),
lies inside the {\bf 15} dimensional adjoint of SU(4) and is accompanied by two SU(3) triplets {$\bf 3 + \bar  3$} ($\tilde{V}_{(3,1)_\frac{2}{3}}+\tilde{V}^*_{(\bar 3,1)_{-\frac{2}{3}}}$) and a singlet $(\tilde{B}_{(1,1)_0})$, 
 where the subscripts of vectors indicate the representations
under the $SU(3)\times SU(2)_L \times U(1)_Y$ subgroup. 
The hypercharge is given by the following combination of group
generators: $Y = \sqrt{\frac{2}{3}} T_{15} + T_R^3 + X$ and under the SM gauge group these fields .
The Lagrangian is the same as in section~\ref{App_NoLeptoquark}, apart from the presence of $\tilde{V}_{(3,1)_\frac{2}{3}}$ and
$\tilde{B}_{(1,1)_0}$ vector bosons. In particular, in the composite sector, leptoquarks couple to fermion currents in which there are 
quark and lepton resonances. Indeed, from Eq.~\ref{L_fermiongauge} one gets also the {interaction}
\begin{equation}
\dfrac{g_*}{\sqrt{2}}\tilde{V}_\mu \overline{\tilde{\mathcal{O}}}_{SM}^q \gamma^\mu \tilde{\mathcal{O}}_{SM}^l 
\end{equation}
where $g_*$ is the strong coupling for the $SU(4)$ of the composite sector. This interaction after integrating out the heavy
fermions reduces to 
\begin{equation}
\frac{g_*}{\sqrt{2}}\bar{\psi}'_q  \left[\hat{s_q}^\dagger \hat{s_l}\right]\gamma^\mu \psi_l'\tilde V_\mu \, .
\end{equation}
Here we focus only on the relevant interaction 
term for $R_{D,D^*}$ anomalies, 
\bea
{\cal L}_{LQ}= { - } g_*\left(\bar q_{L \, i}'\left [\hat s_q^\dagger  \hat s_l\right ]^i_{~j} \gamma_\mu \, l_{L\, j}'\right) V_{LQ}^\mu \, .
\eea
Moving to the mass basis the effective Lagrangian for the $b \to c \, \tau \, \nu$ processes can now be written as, 
\begin{align}
 {\cal L}_{b \to c \, \tau \, \nu}  & = { - }
 \frac{g_*^2}{2M_*^2}  \left(\bar{c}_L\, \left[V_{\text{CKM}}V_L^{d\dagger}\hat{s_q}^\dagger \hat{s_l} V_L^\nu\right]_{~3}^2\, \gamma^\mu  \, \nu_{\tau L} \right)
\left(\bar{\tau}_L\, \left[V_L^{e \dagger}\hat{s_l}^\dagger \hat{s_q}V_L^d\right]_{~3}^3\, \gamma^\mu  \, b_L \right) \, \\
& =^{^{\hspace{-4mm} \rm Fierz}} { - }
 \frac{g_*^2}{2M_*^2}  \, \left[V_{\text{CKM}}V_L^{d\dagger}\hat{s_q}^\dagger \hat{s_l} V_L^\nu\right]_{~3}^2\,  
 \left[V_L^{e \dagger}\hat{s_l}^\dagger \hat{s_q}V_L^d\right]_{~3}^3 \, 
 \left(\bar{c}_L\, \gamma^\mu  \, b_L  \right) \left(\bar{\tau}_L \, \gamma^\mu  \,  \nu_{\tau L} \right) \, . \label{RdMassLep}
\end{align}
In order to find the upper bound on the coefficient of the operator 
$ \left(\bar{c}_L\, \gamma^\mu  \, b_L  \right) \left(\bar{\tau}_L \, \gamma^\mu  \,  \nu_{\tau L} \right) $, we need to find an upper bound on 
$\left[V_{\text{CKM}}V_L^{d\dagger}\hat{s_q}^\dagger \hat{s_l} V_L^\nu\right]_{~3}^2$ consistent with the data on $B$ meson mixing. 
As before, we have used the trivial inequality $ \left[V_L^{e \dagger}\hat{s_l}^\dagger \hat{s_q}V_L^d\right]_{~3}^3 \leq 1$.

Without loss of generality, we now go to the basis of elementary  and composite fields  in which the lepton compositeness matrix 
has the following form: 
\bea
\hat s_l=\left(  
\begin{array}{ccc}
*&0&0\\
*&*&0\\
*&*&*
\end{array}\right) \, ,
\eea
where $*$ stands for non-zero entry.
We now assume that only the third family of leptons has a strong mixing with the composite sector i.e. only $(\hat s_l)_{33} \sim 1$ and
the rest of the elements are much smaller. In this case, the WC in Eq.~\ref{RdMassLep} is controlled by,
\bea
\label{eq:LQcontr}
\left[V_{\text{CKM}}V_L^{d\dagger}\hat{s_q}^\dagger\right]^{2}_{~3}=V_{cd} [V_L^{d\dagger}\hat{s_q}^\dagger]^1_{~3} 
+V_{cs} [V_L^{d\dagger}\hat{s_q}^\dagger]^2_{~3}
+V_{cb}[V_L^{d\dagger}\hat{s_q}^\dagger]^3_{~3} \, .
\eea
Our aim now is to understand how big $\left[V_{\text{CKM}}V_L^{d\dagger}\hat{s_q}^\dagger\right]^{2}_{~3}$ can be, consistently with 
an almost diagonal $\left[V_L^{d\dagger}\hat{s_q}^\dagger \hat{s_q}V_L^d\right]$ (as the off-diagonal elements are constrained 
to be $\lesssim 10^{-3}$, see Eq.~\ref{eq:utfit}).
Similar to the leptonic elementary-composite mixing matrix $\hat{s}_l$, we can also make $\hat{s}_q$ triangular by suitable field 
redefinitions of the elementary fields. Thus, without loss of generality, we can write, 
\bea
\hat s_q=\left(  
\begin{array}{ccc}
s_{11}&0&0\\
s_{21}&s_{22}&0\\
s_{31}&s_{32}&s_{33}
\end{array}\right),
\eea
{ Let us now consider the special case where only the third generation quark mixes strongly with the composite sector so that, 
\bea
\label{eq:hirarch}
 s_{33} \gg s_{i j},~~~i \hbox{ or } j \neq 3 \, ,
\eea}
In this case, while $[V_L^{d\dagger}\hat{s_q}^\dagger]^3_{~3}$ can be close to unity, the other terms in Eq.~\ref{eq:LQcontr}, 
in order to be consistent with a diagonal $\left[V_L^{d\dagger}\hat{s_q}^\dagger \hat{s_q}V_L^d\right]$,  must scale as
{ 
\bea
[V_L^{d\dagger}\hat{s_q}^\dagger]^1_{~3}\sim \frac{s_{31}(s_{ij})^2}{(s_{33})^2},~~
[V_L^{d\dagger}\hat{s_q}^\dagger]^2_{~3}\sim \frac{s_{32}(s_{ij})^2}{(s_{33})^2} \, .
\eea
}
It can be noticed that these elements have an additional suppression of $(s_{ij}/s_{33})$ compared to the naive expectation. This
renders the contributions of the first two terms of Eq.~\ref{eq:LQcontr} subdominant. Thus, adding the contribution of the
electroweak triplet from Eq.~\ref{no-leptoquark}, for the explanation of $R_{D,D^*}$ anomalies at the $1\, \sigma$ level, we must have
{
\bea
 2 V_{cb} &\gtrsim & 0.2 \left(\frac{M_*/g_*}{\text{TeV}} \right)^2  \nonumber \\
\implies M_*/g_* &\lesssim & 0.63 \, \text{TeV} \, , \label{leptoq-final}
\eea
where we have assumed that the electroweak triplet and the leptoquarks have the same mass and coupling.}
Hence, the role of leptoquarks is just to double the contribution to $R_{D,D^*}$ without worsening the other low energy observables. 
This increase of the upper bound on the scale of compositeness by a factor of $\sqrt{2}$ helps ameliorate the constraints
from $S$ and $T$ parameters which are now in agreement at almost $2\,\sigma$ level
\footnote{Note that if we assume $\frac{g_{*4}} {M_{*4}} > \frac{g_{*2}} {M_{*2}}$ i.e., smaller masses of the SU(4) resonances
than those of the SO(4) fields, we can be in the situation where the composite leptoquark contribution dominates in $R_{D, D^*}$
and the tension with electroweak precision observables can be relaxed even further.}.


It is worth emphasising that the result of Eq.~\ref{leptoq-final} was derived assuming the hierarchical nature (see Eq.~\ref{eq:hirarch}) of the
mixing matrix ${\hat s}_q$ and the constraint of Eq.~\ref{leptoq-final} can be relaxed if this assumption is not valid. For example, if we 
assume that the matrix $\hat{s_q}$ is not hierarchical but unitary, then $\left[V_L^{d\dagger}\hat{s_q}^\dagger\right]$ is again unitary and  
$\left[V_L^{d\dagger}\hat{s_q}^\dagger \hat{s_q}V_L^d\right]$ is automatically diagonal. However, this only implies that 
\bea
\left(\left[V_L^{d\dagger}\hat{s_q}^\dagger \right]^1_{~3}\right)^2 
+ \left(\left[V_L^{d\dagger}\hat{s_q}^\dagger \right]^2_{~3}\right)^2 
+ \left(\left[V_L^{d\dagger}\hat{s_q}^\dagger \right]^3_{~3}\right)^2 = 1 \, .
\eea
Now, choosing $\left[V_L^{d\dagger}\hat{s_q}^\dagger \right]^2_{~3} \sim 1$ and the other two elements to be very small, we get from 
Eq.~\ref{eq:LQcontr} that $\left[V_{\text{CKM}}V_L^{d\dagger}\hat{s_q}^\dagger\right]^{2}_{~3} \sim 1$.  In this case, Eq.~\ref{leptoq-final} 
gets modified to
\bea
\label{eq:maxpossible}
&& (1+V_{cb}) \gtrsim 0.2 \left(\frac{M_*/g_*}{\text{TeV}} \right)^2  \nonumber \\
&& \to M_*/g_* \lesssim 2.28 \, \text{TeV} \, .  
\eea
This very conspired scenario could be realised in U(3) symmetric models \cite{Barbieri:2012tu,Redi:2011zi} where $\hat s_q\propto \dblone_{3\times 3}$.
Indeed, if $\left[V_{\rm CKM}V_L^{d\dagger}\hat{s_q}^\dagger \right]^2_{~3}\sim 1$, the constraint on the composite scale becomes that of
Eq.~\ref{eq:maxpossible}. 
However in this case \cite{Barbieri:2012tu} we have to face the constraints from the modification of the Z decays to hadrons requiring (see table 4 of \cite{Barbieri:2012tu}) 
\bea
M_*\gtrsim 6 \sqrt{g_*}\hbox{TeV},
\eea
for the composite fermions masses. Assuming the vector fields  are at the same scale,  fitting the anomalies becomes practically impossible.

\section{ $R_{K,K^*}$ anomalies}
\label{Rk-RKst}

In this section, we investigate very briefly whether the $R_{K,K^*}$ anomalies can be explained within the composite Higgs
framework (see \cite{Gripaios:2014tna,Niehoff:2015iaa,Barbieri:2016las,Niehoff:2016zso,Megias:2016bde,Barbieri:2017tuq,
DAmbrosio:2017wis,Sannino:2017utc,Carmona:2017fsn,Marzocca:2018wcf,Chala:2018igk} for related discussion).
It is known that the discrepancy of the experimental data on $R_{K}$ and $R_{K^*}$ with the SM expectations can be
alleviated by the following operator\cite{DAmico:2017mtc,Capdevila:2017bsm,Ciuchini:2017mik,Geng:2017svp,Altmannshofer:2017yso,
Ghosh:2017ber}\footnote{Actually, the experimental value of $R_{K^*}^{\rm Low}$ cannot be explained simultaneously with
$R_K$ and $R_{K^*}^{\rm Central}$ by this operator (see, for example, the upper left panel of Fig.~6 in \cite{Ghosh:2017ber}), and either 
additional light fields \cite{Ghosh:2017ber} or tensor operators \cite{Bardhan:2017xcc} are required.} 
\bea
{\cal L}_{b \to s \mu \mu} = - \frac{1}{\Lambda^2} \, (\bar s \gamma_\mu P_L \, b)(\bar\mu \gamma^\mu P_L \, \mu)  \, , 
\eea
with $1/\Lambda^2 \gtrsim 1/(38 \, \text{TeV})^2$ at the $1 \, \sigma$ level.

In models with partial compositeness, such an operator can be generated by the exchange of either a neutral $Z'$ vector boson
or a vector leptoquark. We examine the flavour structures of these two cases and identify the features that can explain the data.

 {\bf $Z'$ contribution:} Following the analysis of section \ref{NoLeptoquark}, neutral composite bosons $
 \rho_{L,R}^3, \rho_{X}$ will generate
\bea
\label{eq:bsZ}
\frac{g_{*2}^2}{2M_{*2}^2}  \left( \bar s  \, [V_L^{d \,\dagger} \hat s_q^\dagger \hat s_q V_L^d]^2_{~3} \gamma_\mu P_L \, b \right) 
\left( \bar \mu \, [V_L^{e \,\dagger} \hat s_l^\dagger \hat s_l V_L^e]^2_{~2} \gamma^\mu P_L \, \mu \right) \, 
\eea
which, after implementing the $\bar B_s$ - $B_s$ mixing constraint from Eq.~\ref{eq:utfit},  gives (assuming $V_L^e = \dblone $ and diagonal $\hat s_l$),
{ 
\bea
\label{eq.rkzprime}
\frac{g_*}{(M_*/\text{TeV})} \frac{1}{\sqrt{\rm const}} \, s_\mu^2  \gtrsim 0.35 , 
\eea
}
for the explanation of the $R_{K,K^*}$ anomaly, where `const' is defined in Eq.\ref{df2} and $g_{*,2}=g_*$, $M_{*,2}= M_*$. This inevitably requires  large muon compositeness.
{
Let us compare our results with the discussion in the previous two sections \ref{NoLeptoquark} and \ref{Leptoquark}. Constraints from $\Delta F=2$  processes require an almost diagonal $\left[V_L^{d\dagger}\hat{s}^\dagger \hat{s}V_L^d\right]$ matrix, which forces 
the operator in  Eq. \ref{eq:bsZ} to be small as well. However, note that if $\left[V_L^{d\dagger}\hat{s}^\dagger \hat{s}V_L^d\right]_{23}=\epsilon$ ($\epsilon$ being some small parameter), $\Delta F=2$ observables scale as $\epsilon^2$ and $R_K$ as $\epsilon$.  It is precisely this extra power of $\epsilon$
suppression that can make the explanation for the two measurements consistent \cite{Barbieri:2016las}.
}

\textbf{Leptoquark contribution}: The flavour structure in this case is different from the $Z'$ contribution and the relevant operator in given
by
\bea
\label{bs-lepto}
\frac{g_*^2}{M_*^2}  \left( \bar s  \, [V_L^{d \,\dagger} \hat s_q^\dagger \hat s_l V_L^e]^2_{~2} \gamma_\mu P_L \, \mu \right) 
\left( \bar \mu \, [V_L^{e \,\dagger} \hat s_l^\dagger \hat s_q V_L^d]^2_{~3} \gamma^\mu P_L \, b \right) \, , 
\eea
In this case correlations with the other low energy measurements are less strict and  as an illustration we will consider two
extreme scenarios (for simplicity we will assume diagonal $\hat s _{l,q}$)
\begin{itemize}
\item
{\bf flavour trivial lepton sector:} In this case we assume $V^e_L =\dblone$ and this obviously evades all the constraints from LFV processes
like $\tau \rightarrow 3\mu$. In such a scenario, the main constraint comes from $B_s$ mixing. Using the bound from Eq.~\ref{eq:utfit}, we get
\begin{equation}
{s}_\mu \geq 0.1 (\text{const})^{1/8} \left( \frac{M_*/\text{TeV}}{ g_*}\right)^{3/4},
\end{equation}
where  $s_{s,b}$ are equal to  the maximal possible values allowed by the $\bar{B}_s - B_s$ mixing. Interestingly,
even in this case the bound becomes less strict compared to the one obtained for the $Z'$ contribution (see Eq.~\ref{eq.rkzprime}). 
However, the scale of the muon compositeness must still be quite high.

\item {\bf flavour trivial down quarks\cite{Barbieri:2017tuq}:} If we assume $V^d_L=\dblone $, $R_K$ can be generated solely by
the leptoquark contribution and $Z'$ mediated diagrams vanish. Interestingly, we can correlate the $R_K$ with the flavour violating
$\tau$ decay $\tau \rightarrow 3\mu$ arising from the operator:
\begin{align}
\label{tau3nu}
\text{const} \times \frac{g_*^2}{M_*^2}\left(\bar{\tau}\left[V_L^{e\dagger} \hat{s}_l^\dagger \hat{s}_l V^{e}_L\right]^3_2 \gamma_\mu
P_L \mu\right)\left(\bar{\mu}\left[V_L^{e\dagger} \hat{s}_l^\dagger \hat{s}_l V^{e}_L\right]^2_2 \gamma_\mu P_L \mu\right)
\end{align}
where now
\begin{equation}
\text{const} = \frac{1}{2}\frac{M_*^2}{g_*^2}\left( \frac{g_*^2 }{2 M_{*2}^2}+\frac{3 g_{*3 }^2}{8 M_{*3}^2}+\frac{ g_{*X }^2}{4 M_{*X}^2}\right) ,
\end{equation}
comes from the contributions of the  $\rho^3_L, \rho^3_R, \rho_X$ and $\rho_{T_{15}}$ and we have assumed that the $\lambda_l$ mixing is
the dominant one (see appendix  \ref{sec:Zint}). 
Assuming that the mixing with the first generation is small, we focus only on the $\mu-\tau$ rotations with the mixing angle  $\theta$.
The experimental bound on $\tau\rightarrow 3\mu$ \cite{Patrignani:2016xqp} gives
\begin{equation}
\frac{\text{const}}{2} \frac{g_*^2}{M_*^2} s_\tau^2 \sin 2\theta \left[\cos^2\theta s_\mu^2 + \sin^2\theta s_\tau^2\right] \leq \frac{4\times 10^{-3}}{\text{TeV}^2}
\end{equation}
If the angle $\theta$ is small, say $\theta \sim s_\mu/s_\tau \ll 1$, then the upper bound on the muon compositeness becomes
\begin{equation}
\label{BoundTheta}
s_\mu s_\tau^{1/3}\ \leq 0.1 \left(\text{const}\right)^{-1/3} \left(\frac{M_*/\text{TeV}}{g_*}\right)^{2/3}  \, . 
\end{equation}
On the other hand, the bound on $R_K$ implies
\begin{equation}
s_s s_b s_\mu s_\tau \sin 2\theta \geq 10^{-3}\left(\frac{M_*/\text{TeV}}{g_*}\right)^2
\end{equation}
If $s_b \sim 1$ and $s_\tau \sim 1$ we see that there is no tension between the $R_K$ data and the $\tau\rightarrow 3\mu$ data;
in fact Eq.~(\ref{BoundTheta}) translates into a bound on the compositeness scale of the strange quark
\begin{equation}
s_s \geq 0.02 \left(\frac{M_*/\text{TeV}}{g_*}\right)^{2/3},
\end{equation}
which is similar to the naive expectations for the left-handed strange quark compositeness $s_s\sim V_{ts}$.
\end{itemize}
Thus we can conclude that it is possible to fit $R_K$ as well within the partial compositeness paradigm.
\section{Summary and outlook}

In this paper, we have studied various aspects of the $R_D$ and $R_{D^*}$ anomalies in depth. The main objective of our work has been
to understand potential correlations of $R_{D,D^*}$ to other $\Delta F =1$ and $\Delta F = 2$ processes that give rise to constraints on 
the NP explanations, and thus allowing us to identify the desired properties of the underlying UV theory. 

After reviewing the possible roles of vector, axial-vector, scalar, pseudo-scalar and tensor operators in solving these anomalies (sections
\ref{section2} and \ref{section3}), we have investigated  (section \ref{gauge-invariance}) how the linearly realised $\rm SU(2) \times U(1)$ symmetry 
can give rise to correlations to other well measured $\Delta F =1$ processes, e.g. $B \to K^* \, \nu \, \nu$, $B \to \rho \, \nu \, \nu$, $B \to \tau \, \nu$
and couplings like $Z \, \tau \, \tau$, $Z \, \nu \, \nu$ and $W \, \tau \, \nu$, posing serious difficulties in explaining $R_{D,D^*}$.

We then extend our analysis to composite Higgs paradigm with the partial compositeness mechanism to generate the fermion masses.
In this case, because of an available power-counting rule, the $\Delta F=2$ processes, namely $K$, $B_d$ and $B_s$ mixing measurements
turn out to be extremely constraining. 
We show that generically the models with partial compositeness can offer an explanation of these anomalies only if
the scale of compositeness is below 0.90 (0.64) TeV for scenarios with (without) leptoquarks. While the requirement of
such a low scale is favoured by the electroweak hierarchy problem, it is problematic from direct searches, and also indirect electroweak
precision measurements unless some additional cancellations are involved. 

Finally, in section \ref{Rk-RKst} we also comment on the possibility of explaining the other neutral current B-meson anomalies
$R_K$ and $R_{K^*}$ in this framework.

As the charged current anomalies require a NP scale which is rather low ($\sim$ TeV), they might as well be the harbingers of
new physics at the TeV scale.  It is thus important to critically examine the models that can provide simultaneous solutions to
different problems at the TeV scale. At this point, it seems that the manifestation of New Physics, if any, in the dynamics of flavour
transitions is likely to be quite non-generic and subtle. Thus the interpretation of any NP signal would require a large amount of
data with a high precision. It is encouraging that such a large amount of data are expected to come from both the LHCb and Belle-II
in the near future, and hopefully, we are not far from an exciting discovery.
%
\vspace*{-7mm}
\subsection*{Acknowledgement}
We thank Pritibhajan Byakti for collaboration in the initial stage of this work. 
We would also like to thank D. Marzocca and B. Gripaios for discussions, and
Pouya Asadi and David Shih for pointing out a sign mistake to us.
The research of DB was supported in part by the Israel Science Foundation (grant no. 780/17). 

\begin{appendices}
\section{Decay Width of the $B_c$ meson}
\label{app:Bctaunu}

The differential decay rate for the process $B_c^- (p) \to \tau^- (k_1) + \bar{\nu}_\tau (k_2)$ is given by 
\begin{equation}
\frac{d\Gamma}{d \Omega} = \frac{1}{32 \pi^2} \frac{|{\bf k_1}|}{m_{B_c}^2} \overline{|\mathcal{M}|^2} \nn
\end{equation}
where, ${\bf k_1}$ is the 3-momentum of the $\tau$ in the rest frame of the $B_c$ meson, and 
\begin{equation}
|{\bf k_1}| = \frac{m_{B_c}^2 - m_\tau^2}{2 m_{B_c}} \, . \nn
\end{equation}

The matrix element is given by, 
\begin{eqnarray}
 i\mathcal{M} &=& \frac{2 G_F V_{cb}}{\sqrt{2}} \bigg[C_{\rm AL}^{cb \tau} \, \langle 0 | \bar{c}\gamma^\mu \gamma_5 b | B_c (p)\rangle \, \bar{u}(k_1) (i\gamma_\mu P_L) v(k_2)  + C_{\rm PL}^{cb \tau} \, \langle 0 | \bar{c}\gamma_5 b | B_c (p)\rangle \, \bar{u}(k_1) (iP_L) v(k_2)  \nonumber \\
&&  \hspace{1.35cm} + C_{AR}^{cb \tau} \, \langle 0 | \bar{c}\gamma^\mu \gamma_5 b | B_c (p)\rangle \, \bar{u}(k_1) (i\gamma_\mu P_R) v(k_2) 
+ C_{PR}^{cb \tau} \, \langle 0 | \bar{c}\gamma_5 b | B_c (p)\rangle \, \bar{u}(k_1) (iP_R) v(k_2) \bigg] \nn \\[2mm]
&& \hspace{-1.7cm}\text{ (The operators ${\cal O}^{cb\ell\nu}_{\rm VL, VR}$\, , ${\cal O}^{cb\ell\nu}_{\rm SL, SR}$ and ${\cal O}^{cb\ell\nu}_{\rm TL, TR}$ do not contribute because the 
corresponding} \nn \\
&& \hspace{-1.0cm} \text{matrix elements, $\langle 0 | \bar{c}\gamma^\mu b | B_c \rangle$, $\langle 0 | \bar{c} b | B_c \rangle$ and $\langle 0 | \bar{c} \sigma^{\mu\nu}b | B_c \rangle$ identically vanish)} \nn \\[2mm]
&=& \frac{2 G_F V_{cb}}{\sqrt{2}} \, i f_{B_c} \left[ C_{\rm AL}^{cb \tau}  \, p^\mu \, \bar{u}(k_1) (i\gamma_\mu P_L) v(k_2) - C_{\rm PL}^{cb \tau} \frac{m_{B_c}^2}{m_b + m_c}  
\bar{u}(k_1) (iP_L) v(k_2) \right. \nonumber \\
&& \left. \hspace{2.15cm} + C_{AR}^{cb \tau}  \, p^\mu \, \bar{u}(k_1) (i\gamma_\mu P_R) v(k_2) - C_{PR}^{cb \tau} \frac{m_{B_c}^2}{m_b + m_c}  
\bar{u}(k_1) (iP_R) v(k_2) \right]\nn \\
&& \hspace{-1.0cm} \text{ \bigg( In the above step, we have used $\langle 0 | \bar{c}\gamma^\mu \gamma_5 b | B_c (p)\rangle = i f_{B_c} p^\mu$, and } \nn \\
&& \hspace{5.9cm}\text{$\langle 0 | \bar{c} \gamma_5 b | B_c (p)\rangle = - i f_{B_c} \frac{m_{B_c}^2}{m_b + m_c}$ \bigg)} \nn \\
&=& \frac{2 G_F V_{cb}}{\sqrt{2}} \,  i f_{B_c} \left[ C_{\rm AL}^{cb \tau}  \, m_\tau \, \bar{u}(k_1) (iP_L) v(k_2) - C_{\rm PL}^{cb \tau} \frac{m_{B_c}^2}{m_b + m_c}  
\bar{u}(k_1) (iP_L) v(k_2) \right. \nonumber \\
&& \hspace{2.15cm} \left. + C_{AR}^{cb \tau}  \, m_\tau \, \bar{u}(k_1) (iP_R) v(k_2) - C_{PR}^{cb \tau} \frac{m_{B_c}^2}{m_b + m_c}  
\bar{u}(k_1) (iP_R) v(k_2) \right] \nn \\
&=& \frac{2 G_F V_{cb}}{\sqrt{2}} \,  i m_\tau f_{B_c}  \left[ \left( C_{\rm AL}^{cb \tau}  - C_{\rm PL}^{cb \tau} \frac{m_{B_c}^2}{m_\tau (m_b + m_c)} 
 \right) \bar{u}(k_1) (iP_L) v(k_2) +  \right. \nonumber \\
 &&\hspace{3cm}\left.  \left( C_{AR}^{cb \tau}  - C_{PR}^{cb \tau} \frac{m_{B_c}^2}{m_\tau (m_b + m_c)} 
 \right) \bar{u}(k_1) (iP_R) v(k_2) \right] \nn
 \end{eqnarray}
This gives, 
\begin{eqnarray}
\overline{|\mathcal{M}|^2} &=& \bigg[2 G_F^2 |V_{cb}|^2\bigg] \bigg[ m_\tau^2  f_{B_c}^2 \bigg] \bigg[ m_{B_c}^2 
\left(1 - \frac{m_\tau^2}{m_{B_c}^2}\right) \bigg] \times \nn \\ 
&& \left( \left|C_{\rm AL}^{cb\tau} - \frac{m_{B_c}^2}{m_\tau(m_b + m_c)} C_{\rm PL}^{cb\tau} \right|^2  + 
\left|C_{AR}^{cb\tau} - \frac{m_{B_c}^2}{m_\tau(m_b + m_c)} C_{PR}^{cb\tau} \right|^2 \right)  \nn
\end{eqnarray}

\begin{figure}[t!]
\begin{center}
\begin{tabular}{cc}
\includegraphics[scale=0.55]{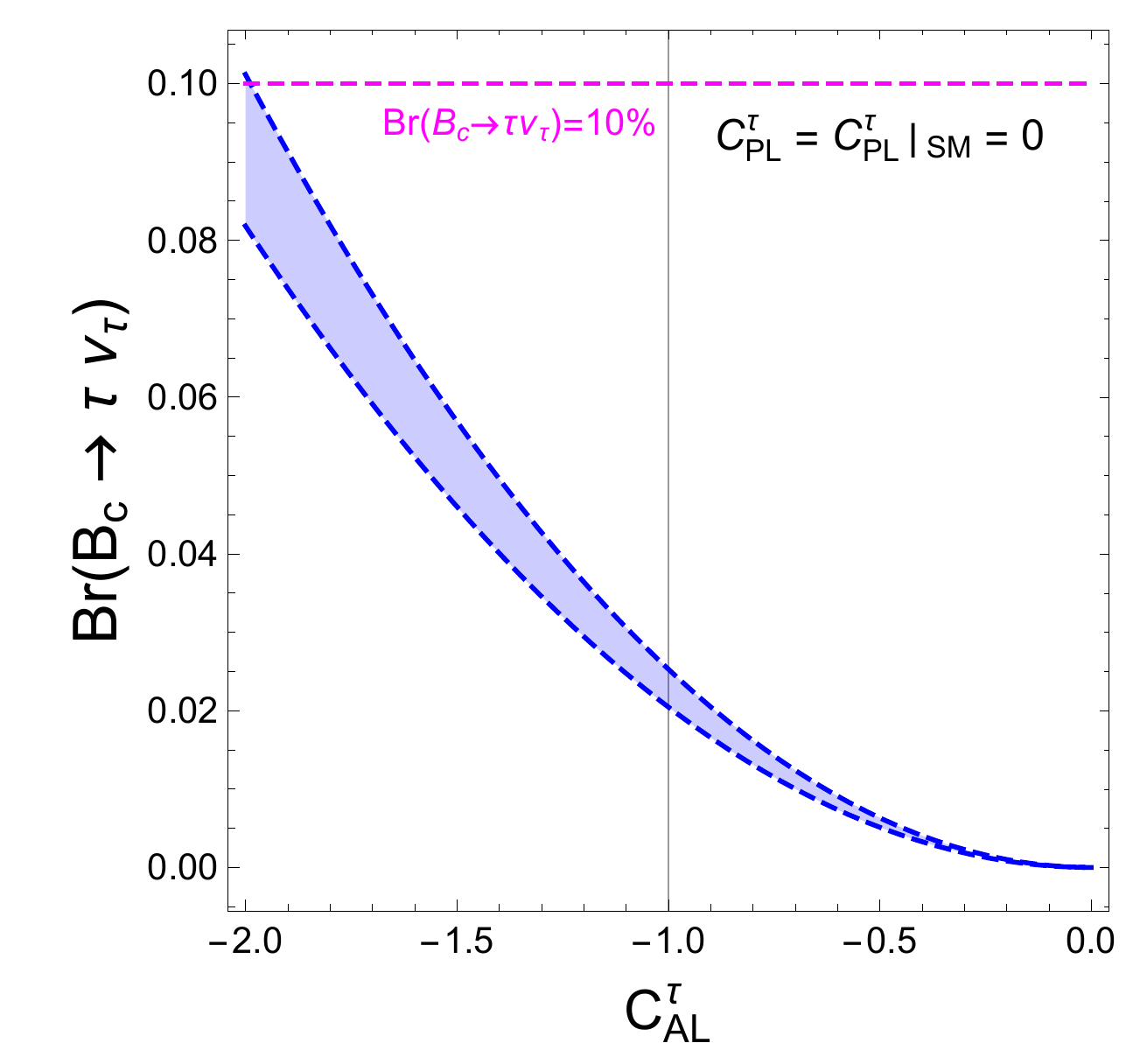} & \includegraphics[scale=0.55]{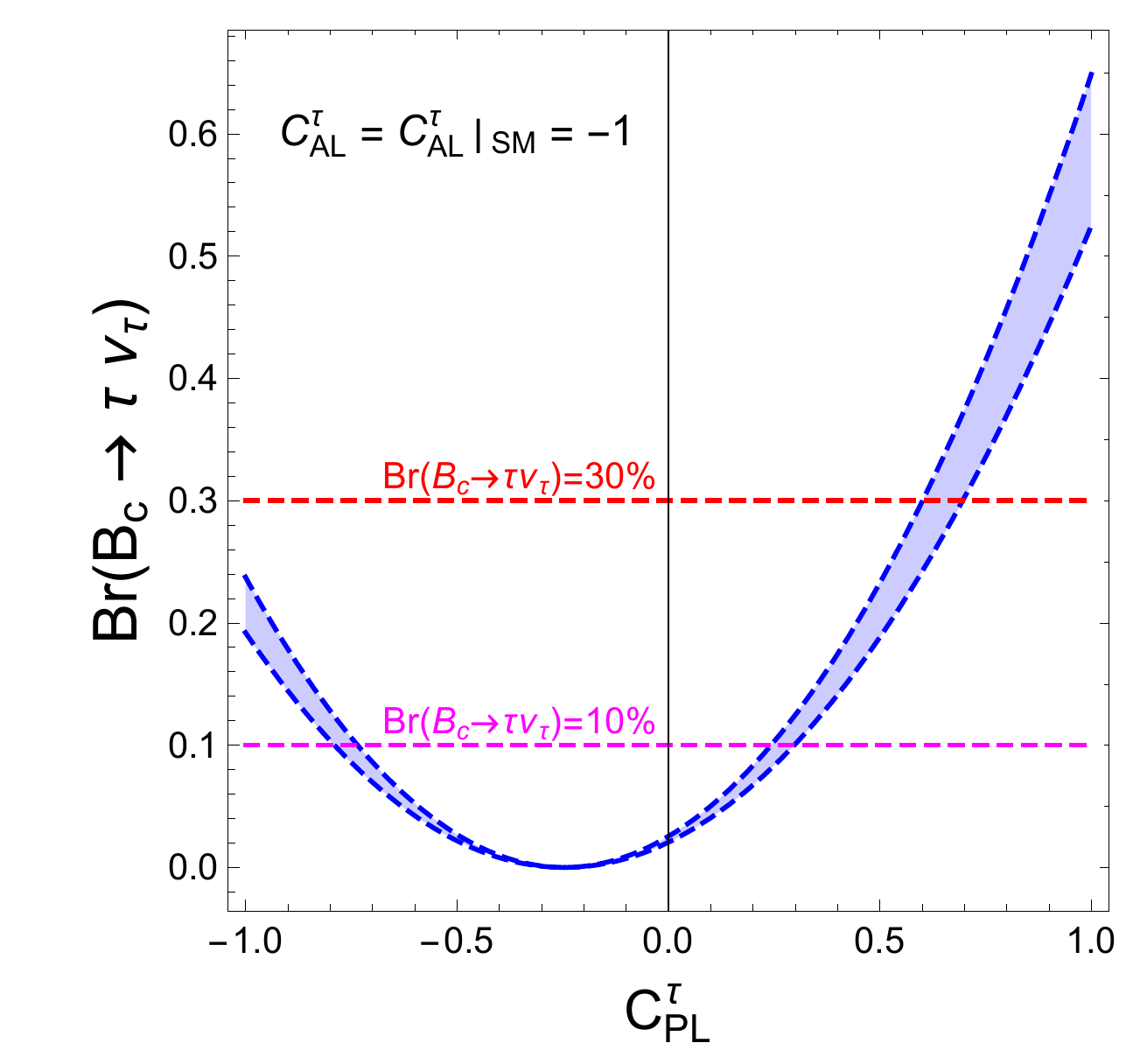}
\end{tabular}
\caption{\sf The Br($B_c \to \tau \, \nu_\tau$) as a function of $C_{\rm AL}^\tau$ and $C_{\rm PL}^\tau$. The upper bounds, 30\% and 10\%
on this branching fraction from \cite{Alonso:2016oyd} and \cite{Akeroyd:2017mhr} respectively are also shown.
The SM branching ratio is $\approx 2\%$.  We have used $f_{B_c} = 0.434 \pm 0.015$ GeV \cite{Colquhoun:2015oha} in our calculation. \label{Bctaunu-vs-CAL-CPL}}
\end{center}
\end{figure}

Thus, the partial decay rate is given by, 
\begin{eqnarray}
\Gamma_{B_c \to \tau \nu} &=& \frac{1}{16 \, \pi \, m_{B_c}} \left(1 - \frac{m_\tau^2}{m_{B_c}^2}\right) \overline{|\mathcal{M}|^2} \nonumber \\
&=& \frac{1}{8 \pi} G_F^2 |V_{cb}|^2 f_{B_c}^2 m_\tau^2 m_{B_c} \left(1 - \frac{m_\tau^2}
{m_{B_c}^2}\right)^2 \times \nn \\
&& \left( \left|C_{\rm AL}^{cb\tau} - \frac{m_{B_c}^2}{m_\tau(m_b + m_c)} C_{\rm PL}^{cb\tau} \right|^2  + 
\left|C_{AR}^{cb\tau} - \frac{m_{B_c}^2}{m_\tau(m_b + m_c)} C_{PR}^{cb\tau} \right|^2 \right) \nn 
\end{eqnarray}
which gives, for the branching ratio,
\begin{eqnarray}
\mathcal{B}(B_c^- \to \tau^- \bar{\nu}_\tau) &=& \frac{1}{8 \pi} G_F^2 |V_{cb}|^2 f_{B_c}^2 m_\tau^2 
m_{B_c} \tau_{B_c} \left(1 - \frac{m_\tau^2} {m_{B_c}^2}\right)^2 \times \nn \\
&& \hspace{-1cm}\left( \left|C_{\rm AL}^{cb\tau} - \frac{m_{B_c}^2}{m_\tau(m_b + m_c)} C_{\rm PL}^{cb\tau} \right|^2  + 
\left|C_{AR}^{cb\tau} - \frac{m_{B_c}^2}{m_\tau(m_b + m_c)} C_{PR}^{cb\tau} \right|^2 \right)
\label{bc_branching}
\end{eqnarray}

The variation of Br($B_c \to \tau \, \nu_\tau$) as a function of $C_{\rm AL}^\tau$ or $C_{\rm PL}^\tau$ is
shown in Fig.~\ref{Bctaunu-vs-CAL-CPL}. 

\vspace*{-5mm}
\section{Form Factors for $B_c \to J/ \psi$ and $B_c \to \eta_c$ decay processes} 
\label{Bc2Jpsi-FF}
\subsection{Vector and axial-vector form-factors}
\subsubsection{$B_c \to \eta_c$}

\begin{figure}[h!]
\begin{center}
\includegraphics[scale=0.55]{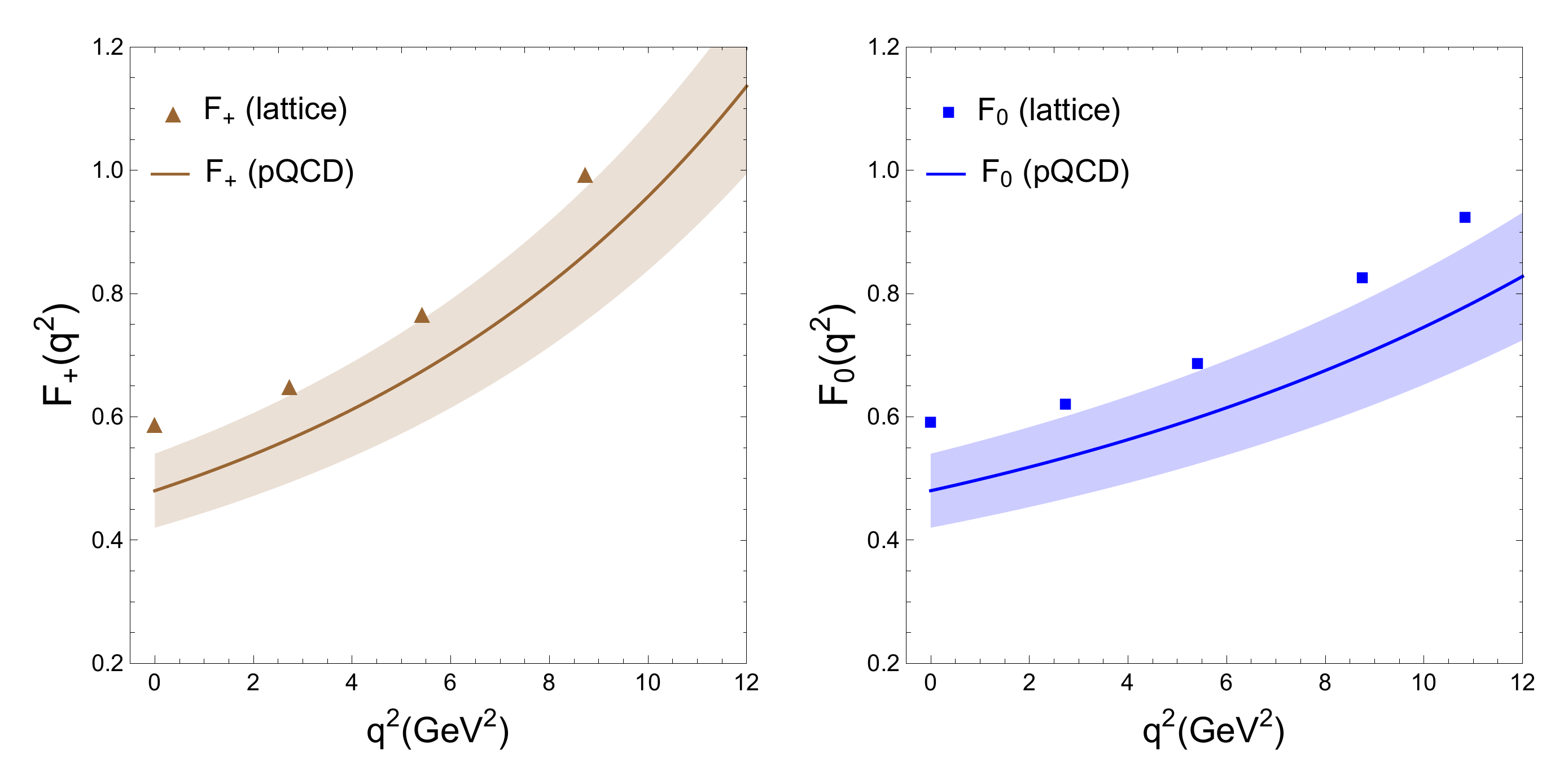}
\end{center}
\caption{\sf Form factors $F_0 (q^2)^{B_c \to \eta_c}$ and $F_+ (q^2)^{B_c \to \eta_c}$ from pQCD \cite{Wen-Fei:2013uea} and 
lattice \cite{Colquhoun:2016osw}.}
\label{fig:EtacFF}
\end{figure}

The $B_c \to \eta_c$ matrix elements are parametrised in the same way as the $B \to D$ matrix elements, see for example section-4 of 
\cite{Bardhan:2016uhr}. Unfortunately, only preliminary lattice results are available for $B_c \to \eta_c$ matrix elements \cite{Colquhoun:2016osw}. 
In Fig.~\ref{fig:EtacFF}, we show the pQCD estimates of the $F_+$ and $F_0$ form-factors from \cite{Wen-Fei:2013uea}. 
The preliminary lattice results from \cite{Colquhoun:2016osw} are also overlaid. 

The functional form of $F_0$ and $F_+$ is given by  
\begin{equation}
f = f_0 \exp \left(a \, q^2 + b \, (q^2)^2\right)
\end{equation}
where $f$ can be either $F_0$ or $F_+$. Then $f_0 = 0.48 \pm 0.06$, $a_{0(+)} = 0.037 (0.055),
 \ \ b_{0(+)} = 0.0007 (0.0014)$.  

\bea
\bra \eta_c(p_{\eta_c}, M_{\eta_c})| \bar c \gamma^\mu b |\bar B(p_{B_c}, M_{B_c})\ket &=& F_+(q^2) \Big[(p_{B_c}+p_{\eta_c})^\mu -\frac{M_{B_c}^2 - M_{\eta_c}^2}{q^2} q^\mu \Big] \nonumber \\
&&  + F_0(q^2) \frac{M_{B_c}^2- M_{\eta_c}^2}{q^2} q^\mu \label{eqn:bc_eta_v} \\
\bra \eta_c(p_{\eta_c}, M_{\eta_c})| \bar c \gamma^\mu \gamma_5 b |\bar B(p_{B_c}, M_{B_c})\ket &=& 0  \label{eqn:bc_eta_a}
\eea

\subsubsection{$B_c \to J/\psi$}

\begin{figure}[h!]
\begin{center}
\includegraphics[scale=0.50]{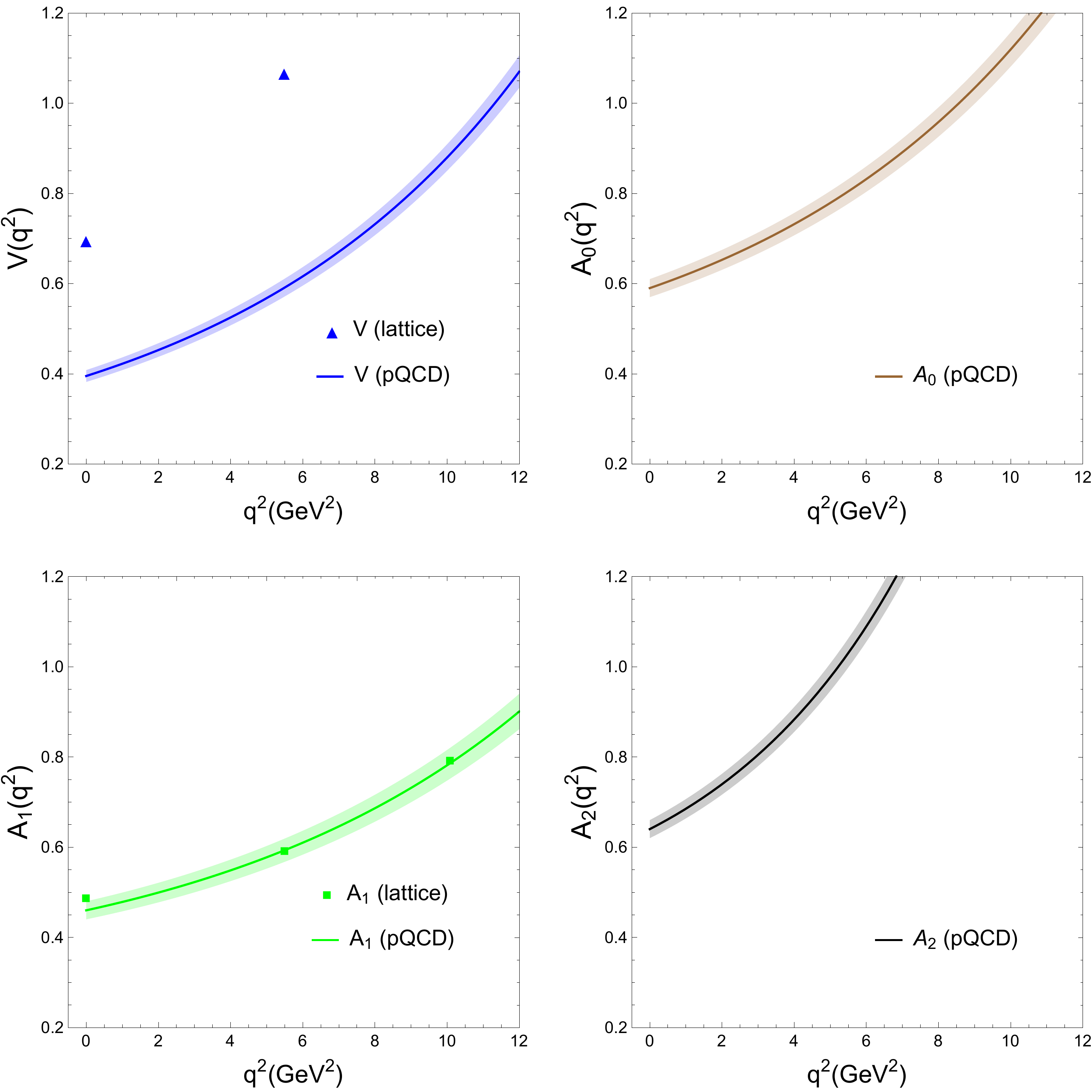}
\end{center}
\caption{\sf Form factors $V(q^2)^{B_c \to J/\psi}$, $A_0(q^2)^{B_c \to J/\psi}$, $A_1(q^2)^{B_c \to J/\psi}$ and 
$A_2(q^2)^{B_c \to J/\psi}$ from pQCD and lattice.}
\label{fig:JpsiFF}
\end{figure}

Similarly, parametrisation of the different $B_c \to J/\psi$ matrix elements are the same as those for $B \to D^\ast$ 
matrix elements, see again \cite{Bardhan:2016uhr} for the notations. 
The pQCD estimates \cite{Wen-Fei:2013uea} for the $B_c \to J/\psi$ form-factors: $V$, $A_0$, $A_1$ and $A_2$, 
are shown in Fig.~\ref{fig:JpsiFF}.  
Preliminary lattice results for $V$ and $A_1$ from \cite{Colquhoun:2016osw} are also shown. 

\subsection{Tensor form-factors}

As no estimate of the tensor form-factors exists in the literature, we use the quark level equations of motion to relate them 
to the other form-factors. We show this explicitly below. 

\subsubsection{$B_c \to \eta_c$ }

The tensorial form factors are given by
\bea
\label{eqn:bc_eta_t}
{\bra \eta_c(p_{\eta_c}, M_{\eta_c})| \bar c \, i\sigma^{\mu\nu}  b |\bar B(p_{B_c}, M_{B_c})\ket} &=&  
(p_{B_c}^\mu p_{\eta_c}^\nu - p_{B_c}^\nu p_{\eta_c}^\mu) \frac{2F_T(q^2)}{ M_{B_c} + M_{\eta_c}}  \label{tmp-tensor}\\
\label{eqn:bc_eta_t5}
{\bra \eta_c(p_{\eta_c}, M_{\eta_c})| \bar c \sigma^{\mu\nu} \gamma_5 b |\bar B(p_{B_c}, M_{B_c})\ket} 
&=& \varepsilon^{\mu\nu\rho\sigma}  
p_{{B_c}\rho} p_{{\eta_c}\sigma} \frac{2F_T(q^2)}{ M_{B_c} + M_{\eta_c}} 
\eea

Multiplying the LHS of Eq.~\ref{tmp-tensor} by $q_\mu$ and using $ i \sigma^{\mu\nu} = \eta^{\mu\nu} - \gamma^\mu \gamma^\nu$ we get,
\bea
q_\mu \bra \eta_c | \bar{c} i \sigma^{\mu\nu} b | \bar{B}_c\ket &=& q^{\nu} \bra \eta_c | \bar{c} b | \bar{B}_c\ket - \bra \eta_c | \bar{c} \, \slashed{q} \, \gamma^\nu b | \bar{B}_c\ket \label{eqn:tensor-q} \\
&=& \bra \eta_c | \bar{c} b | \bar{B}_c\ket q^\nu + (m_b + m_c) \bra \eta_c | \bar{c} \gamma^\nu b | \bar{B}_c\ket  - 2 p_B^\nu \bra \eta_c | \bar{c} b | \bar{B}_c\ket \nn  \\
&=& - \bra \eta_c | \bar{c} b | \bar{B}_c\ket (p_{B_c} + p_{\eta_c})^\nu + (m_b + m_c) \bra \eta_c | \bar{c} \gamma^\nu b | \bar{B}_c\ket \nn \\
&=& - F_0 \frac{M_{B_c}^2 - M_{\eta_c}^2}{m_b - m_c} (p_{B_c} + p_{\eta_c})^\nu \nn \\
& & \quad \quad +  (m_b + m_c) \left[F_+ (p_{B_c}+p_{\eta_c})^\nu - (F_+ - F_0) \frac{M_{B_c}^2 - M_{\eta_c}^2}{q^2}q^\nu  \right]\nn
\eea

The LHS of Eqn \ref{eqn:tensor-q}, using the tensor form factor, is then
\beq
\left[(p_{B_c} - p_{\eta_c})^\nu \left( -\frac{M_{B_c}^2 - M_{\eta_c}^2}{2}  \right) +  (p_{B_c} + p_{\eta_c})^\nu \left( \frac{M_{B_c}^2 + M_{\eta_c}^2}{2} - p_{B_c} . p_{\eta_c} \right)\right] \frac{2 F_T}{M_{B_c} + M_{\eta_c}}
\eeq

Noting that $q = p_{B_c} - p_{\eta_c}$, we can equate the coefficients of $q^\nu$ on either side of Eqn. \ref{eqn:tensor-q}. This gives us the relation between the tensor and the vector form factors to be 
\bea
-\frac{M_{B_c}^2 - M_{\eta_c}^2}{M_{B_c} + M_{\eta_c}} F_T = - (m_b + m_c) \frac{M_{B_c}^2 - M_{\eta_c}^2}{q^2} \left(F_+ - F_0\right) 
\eea
\bea
\implies \boxed{F_T = (m_b + m_c)\frac{M_{B_c} + M_{\eta_c}}{q^2} \left(F_+ - F_0\right)}
\eea

\subsubsection{$B_c \to J/\psi$}

The hadronic matrix elements for $\bar{B} \to V$ transition are parametrised by
\bea
\bra V(p_V, \epsilon, M_V)| \bar c \gamma_\mu b |\bar B(p_B, M_B)\ket &=&  
i \varepsilon_{\mu\nu\rho\sigma} \epsilon^{ \nu \ast} p_B^\rho p_V^\sigma  \, \frac{2 V(q^2)}{M_B+M_V} \label{rds-ff-v} \\[3mm]
\bra V(p_V, \epsilon, M_V)| \bar c \gamma_\mu \gamma_5 b |\bar B(p_B,
M_B)\ket &=&  2 M_V  \frac{\epsilon^\ast.q }{q^2 } q_\mu A_0(q^2) +
(M_B+M_V)  \Big[\epsilon_{\mu}^\ast - \frac{\epsilon^\ast.q}{q^2 } q_\mu   \Big] A_1(q^2) \nonumber\\
&& \hspace*{-2mm} - \frac{\epsilon^\ast.q}{M_B+M_V} \Big[(p_B + p_V)_\mu  -\frac{M_B^2-M_V^2}{q^2} q_\mu  \Big] A_2(q^2) \label{rds-ff-a}\\[3mm]
\bra V(p_V, \epsilon, M_V)| \bar c \gamma_5 b |\bar B(p_B, M_B)\ket &=&  
- \epsilon^\ast.q \, \frac{2 M_V}{m_b + m_c}  \, A_0(q^2)  \label{rds-ff-p} \\[3mm]
\bra V(p_V, \epsilon, M_V)| \bar c \ i\sigma_{\mu\nu}  b |\bar B(p_B,
M_B)\ket &=&    - i\varepsilon_{\mu \nu \alpha\beta } 
\Big[-\epsilon^{\alpha\ast}(p_V + p_B)^\beta T_1(q^2) \nonumber\\
&& + \frac{M_B^2-M_V^2}{q^2}\epsilon^{\ast\alpha} q^\beta \left(T_1(q^2) - T_2(q^2)\right) \\
 && +  2 \frac{\epsilon^\ast.q }{q^2 }   p_B^\alpha p_V^\beta 
\left(T_1(q^2) -T_2(q^2) - \frac{q^2}{M_B^2-M_{D^\ast}^2} T_3(q^2)\right) \Big] \nonumber \label{rds-ff-t}\\[3mm]
\bra V(p_V, \epsilon, M_V)| \bar c \ i\sigma_{\mu\nu} q^\nu b |\bar B(p_B,M_B)\ket &=& 
- 2 i \varepsilon_{\mu \nu \rho \sigma} \epsilon^{\ast \nu} p_{B}^\rho p_V^\sigma T_1(q^2) \label{rds-ff-t5}
\eea

Using, 
\begin{equation}
i \sigma_{\mu \nu} = \eta_{\mu \nu} - \gamma_\mu \gamma_\nu
\end{equation}
the LHS of Eqn. \ref{rds-ff-t5} yields,
\begin{eqnarray}
\bra V \left( p_V, \epsilon, M_V\right)|\bar c \ i\sigma_{\mu\nu} q^\nu b |\bar B (p_B,M_B)\ket &=& \bra V \left( p_V,\epsilon, M_V\right)|
\bar c \left(q_\mu - \gamma_\mu \not{q}\right) b | \bar B (p_B)\ket\nn \\
&=& q_\mu \bra V \left( p_V, \epsilon, M_V\right)|
\bar c b | \bar B (p_B,M_B)\ket \nn \\
&& \hspace{-2cm} - (m_b + m_c) \bra V \left( p_V,\epsilon, M_V\right)|\bar c \ \gamma_\mu b |\bar B (p_B,M_B)\ket
\end{eqnarray}
The first term vanishes, and after using Eqn. \ref{rds-ff-v} and \ref{rds-ff-t5}, leaves 
\begin{eqnarray}
-2 i \varepsilon_{\mu\nu\rho\sigma} \epsilon^{\ast \nu} p_{B}^\rho p_V^\sigma T_1 (q^2) &=& -(m_b + m_c) \times 2 i \varepsilon_{\mu\nu\rho\sigma} \epsilon^{ \nu \ast} p_B^\rho p_V^\sigma  \, \frac{V(q^2)}{M_B+M_V} \nn 
\end{eqnarray}
which give us
\begin{equation}
\boxed{T_1 (q^2) = \frac{m_b + m_c}{M_B + M_V} V(q^2)}
\label{eqn:jpsi_t1}
\end{equation}

Consider the term $\bra V \left( p_V, \epsilon, M_V\right)|\bar c \ i\sigma_{\mu\nu} q^\nu \gamma_5 b |\bar B (p_B,M_B)\ket$. Using 
$$ \sigma_{\mu \nu} \gamma_5 = \frac{i}{2} \varepsilon_{\mu \nu \rho \sigma} \sigma^{\rho \sigma} ~~~~ \text{(we use $\epsilon^{0123}=1$, which implied that $\epsilon_{0123}=-1$)}$$
we have
\begin{equation}
\bra V \left( p_V, \epsilon, M_V\right)|\bar c \ i\sigma_{\mu\nu} q^\nu \gamma_5 b |\bar B (p_B)\ket = \frac{i}{2} \varepsilon_{\mu \nu \rho \sigma} \bra V \left( p_V, M_V \epsilon\right)|\bar c \ i\sigma^{\rho\sigma} q^\nu  b |\bar B (p_B)\ket \nn
\end{equation}
Simplification of the RHS using Eqn. \ref{rds-ff-t}, we get, 
\begin{eqnarray}
&=& \frac{i}{2} \varepsilon_{\mu \nu \rho \sigma} q^\nu \left[  - i\varepsilon^{\rho \sigma \alpha\beta } 
\left( -\epsilon^\ast_\alpha (p_{B_c} + p_{J/\psi})_\beta T_1  + \frac{M_{B_c}^2-M_{J/\psi}^2}{q^2}\epsilon^{\ast}_\alpha q_\beta \left(T_1
- T_2\right) \right. \right. \nn \\ 
&& \left. \left.-  \frac{\epsilon^\ast.q }{q^2 }   (p_{B_c} + p_{J/\psi})_\alpha (p_{B_c} - p_{J/\psi})_\beta 
\left(T_1 -T_2 - \frac{q^2}{M_{B_c}^2-M_{J/\psi}^2} T_3\right) \right)\right] \nn \\
&=& - \left(\delta^\alpha_\mu \delta^\beta_\nu - \delta^\alpha_\nu \delta^\beta_\mu \right) \left[ -\epsilon^\ast_\alpha (p_{B_c} + p_{J/\psi})_\beta T_1  + \frac{M_{B_c}^2-M_{J/\psi}^2}{q^2}\epsilon^{\ast}_\alpha q_\beta \left(T_1 - T_2\right) \right. \nn \\ 
&& \left. -  \frac{\epsilon^\ast.q }{q^2 }   (p_{B_c} + p_{J/\psi})_\alpha q_\beta 
\left( T_1 -T_2 - \frac{q^2}{M_{B_c}^2-M_{J/\psi}^2} T_3\right) \right] \\
&=& \left[(p_{B_c} + p_{J/\psi}).q \ \epsilon_\mu^\ast - \epsilon^\ast . q (p_{B_c} + p_{J/\psi})_\mu \right] T_1 -\left(M_{B_c}^2 - M_{J/\psi}^2\right) \left(\epsilon_\mu^\ast - \frac{\epsilon^\ast .q}{q^2}q_\mu\right) \left(T_1 - T_2\right)\nn \\
&+& \epsilon^\ast . q \left[(p_{B_c} + p_{J/\psi})_\mu - \frac{(p_{B_c} + p_{J/\psi}).q}{q^2} q_\mu\right] \left(T_1  - T_2 - \frac{q^2}{M_{B_c}^2 - M_{J/\psi}^2}T_3\right) \nn \\
&=& \epsilon^\ast_\mu \left[(M_{B_c}^2 - M_{J/\psi}^2) T_2 \right] - \epsilon^\ast. q (p_{B_c} + p_{J/\psi})_\mu
\left[T_2 + \frac{q^2}{M_{B_c}^2-M_{J/\psi}^2} T_3 \right]+\left(\frac{\epsilon^\ast.q}{q^2} q_\mu\right) \left[  q^2\  T_3 \right] \nn 
\end{eqnarray}

The LHS, simplified using the equation of motion is, 
\begin{eqnarray}
\bra {J/\psi} |\bar c \ i\sigma_{\mu\nu} q^\nu \gamma_5 b |\bar{B}_c \ket &=& (m_b - m_c) \bra {J/\psi} |\bar c \ \gamma_\mu \gamma_5 b |\bar{B}_c \ket  \nn \\
&=& (m_b - m_c)\left[ 2 M_{J/\psi}  \left(\frac{\epsilon^\ast.q }{q^2 } q_\mu\right) A_0 \right. \left.+ (M_{B_c}+M_{J/\psi})  \left(\epsilon_{\mu}^\ast - \frac{\epsilon^\ast.q}{q^2 } q_\mu   \right) A_1 \nn \right. \\
&&\left. - \frac{\epsilon^\ast.q}{M_{B_c}+M_{J/\psi}} \left((p_{B_c} + p_{J/\psi})_\mu  -\frac{M_{B_c}^2-M_{J/\psi}^2}{q^2} q_\mu  \right) A_2\right]
\end{eqnarray}

Comparing the coefficients of $\epsilon^\ast_\mu$ from
either side, we get
\begin{equation}
\boxed{T_2 = \frac{m_b - m_c}{M_{B_c} - M_{J/\psi}} A_1 }
\label{eqn:jpsi_t2}
\end{equation} 

Comparing the coefficients of $\epsilon^\ast.q\ q_\mu / q^2 $ from
either side, we get
\begin{equation}
\boxed{T_3 = - \left(\frac{m_b - m_c}{q^2}\right) \left(M_{B_c}( A_1 - A_2)+ M_{J/\psi} (A_1 + A_2 - 2 A_0 \right)}
\label{eqn:jpsi_t3}
\end{equation}

 The equations \ref{eqn:jpsi_t1}, \ref{eqn:jpsi_t2} and \ref{eqn:jpsi_t3} agree with those given in \cite{Watanabe:2017mip}.

\section{Formulae for calculating branching ratios}
\label{app-formulas}

The double differential branching fractions for the decays $B \to D \ell \nu_\ell$ and 
$B \to D^\ast \ell \nu_\ell$ can be written as
\bea
\frac{d^2 {\mathcal B}^{D^{(*)}}_\ell}{d q^2 \, d(\cos\theta)} &=&  {\mathcal N} \, |p_{D^{(*)}}| \, \left( a_\ell^{D^{(*)}} +
 b_\ell^{D^{(*)}} \cos\theta +  c_\ell^{D^{(*)}} \cos^2\theta \right) \, .
\eea
The normalisation factor, $\cal{N}$ and the absolute value of the $D^{(*)}$-meson momentum, $|p_{D^{(*)}}|$ are  given by,
\bea
{\cal N}  = \frac{\tau_B \, G_F^2 |V_{cb}|^2q^2}{256 \pi^3 M_B^2}  \, \left( 1 - \frac{m_\ell^2}{q^2} \right)^2 \, \, , \, \, 
|p_{D^{(*)}}| = \frac{\sqrt{\lambda(M_B^2, M_{D^{(*)}}^2,q^2)}}{2 M_B},
\eea
where $\lambda(a,b,c)= a^2 + b^2 +c^2 -2 (ab + bc + ca)$.
The angle $\theta$ is defined as the angle between the lepton and $D^{(*)}$-meson in the lepton-neutrino centre-of-mass frame,
and $q^2$ is the invariant mass squared of the lepton-neutrino system. 

\subsection{Analytic formulas for $B\to D$ decay}
The quantities $a_\ell^D$, $b_\ell^D$ and $c_\ell^D$ for negative and positive helicity lepton are given by \cite{Bardhan:2016uhr} : 

{\bf Negative helicity lepton:}
\bea
a^D_\ell (-) &=& \frac{ 8 M_B^2 {\PDSQ}  }{q^2}  \,  {\bf \CVLSq  
\FoneSQ} 
+ m_\ell \left[ \frac{32 M_B^2 \PDSQ}{q^2\left(M_B+M_D \right)}
 \mathcal{R} \left({\bf C_{ \bf TL}^\ell} {\bf C_{\rm VL}^{\ell*}}\right) 
{\bf F_+} {\bf F_T}     \right]\nonumber\\
&+& m_\ell^2 \left[  \frac{32 {\PDSQ}M_B^2}{q^2\left(M_B+M_D \right)^2} {\CTLSq}  
{\FtwoSQ}   \right] \nn \\
 b^D_\ell (-) &=& 0 \nn\\
%
%
%
%
c^D_\ell (-) &=&- \frac{ 8 M_B^2 {\PDd}^2}{q^2} \CVLSq \FoneSQ 
-  m_\ell  \left[ \frac{32  {\PDd}^2 M_B^2} {q^2\left(M_B + M_D \right)} {\bf \re \left( \cvl  \ctl\right)}  {\bf F_+ F_T}\right]  \nn \\
&-& m_\ell^2  \left[\frac{32 {\PDd}^2 M_B^2}{\left(M_B + M_D \right)^2 q^2}    \CTLSq \FtwoSQ \right]
\eea

{\bf Positive helicity lepton:}
\bea
a^D_\ell(+) &=&  \frac{ 2 \left(M_B^2-M_D^2\right)^2}{\left(m_b-m_c\right){}^2}  \, {\bf {\CSLSq} {\FzeroSQ}}
+ m_\ell \left[ \frac{4 \left(M_B^2-M_D^2\right)^2}{q^2 (m_b-m_c)} \, {\bf \re 
\left({\bf C_{\rm SL}^\ell} {\bf C_{\rm VL}^{\ell*}}\right){\FzeroSQ}}    \right]\nonumber\\
&+& m_\ell^2 \left[ \frac{2 \left(M_B^2-M_D^2\right)^2}{q^4} \, {\bf \CVLSq \FzeroSQ} \right] \nn \\
b^D_\ell(+) &=& \left[ -16 M_B \PDd \frac{M_B -M_D }{m_b - m_c}  
 \re \left({\bf C_{\bf SL}^\ell} {\bf C_{\bf TL}^{\ell 
*}}\right) {\bf F_0} {\bf F_T} \right]\nonumber\\
&& \hspace*{-2cm} - m_\ell \left[ \frac{16 {\PDd} \left(M_B-M_D\right) M_B }{q^2} 
\re \left({\bf C_{\rm VL}^\ell} {\bf C_{ \bf TL}^{\ell *}}\right) {\bf F_0} 
{\bf F_T} 
+ \frac{8{\PDd} M_B \left(M_B^2-M_D^2\right)  }{q^2 
\left(m_b-m_c\right)} \re \left({\bf C_{\rm SL}^\ell C_{\rm VL}^{\ell 
*} }\right) {\bf 
F_0} {\bf F_+}   \right] \nonumber\\
&-& m_\ell^2 \left[ \frac{ 8 {\PDd} M_B
\left(M_B^2-M_D^2\right)}{q^4}  { \CVLSq} {\bf F_0}  {\bf F_+} \right] \nn \\
c^D_\ell (+) &=&
 \left[\frac{{32 \PDd}^2 M_B^2}{\left(M_B + M_D \right)^2 }{\CTLSq} {\FtwoSQ} 
\right] - \frac{32 M_B^2 \PDSQ}{\left(M_B + M_D \right)q^2} \re {\bf\left(
C_{\rm VL}^\ell  C_{\rm TL}^{\ell *}\right)}{\bf  F_+ F_T}   \nonumber\\
&+& m_\ell^2 \left[\frac{8 {\PDd}^2 M_B^2}{q^4} \CVLSq
{\FoneSQ}   \right]
\eea

\subsection{Semi-numerical formulas for $R_{D}$}

We now provide semi-numerical formulae for the branching ratios and $R_D$ in terms of the Wilson coefficients (WCs)
(from now onwards, instead of using ``$cb\ell\nu$" we will just use ``$\ell \, $" in the superscript of the operators and WCs) :
\bea
\mathcal{B}\left(B\to D \tau \nu_\tau\right) &=& \bigg( 6.9 + 13.9 \ {\bf \Delta C_{\rm VL}^\tau} 
+ 11.9\  {\bf \Delta C_{\rm SL}^\tau} + 3.5 \ {\bf \Delta C_{\rm TL}^\tau}  \nn \\
&+& 6.9 \ {\bf (\Delta C_{\rm VL}^\tau)^2} + 9.4\  {\bf (\Delta C_{\rm SL}^\tau)^2} + 1.2\  {\bf ( \Delta C_{\rm TL}^\tau)^2}   \nn \\ 
&+& 11.9\  {\bf \Delta C_{\rm VL}^\tau} {\bf  \Delta C_{\rm SL}^\tau}  + 3.5\  {\bf \Delta C_{\rm VL}^\tau} {\bf \Delta C_{\rm TL}^\tau} \bigg) \times 10^{-3} \\[2mm]
\mathcal{B}\left(B\to D \llx \, \nu_{\llx}\right) &=& \bigg( 23.3 + 46.6 \  {\bf \Delta C_{\rm VL}^{\llx}}  + 
2.0 \  {\bf \Delta C_{\rm SL}^{\llx}} + 1.0\  {\bf \Delta C_{\rm TL}^{\llx}} \nn \\
&+& 23.3\  {\bf (\Delta C_{\rm VL}^{\llx})^2} + 33.5\  {\bf (\Delta C_{\rm SL}^{\llx})^2} + 3.5\  {\bf (\Delta C_{\rm TL}^{\llx})^2}  \nn \\ 
&+& 2.0\  {\bf \Delta C_{\rm VL}^{\llx}}  {\bf \Delta C_{\rm SL}^{\llx}} 
+ 1.0\  {\bf \Delta C_{\rm VL}^{\llx}} {\bf \Delta C_{\rm TL}^{\llx}} \bigg) \times 10^{-3}  \, 
\eea
Here, $\rm \Delta C^\ell_i$ correspond to the NP WCs $g_i^\ell$ of Eq.~\ref{eff-lag}. 
The above formulas are based on the analytic expressions of the decay amplitudes given above which are based
on ref \cite{Bardhan:2016uhr}. In order to obtain the various numerical coefficients, central values of the 
form-factors (see \cite{Bardhan:2016uhr} for more details) and other parameters have been used. 

If NP is assumed to exist only in the decay to $\tau$ leptons, one gets from the above formulas
\bea
\label{eq:rdappr}
R_D &=& 0.30 + 0.60 \, {\bf \Delta C_{\rm VL}^\tau} + 0.51 \, {\bf \Delta C_{\rm SL}^\tau} +  0.15 \, {\bf \Delta C_{\rm TL}^\tau}  \nn \\ 
&+& 0.30  \, {\bf (\Delta C_{\rm VL}^\tau)^2} + 0.40 \, { \bf (\Delta C_{\rm SL}^\tau)^2}  + 0.05 \, { \bf (\Delta C_{\rm TL}^\tau)^2}  \nn \\
&+&  0.51 \, {\bf \Delta C_{\rm VL}^\tau \Delta C_{\rm SL}^\tau} +  0.15 \, {\bf \Delta C_{\rm VL}^\tau \Delta C_{\rm TL}^\tau}  \, 
\eea

\subsection{Analytic formulas for $B \to D^\ast$ decay}
{\bf Negative helicity lepton} : 
\begin{eqnarray}
&& \hspace{-1cm} a_\ell^{D^\ast} (-) = \frac{8 M_B^2 \pds}{\mbmd^2} {\bf \cvlsq \vsq} + 
\frac{\mbmd^2 ( 8 M_{D^*}^2 q^2 + \lambda)}{2 M_{D^*}^2 q^2} {\bf \calsq \aosq}
\nonumber \\
&& + \frac{8 M_B^4 |p_{D^\ast}|^4 }{\mds \mbmd ^2 q^2} {\bf \calsq \atsq} 
- \frac{4\pds M_B^2 \mbmdqsq }{\mds q^2} {\bf \calsq  A_1 A_2}
\nn\\
&& +  m_\ell \left[\frac{32 M_B^2 \pds}{q^2 \mbmd} {\bf \re\left(\cvl \ctlconj \right)
V T_1} \nn \right. \\
&& + \left.  \frac{8 \mbmd \left(2 \mds \mbmdsq + M_B^2 \pds \right)}{q^2 \mds}
{\bf \re\left(\call \ctlconj \right) A_1 T_2} \nn \right. \\
&& \hspace{-10mm} - \left.  \frac{8 M_B^2 \mbmdqsq \pds}{q^2 \left(M_B - M_{D^\ast}\right) \mds}
{\bf \re \left(\call \ctlconj \right) A_1 T_3} \nonumber \right. 
 - \left.  \frac{8 M_B^2 \mbmdqthsq \pds}{q^2 \mbmd \mds} {\bf \re
\left(\call \ctlconj \right) A_2 T_2} \right. \nn \\
&& + \left.  \frac{32 M_B^4 |p_{D^\ast}|^4}{q^2 \mds \mbmd \mbmdsq} {\bf \re
\left(\call \ctlconj \right) A_2 T_3} \right] \nn \\
&& + m_\ell^2 \left[\frac{32 M_B^2 \pds}{q^4} {\bf \ctlsq T_1^2}  +
  \frac{2 \left(8 \mds\left(2 \left(M_B^2 + M_{D^\ast}^2\right) -
q^2\right) q^2 + \left(4 \mds + q^2\right)\lambda\right)}{q^4 \mds} {\bf \ctlsq
T_2^2} \nonumber \right. \nn\\
&& +  \left. \frac{32 M_B^4 |p_{D^\ast}|^4}{q^2 M_{D^\ast}^2 \mbmdsq^2}
{\bf \ctlsq T_3^2} - \frac{16 M_B^2 \pds \mbmdqthsq}{q^2 \mds \mbmdsq} {\bf
\ctlsq T_2 T_3} \right] \nn \\[10mm]
&&  \hspace{-1cm} b_\ell^{D^\ast} (-) = 
-16 |p_{D^\ast}| M_B {\bf \re \left(\cvl \calconj \right)  V A_1 }
-  m_\ell \left[\frac{32 M_B \left(M_B -
M_{D^\ast}\right)|p_{D^\ast}|}{q^2}
{\bf \re \left(\cvl \ctlconj \right) V T_2} \right. \nn \\ 
&& \hspace*{0cm} + \left.  \frac{32 M_B \left(M_B + M_{D^\ast}\right) |p_{D^\ast}| }{q^2} {\bf
\re \left(\call \ctlconj\right) A_1 T_1} \right] - m_\ell^2 \left[\frac{64 M_B \mbmdsq |p_{D^\ast}|}{q^4} {\bf \ctlsq T_1
T_2}\right]  \nn \\ \nn
\eea
\bea
&&  \hspace{-1cm} c_\ell^{D^\ast} (-) = \frac{8 \pds M_B^2}{\mbmd^2} {\bf{\cvlsq \vsq}} - 
\frac{\mbmd^2 \lambda}{2 \mds q^2} {\bf \calsq \aosq} \nn - \frac{8 |p_{D^\ast}|^4 M_B^4}{\mbmd^2 \mds q^2} {\bf \calsq \atsq} \nn\\
&& +  \frac{4 \pds M_B^2 \mbmdqsq}{\mds q^2} {\bf \calsq A_1 A_2} \nn \\
&& +  m_\ell \left[\frac{32 M_B^2 \pds}{q^2 \mbmd} {\bf \re\left(\cvl \ctlconj\right)
V T_1}  \nn \right.  \left. -\frac{8 M_B^2 \left(M_B + M_{D^\ast}\right)\pds }{q^2 \mds}{\bf
\re\left(\call \ctlconj \right) A_1 T_2} \right. \nn \\
&&  \hspace{-1cm} +  \left. \frac{8 M_B^2 \mbmdqsq \pds}{q^2 \mds \left(M_B - M_{D^\ast}\right) }
{\bf \re\left(\call \ctlconj \right) A_1 T_3 }  \right.  + \left.  \frac{8 M_B^2 \mbmdqthsq \pds}{q^2 \mds \mbmd }{\bf \re\left(\call
\ctlconj \right) A_2 T_2} \nn \right. \\ 
&& - \left.  \frac{32 M_B^4 |p_{D^\ast}|^4}{q^2 \mds \mbmd \mbmdsq} {\bf
\re\left(\call \ctlconj \right) A_2 T_3} \right. \nn \\
&& +   m_\ell^2 \left[\frac{32 M_B^2 \pds}{q^4} {\bf \ctlsq T_1^2} + \frac{2
\left(4\mds - q^2\right)\lambda}{\mds q^4} {\bf \ctlsq T_2^2} \nn \right. \\
&& -  \left.  \frac{32 M_B^4 |p_{D^\ast}|^4 }{q^2 \mds \mbmdsq^2} {\bf \ctlsq
T_3^2}  \right. 
 \left. + \frac{16 M_B^2 \pds \mbmdqthsq}{q^2 \mds \mbmdsq} {\bf \ctlsq
T_2 T_3} \right] 
\eea

{\bf Positive helicity lepton:} \\
\begin{eqnarray}
&& \hspace{-1cm} a_\ell^{D^\ast} (+) = \frac{8 \pds M_B^2}{\mbmc^2} {\bf \cplsq \azsq} +
\frac{32 M_B^2 \pds}{q^2} {\bf \ctlsq T_1^2} + \frac{8 \mbmdsq^2}{q^2} {\bf
\ctlsq T_2^2} \nn \\
&& - m_\ell \bigg[\frac{16 \pds M_B^2}{\mbmc q^2}{\bf \re \left(\call \cplconj
\right)\azsq}   - \frac{32 M_B^2 \pds}{q^2 \mbmd} {\bf \re\left(\cvl \ctlconj\right) V
T_1} \nonumber\\
&& -   \frac{8(M_B + M_{D^\ast})\mbmdsq}{q^2}{\bf \re \left(\call
\ctlconj\right) A_1 T_2} \bigg]\nn\\ 
&& +   m_\ell^2 \left[\frac{8 \pds M_B^2}{q^4} {\bf \calsq \azsq} 
+\frac{8 \pds M_B^2}{\mbmd^2 q^2} {\bf \cvlsq \vsq} \right. + \left.  \frac{2 \mbmd^2}{q^2} {\bf \calsq \aosq} \right.\nn\bigg]
\eea
\begin{eqnarray}
&& \hspace{-1cm} b_\ell^{D^\ast} (+) = \frac{8 M_B \mbmdqthsq |p_{D^\ast}|}{\mbmc M_{D^\ast}}
{\bf \re \left(\cpl \ctlconj\right) A_0 T_2}  \nn \\ 
&& -  \frac{32 M_B^3 |p_{D^\ast}|^3}{\mbmc M_{D^\ast} \mbmdsq} {\bf \re
\left(\cpl \ctlconj\right) A_0 T_3} \nonumber \\
&& +  m_\ell \left[\frac{ 4 |p_{D^\ast}| M_B \mbmd\mbmdqsq}{M_{D^\ast} \mbmc q^2}
{\bf \re \left( \call \cplconj\right) A_0 A_1} \right.\nn\\
&& - \left.  \frac{16}{\mbmc}\frac{|p_{D^\ast}|^3 M_B^3}{\mbmd M_{D^\ast} q^2}
{\bf \re \left(\call \cplconj\right) A_0 A_2} \right. \nn \\
&& - \left.  \frac{8 M_B \mbmdqthsq |p_{D^\ast}|}{M_{D^\ast} q^2} {\bf \re
\left(\call \ctlconj\right) A_0 T_2} \nonumber \right. + \left.  \frac{32 M_B^3 |p_{D^\ast}|^3}{q^2 M_{D^\ast} \mbmdsq} {\bf \re
\left(\call \ctlconj\right) A_0 T_3} \right. \nn \bigg]\nn\\
&& \hspace{-1cm} +   m_\ell^2 \left[- \frac{4|p_{D^\ast}| M_B \mbmd}{M_{D^\ast} q^4} \mbmdqsq
{\bf \calsq A_0 A_1} \right.  + \left.  \frac{16 |p_{D^\ast}|^3 M_B^3}{\mbmd M_{D^\ast} q^4}{\bf \calsq A_0
A_2} \right. \bigg] \nn 
\end{eqnarray}
\begin{eqnarray}
&&  \hspace{-1cm} c_\ell^{D^\ast} (+) = -\frac{32 M_B^2 \pds}{q^2} {\bf \ctlsq T_1^2} - \frac{2
\left(4 \mds - q^2\right)\lambda}{\mds q^2} {\bf \ctlsq T_2^2} \nn \\
&& +  \frac{32 M_B^4
|p_{D^\ast}|^4 }{\mds \mbmdsq^2 } {\bf \ctlsq T_3^2} 
-  \frac{16 M_B^2 \pds \mbmdqthsq}{\mds \mbmdsq} {\bf \ctlsq T_2 T_3} \nn
\\
&& -  m_\ell \left[\frac{32 M_B^2 \pds}{q^2 \mbmd} {\bf \re\left(\cvl \ctlconj\right)
V T_1}  - \frac{8 M_B^2 \left(M_B + M_{D^\ast}\right)\pds }{q^2 \mds}{\bf
\re\left(\call \ctlconj\right) A_1 T_2} \right. \nn \\
&& \hspace{-1cm} + \left. \frac{8 M_B^2 \mbmdqsq \pds}{q^2 \mds \left(M_B - M_{D^\ast}\right) }
{\bf \re\left(\call \ctlconj\right) A_1 T_3 } \nn \right.  + \left.  \frac{8 M_B^2 \mbmdqthsq \pds}{q^2 \mds \mbmd }{\bf \re\left(\call
\ctlconj\right) A_2 T_2} \nn \right. \\ 
&& - \left.  \frac{32 M_B^4 |p_{D^\ast}|^4}{q^2 \mds \mbmd \mbmdsq} {\bf
\re\left(\call \ctlconj\right) A_2 T_3} \right. \nn \\
&& +  m_\ell^2 \left[ - \frac{8 \pds M_B^2}{\mbmd^2 q^2} {\bf \cvlsq \vsq} +
\frac{\mbmd^2 \lambda}{2 \mds q^4} {\bf \calsq \aosq} \right. \nn\\
&& + \left.  \frac{8 |p_{D^\ast}|^4 M_B^4}{\mds \mbmd^2 q^4} {\bf \calsq \atsq}
\right. \left.- \frac{4 \pds M_B^2}{\mds q^4} \mbmdqsq {\bf \calsq A_1
A_2}\right]
\end{eqnarray}

\newpage

\subsection{Semi-numerical formulas for $R_{D^*}$}

\bea
\mathcal{B}\left(B\to D^\ast \tau \nu_\tau\right) &=& \bigg( 13.8 + 1.6 \ {\bf \Delta C_{\rm VL}^\tau} -26.1 \ {\bf \Delta C_{\rm AL}^\tau} + 1.6 \ {\bf \Delta C_{\rm PL}^\tau} \nn \\
&-& 28.8  \ {\bf \Delta C_{\rm TL}^\tau} + 0.8 \ \left(\bf \Delta C_{\rm VL}^\tau\right)^2 + 13.0 \ \left(\bf \Delta C_{\rm AL}^\tau \right)^2 \nn \\
&+& 0.6 \ \left(\bf \Delta C_{\rm PL}^\tau \right)^2 + 42.1 \ \left(\bf \Delta C_{\rm TL}^\tau\right)^2 
+ 5.4\  {\bf \Delta C_{\rm VL}^\tau}{\bf \Delta C_{\rm TL}^\tau} \nn\\
&-& 1.6\  {\bf \Delta C_{\rm AL}^\tau}{\bf \Delta C_{\rm PL}^\tau} + 34.2 \ {\bf \Delta C_{\rm AL}^\tau}{\bf \Delta C_{\rm TL}^\tau}\bigg) \times 10^{-3}    \nn     \\[3mm]
\mathcal{B}\left(B\to D^\ast \llx \nu_{\llx}\right) &=& 
\bigg( 54.9+ 11.9 \ {\bf \Delta C_{\rm VL}^{\llx}} -151.5 \ {\bf \Delta C_{\rm AL}^{\llx}} + 0.5 \ {\bf \Delta C_{\rm PL}^{\llx}} \nn \\
&-&  6.8  \ {\bf \Delta C_{\rm TL}^{\llx}} + 3.8 \ \left(\bf \Delta C_{\rm VL}^{\llx}\right)^2 + 51.1 \ \left(\bf \Delta C_{\rm AL}^{\llx}\right)^2 \nn \\
&+& 3.3 \ \left(\bf \Delta C_{\rm PL}^{\llx}\right)^2 + 163.4 \ \left(\bf \Delta C_{\rm TL}^{\llx}\right)^2 + 
1.9\  {\bf \Delta C_{\rm VL}^{\llx}}{\bf \Delta C_{\rm TL}^{\llx}} \nn \\
&-& 0.5\  {\bf \Delta C_{\rm AL}^{\llx}}{\bf \Delta C_{\rm PL}^{\llx}} + 6.6\  {\bf \Delta C_{\rm AL}^{\llx}}{\bf \Delta C_{\rm TL}^{\llx}} \bigg) \times 10^{-3}      \nn   \\[3mm]
R_{D^\ast} &=&  0.25 + 0.03 \ {\bf \Delta C_{\rm VL}^\tau} -0.48 \ {\bf \Delta C_{\rm AL}^\tau}  + 0.03 \ {\bf \Delta C_{\rm PL}^\tau}  - 0.52  \ {\bf \Delta C_{\rm TL}^\tau} \nn \\
&+& 0.01 \ \left(\bf \Delta C_{\rm VL}^\tau \right)^2 + 0.24 \ \left(\bf \Delta C_{\rm AL}^\tau \right)^2 + 0.01 \ \left(\bf \Delta C_{\rm PL}^\tau \right)^2 + 0.77 \ \left(\bf \Delta C_{\rm TL}^\tau \right)^2 \nn \\
&+& 0.10\ {\bf \Delta C_{\rm VL}^\tau}{\bf \Delta C_{\rm TL}^\tau} - 0.03\  {\bf \Delta C_{\rm AL}^\tau}{\bf \Delta C_{\rm PL}^\tau} + 0.62 \ {\bf \Delta C_{\rm AL}^\tau}{\bf \Delta C_{\rm TL}^\tau} \,    \nn
%
\label{eq:rdstappr}
\eea

\subsection{Semi-numerical formulas for $R_{\eta_c}$}
\label{app:formula-Bc-etac}
\vspace*{-7mm}
\bea
\mathcal{B}\left(B_c\to \eta_c \tau \nu_\tau\right) &=& \bigg( 1.4 + 2.9 \ {\bf \Delta C_{\rm VL}^\tau} 
+ 2.5\  {\bf \Delta C_{\rm SL}^\tau} + 0.6 \ {\bf \Delta C_{\rm TL}^\tau}  \nn \\
&+& 1.4 \ {\bf (\Delta C_{\rm VL}^\tau)^2} + 1.9\  {\bf (\Delta C_{\rm SL}^\tau)^2} + 0.2\  {\bf ( \Delta C_{\rm TL}^\tau)^2}   \nn \\ 
&+& 2.5\  {\bf \Delta C_{\rm VL}^\tau} {\bf  \Delta C_{\rm SL}^\tau}  + 0.6\  {\bf \Delta C_{\rm VL}^\tau} {\bf \Delta C_{\rm TL}^\tau} \bigg) 
\times 10^{-3} \\[2mm]
\mathcal{B}\left(B_c \to \eta_c \llx \, \nu_{\llx}\right) &=& \bigg( 4.6 + 9.3 \  {\bf \Delta C_{\rm VL}^{\llx}}  + 
0.4 \  {\bf \Delta C_{\rm SL}^{\llx}} + 0.2\  {\bf \Delta C_{\rm TL}^{\llx}} \nn \\
&+& 4.6\  {\bf (\Delta C_{\rm VL}^{\llx})^2} + 7.0\  {\bf (\Delta C_{\rm SL}^{\llx})^2} + 0.4\  {\bf (\Delta C_{\rm TL}^{\llx})^2}  \nn \\ 
&+& 0.4\  {\bf \Delta C_{\rm VL}^{\llx}}  {\bf \Delta C_{\rm SL}^{\llx}} 
+ 0.2\  {\bf \Delta C_{\rm VL}^{\llx}} {\bf \Delta C_{\rm TL}^{\llx}} \bigg) \times 10^{-3}  \\[2mm]
R_{\eta_c}&=&0.30 + 0.62 \ {\bf \Delta C_{\rm VL}^\tau} + 0.53 \ {\bf \Delta C_{\rm SL}^\tau}  + 0.13  \ {\bf \Delta C_{\rm TL}^\tau} \nn \\
&+& 0.30 \ \left(\bf \Delta C_{\rm VL}^\tau \right)^2 + 0.41 \ \left(\bf \Delta C_{\rm SL}^\tau \right)^2  + 0.04 \ 
\left(\bf \Delta C_{\rm TL}^\tau \right)^2 \nn \\
&+& 0.53\ {\bf \Delta C_{\rm VL}^\tau}{\bf \Delta C_{\rm SL}^\tau} + 0.13\ {\bf \Delta C_{\rm VL}^\tau}{\bf \Delta C_{\rm TL}^\tau}  \, . 
\eea

\subsection{Semi-numerical formulas for $R_{J/\psi}$}
\label{app:formula-Bc-Jpsi}
\vspace*{-7mm}
\bea
\mathcal{B}\left(B_c\to J/\psi \ \tau \nu_\tau\right) &=& \bigg( 3.1 + 0.1 \ {\bf \Delta C_{\rm VL}^\tau} - 6.2 \ {\bf \Delta C_{\rm AL}^\tau} 
+ 0.4 \ {\bf \Delta C_{\rm PL}^\tau} \nn \\
&-& 7.7  \ {\bf \Delta C_{\rm TL}^\tau} 
+ 3.1 \ \left(\bf \Delta C_{\rm AL}^\tau \right)^2 \nn \\
&+& 0.2 \ \left(\bf \Delta C_{\rm PL}^\tau \right)^2 + 8.2\ {\bf \Delta C_{\rm TL}^\tau}^2 
+ 0.3 \ {\bf \Delta C_{\rm VL}^\tau}{\bf \Delta C_{\rm TL}^\tau} \nn\\
&-& 0.4 \  {\bf \Delta C_{\rm AL}^\tau}{\bf \Delta C_{\rm PL}^\tau}  + 8.0 \  {\bf \Delta C_{\rm AL}^\tau}{\bf \Delta C_{\rm TL}^\tau} \bigg) 
\times 10^{-3} \\[2mm]
\mathcal{B}\left(B_c\to J/\psi \  \llx \nu_{\llx}\right) &=& 
\bigg( 10.8 + 0.7 \ {\bf \Delta C_{\rm VL}^{\llx}} -34.2 \ {\bf \Delta C_{\rm AL}^{\llx}} + 0.2 \ {\bf \Delta C_{\rm PL}^{\llx}} \nn \\
&-&  1.8  \ {\bf \Delta C_{\rm TL}^{\llx}} + 0.2 \ \left(\bf \Delta C_{\rm VL}^{\llx}\right)^2 + 10.6 \ \left(\bf \Delta C_{\rm AL}^{\llx}\right)^2 \nn \\
&+& 1.0 \ \left(\bf \Delta C_{\rm PL}^{\llx}\right)^2 
+ 36.5 \ \left(\bf \Delta C_{\rm TL}^{\llx}\right)^2 + 0.1 \  {\bf \Delta C_{\rm VL}^{\llx}}{\bf \Delta C_{\rm TL}^{\llx}} \nn \\
&-&0.2\  {\bf \Delta C_{\rm AL}^{\llx}}{\bf \Delta C_{\rm PL}^{\llx}} + 0.6\  {\bf \Delta C_{\rm AL}^{\llx}}{\bf \Delta C_{\rm TL}^{\llx}} \bigg) 
\times 10^{-3}\\[3mm]
R_{J/\psi} &=&  0.29 + 0.01 \ {\bf \Delta C_{\rm VL}^\tau} -0.57 \ {\bf \Delta C_{\rm AL}^\tau}  + 0.04 \ {\bf \Delta C_{\rm PL}^\tau} -0.71 
\ {\bf \Delta C_{\rm TL}^\tau} \nn \\
&+& 0.29 \ \left(\bf \Delta C_{\rm AL}^\tau \right)^2 + 
0.02 \ \left(\bf \Delta C_{\rm PL}^\tau \right)^2 + 0.76 \ \left(\bf \Delta C_{\rm TL}^\tau \right)^2 \nn \\
&+& 0.03\ {\bf \Delta C_{\rm VL}^\tau}{\bf \Delta C_{\rm TL}^\tau} -0.04\  {\bf \Delta C_{\rm AL}^\tau}{\bf \Delta C_{\rm PL}^\tau} 
+ 0.74 \ {\bf \Delta C_{\rm AL}^\tau}{\bf \Delta C_{\rm TL}^\tau} \, . \nn \\ 
\eea

\subsection{Combination of vector and scalar operators}
\label{sec:scalar-vector}

In this appendix, we briefly comment on the scenario where both vector and scalar operators are present (see, for example \cite{Bordone:2017bld}
for a model). In Fig.~\ref{fig:V-S}, we show
the allowed regions in the $C_{\rm VL}^\tau$  - $C_{\rm SL}^\tau$ plane assuming $C_{\rm VL}^\tau = - C_{\rm AL}^\tau$ and
$C_{\rm SL}^\tau = \pm C_{\rm PL}^\tau$. 

\begin{figure}[h!]
\centering
\begin{tabular}{cc}
\includegraphics[scale=0.55]{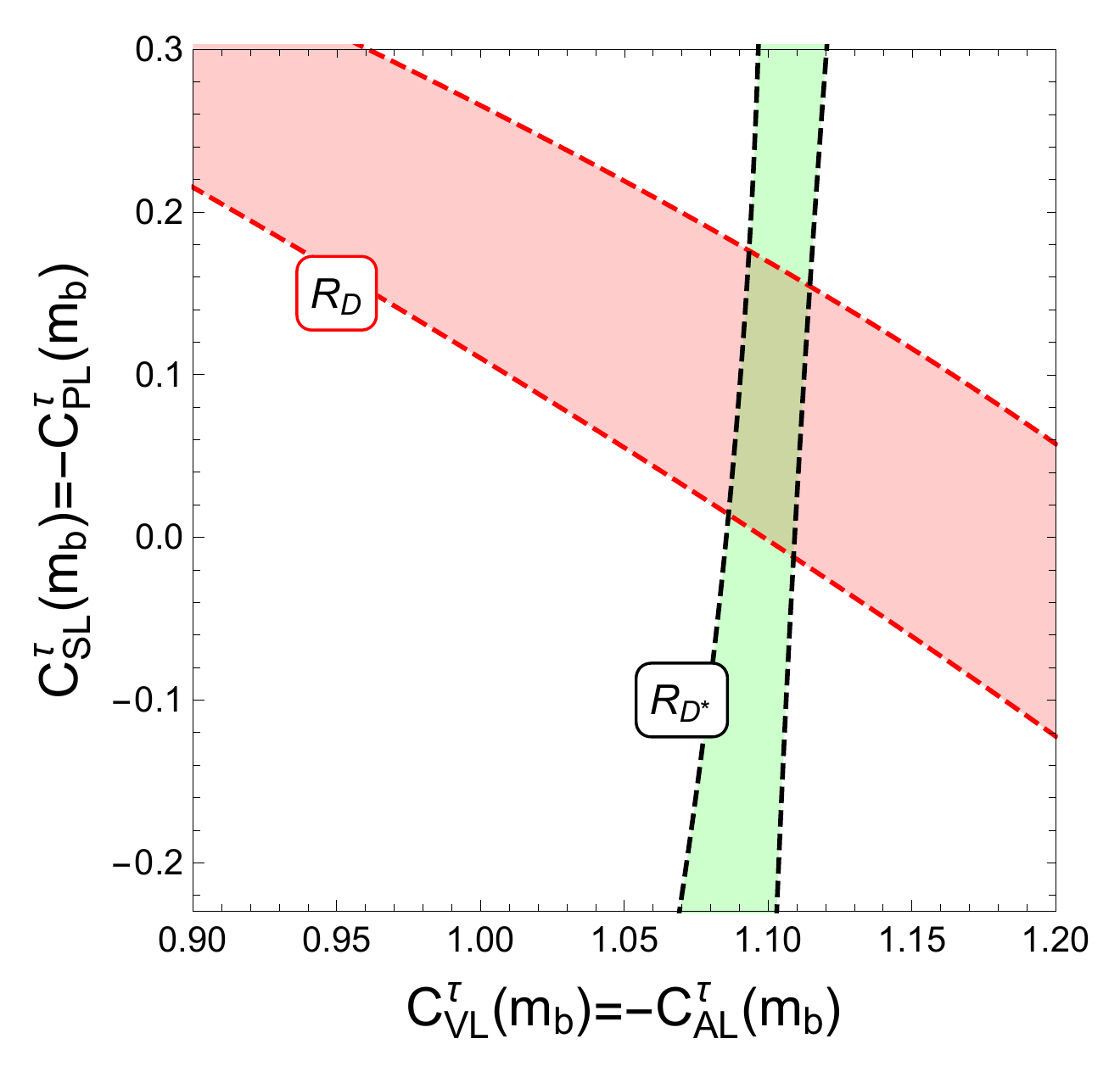} &   \includegraphics[scale=0.55]{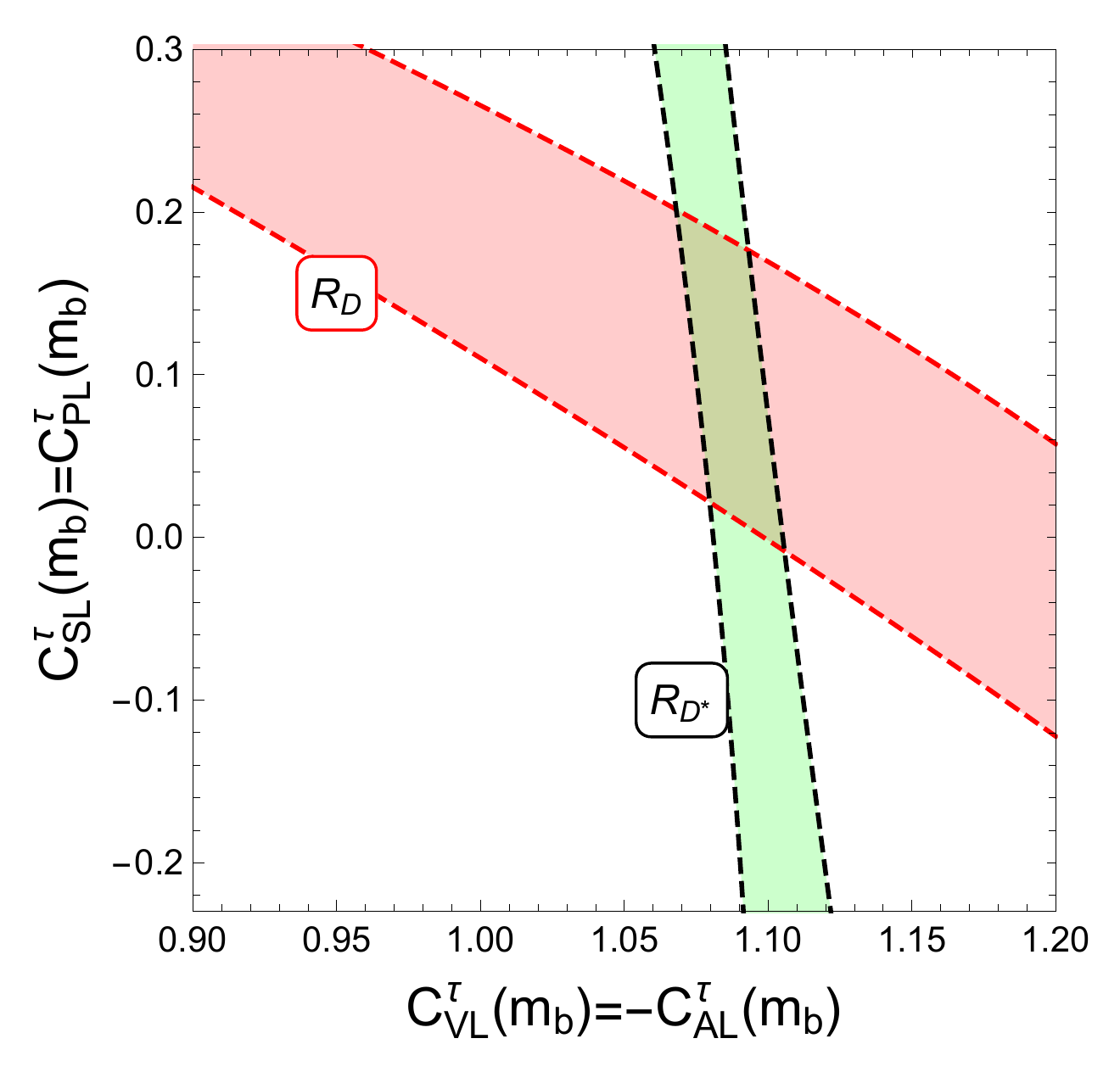} 
\end{tabular}
\caption{\sf The red and green shaded regions correspond to the values of $C_{\rm VL}^\tau$ $(= - C_{\rm AL}^\tau)$ and
$C_{\rm SL}^\tau$ ($=-C_{\rm PL}^\tau$ for the left panel and $=C_{\rm PL}^\tau$  for the right panel) that satisfy the
experimental measurement of $R_D$ and $R_{D^*}$ within $1\sigma$ respectively. 
\label{fig:V-S} }
\end{figure}

It can be seen that the overlap of the red and green regions (that corresponds to the simultaneous solution of $R_D$ and $R_{D^*}$)
touches the $C_{\rm SL}^\tau = \pm C_{\rm PL}^\tau = 0$ point. 
Thus, a combination of the vector and scalar operators extends the solution with only the vector operator discussed in section \ref{res:VA}.
Interestingly, if the red shaded region shrinks in the future due to more precise measurement of $R_D$ (without affecting the current
central value much), the combination of scalar and vector operators may lead to a better fit than with only vector operators.

\vspace*{-8mm}
\section{From the gauge to the mass eigenstates}
\label{app:gauge2mass}

\subsubsection*{$\boxed{ [C_{lq}^{(3)}]^\prime_{p'r's't'}  \left(\bar{l^\prime}_{p'} \gamma_\mu \sigma^I l^\prime_{r'} \right) 
\left( \bar{q^\prime}_{s'} \gamma^\mu \sigma^I q^\prime_{t'} \right)}$ :} 
Using Eq.~\ref{break-1} and the definitions from Eq.~\ref{rot-matrix}, we get, 
\bea
 &&\sum_{p',r',s',t'}\bigg[[C_{lq}^{(3)}]^\prime_{p'r's't'} 
\left(\bar{l^\prime}_{p'} \gamma_\mu \sigma^I 
l^\prime_{r'} \right) \left( \bar{q^\prime}_{s'} \gamma^\mu \sigma^I q^\prime_{t'} 
\right) \bigg]\nn\\
&=& \sum_{p,r,s,t}\Bigg[ \sum_{p',r',s',t'} [C_{lq}^{(3)}]^\prime_{p'r's't'} 
\bigg[
\vnd{L}{p} \vn{L}{r} \vud{L}{s} \vu{L}{t}
\left(\bar{\nu}_{p} \gamma ^{\mu } P_L \nu_{r}\right) \left(\bar{u}_{s} 
\gamma_{\mu  } P_L u_{t}\right) \nn\\
&& \hspace*{4.2cm}+  \ved{L}{p} \ve{L}{r} \vdd{L}{s} \vd{L}{t}
\left(\bar{e}_{p} \gamma ^{\mu} P_L e_{r}\right) 
\left(\bar{d}_{s} \gamma _{\mu} P_L d_{t}\right)
 \nn \\
&& \hspace*{4.2cm} - 
 \ved{L}{p} \ve{L}{r} \vud{L}{s} \vu{L}{t} 
\left(\bar{e}_{p} \gamma ^{\mu } P_L e_{r}\right) \left(\bar{u}_{s} \gamma _{\mu 
} P_L u_{t} \right)\nn\\
&& \hspace*{4.2cm} -   \vnd{L}{p} \vn{L}{r} \vdd{L}{s} \vd{L}{t}
\left(\bar{\nu}_{p} \gamma ^{\mu } P_L \nu_{r} 
\right) \left(\bar{d}_{s} \gamma_{\mu } P_L d_{t}\right) \nn \\
&& \hspace*{4.2cm} + 
 2 \vnd{L}{p} \ve{L}{r} \vdd{L}{s} \vu{L}{t}
 \left(\bar{\nu}_{p} \gamma ^{\mu } P_L e_{r}\right) 
\left(\bar{d}_{s} \gamma_{\mu} P_L u_{t} \right) \nn\\
&& \hspace*{4.2cm} + 2 \ved{L}{p} \vn{L}{r} \vud{L}{s} \vd{L}{t}
\left(\bar{e}_{p} \gamma ^{\mu } P_L 
\nu_{r}\right) \left(\bar{u}_{s} \gamma _{\mu} P_L 
d_{t}\right) \bigg] \Bigg]\nn\\
 &=& \sum_{p,r,s,t}\Bigg[ 
 \clqT{\nu}{\nu}{u}{u}
\left(\bar{\nu}_{p} \gamma ^{\mu } P_L \nu_{r}\right) \left(\bar{u}_{s} 
\gamma_{\mu  } P_L u_{t}\right)+
 \clqT{e}{e}{d}{d}
\left(\bar{e}_{p} \gamma ^{\mu} P_L e_{r}\right) 
\left(\bar{d}_{s} \gamma _{\mu} P_L d_{t}\right)
 \nn \\ 
&-&
 \clqT{e}{e}{u}{u} 
\left(\bar{e}_{p} \gamma ^{\mu } P_L e_{r}\right) \left(\bar{u}_{s} \gamma 
_{\mu 
} P_L u_{t} \right)
-
 \clqT{\nu}{\nu}{d}{d}
\left(\bar{\nu}_{p} \gamma ^{\mu } P_L \nu_{r} 
\right) \left(\bar{d}_{s} \gamma_{\mu } P_L d_{t}\right) \nn \\
&+ & \hspace*{-3.5mm}
2 \clqT{\nu}{e}{d}{u}
 \left(\bar{\nu}_{p} \gamma ^{\mu } P_L e_{r}\right) 
\left(\bar{d}_{s} \gamma_{\mu} P_L u_{t} \right) 
+
2 \clqT{e}{\nu}{u}{d}
\left(\bar{e}_{p} \gamma ^{\mu } P_L 
\nu_{r}\right) \left(\bar{u}_{s} \gamma _{\mu} P_L 
d_{t}\right)  \Bigg], \label{op1-massbasis}
\eea

where
\bea
\sum_{p',r',s',t'} [C_{lq}^{(3)}]^\prime_{p'r's't'} \vnd{L}{p} \vn{L}{r} 
\vud{L}{s} \vu{L}{t}  &\equiv& \clqT{\nu}{\nu}{u}{u} \nn\\
\sum_{p',r',s',t'} [C_{lq}^{(3)}]^\prime_{p'r's't'} \ved{L}{p} \ve{L}{r} 
\vdd{L}{s} \vd{L}{t} &\equiv& \clqT{e}{e}{d}{d}  \nn\\
\sum_{p',r',s',t'} [C_{lq}^{(3)}]^\prime_{p'r's't'} \ved{L}{p} \ve{L}{r} 
\vud{L}{s} \vu{L}{t} &\equiv&  \clqT{e}{e}{u}{u}  \label{op1-massbasis-def} \\
\sum_{p',r',s',t'} [C_{lq}^{(3)}]^\prime_{p'r's't'} \vnd{L}{p} \vn{L}{r} 
\vdd{L}{s} \vd{L}{t} &\equiv& \clqT{\nu}{\nu}{d}{d}  \nn\\
\sum_{p',r',s',t'} [C_{lq}^{(3)}]^\prime_{p'r's't'} \vnd{L}{p} \ve{L}{r} 
\vdd{L}{s} \vu{L}{t} &\equiv& \clqT{\nu}{e}{d}{u} \nn\\
\sum_{p',r',s',t'} [C_{lq}^{(3)}]^\prime_{p'r's't'} \ved{L}{p} \vn{L}{r} 
\vud{L}{s} \vd{L}{t} &\equiv& \clqT{e}{\nu}{u}{d} \nn 
\eea

\subsubsection*{ $\boxed{ [C_{\phi l}^{(3)}]^\prime_{p'r'} \left(\phi^\dagger i 
\overleftrightarrow{D}_\mu^I \phi \right) 
\left( \bar{l^\prime}_{p'} \, \sigma^I \gamma^\mu \, l^\prime_{r'} 
\right) }$ : }

\begin{eqnarray}
&& \sum_{p',r'}[C_{\phi l}^{(3)}]^\prime_{p'r'} \left(\phi^\dagger i 
\overleftrightarrow{D}_\mu^I \phi \right) 
\left( \bar{l^\prime}_{p'} \, \sigma^I\gamma^\mu \, l^\prime_{r'} 
\right)  \nn\\
&=&  \left( v^2+  2v h +h^2 \right)   \sum_{p,r} \Bigg[ \sum_{p',r'}[C_{\phi 
l}^{(3)}]^\prime_{p'r'}\bigg[ - \frac12 
\frac{g_2}{{\rm cos}\theta_W} \vnd{L}{p} \vn{L}{r} Z_\mu 
\left(\overline{\nu}_{p} \gamma ^{\mu } 
P_L \nu _{r}\right) \nn\\
&+& \frac12   \frac{g_2}{{\rm cos}\theta_W} \ved{L}{p} \ve{L}{r}  Z_\mu 
\left(\overline{e}_{p} 
\gamma ^{\mu } P_L e _{r}\right) 
- \frac{g_2}{ \sqrt{2}} \vnd{L}{p} \ve{L}{r}   W^+_\mu 
\left(\overline{\nu}_{p} \gamma ^{\mu } P_L e _{r}\right) \nn\\
&-& \frac{g_2}{ \sqrt{2}} \ved{L}{p} \vn{L}{r}   W^-_\mu 
\left(\overline{e}_{p} \gamma ^{\mu } P_L \nu _{r}\right)  \bigg]\Bigg]  \nn\\
&=&  \left( v^2+  2v h +h^2 \right)   \sum_{p,r} \Bigg[  - \frac12 
\frac{g_2}{{\rm cos}\theta_W} \cplT{\nu}{\nu} Z_\mu 
\left(\overline{\nu}_{p} \gamma ^{\mu } 
P_L \nu _{r}\right) \nn\\
&+& \frac12   \frac{g_2}{{\rm cos}\theta_W} \cplT{e}{e}   
Z_\mu 
\left(\overline{e}_{p} 
\gamma ^{\mu } P_L e _{r}\right) 
- \frac{g_2}{ \sqrt{2}} \cplT{\nu}{e}    W^+_\mu 
\left(\overline{\nu}_{p} \gamma ^{\mu } P_L e _{r}\right) \nn\\
&-& \frac{g_2}{ \sqrt{2}} \cplT{e} {\nu}  W^-_\mu 
\left(\overline{e}_{p} \gamma ^{\mu } P_L \nu _{r}\right)  \Bigg],
\end{eqnarray}
where
\begin{eqnarray}
  \sum_{p',r'} [C_{\phi l}^{(3)}]^{\prime}_{p'r'} \vnd{L}{p} \vn{L}{r}  &=& 
\cplT{\nu}{\nu} \nn\\
\sum_{p',r'} [C_{\phi l}^{(3)}]^{\prime}_{p'r'} \ved{L}{p} \ve{L}{r} 
&=& \cplT{e}{e} \nn\\
\sum_{p',r'} [C_{\phi l}^{(3)}]^{\prime}_{p'r'} \vnd{L}{p} \ve{L}{r} 
&=& \cplT{\nu}{e} \nn\\
\sum_{p',r'} [C_{\phi l}^{(3)}]^{\prime}_{p'r'} \ved{L}{p} \vn{L}{r} 
&=& \cplT{e}{\nu} 
\end{eqnarray}

\subsubsection*{$\boxed{[C_{ledq}]^\prime_{p'r's't'}  \left(\bar{l^\prime}_{p'}^j e^\prime_{r'} \right) 
\left( \bar{d^\prime}_{s'} q_{t'}^{\prime j} \right)}$ : }

\begin{eqnarray}
&&\sum_{p',r',s',t'} [C_{ledq}]^\prime_{p'r's't'}  \left(\bar{l^\prime}_{p'}^j 
e^\prime_{r'} \right) 
\left( \bar{d^\prime}_{s'} q_{t'}^{\prime j} \right) \nn\\
&=& 
\sum_{p,r,s,t}\Bigg[ \sum_{p',r',s',t'} [C_{ledq}]^\prime_{p'r's't'} \bigg[
\vnd{L}{p} \ve{R}{r} \vdd{R}{s} \vu{L}{t} 
\left(\bar{\nu}_{p} P_R e_{r}\right) 
\left(\bar{d}_{s} P_L u_{t}\right) \nn\\
&+& \ved{L}{p} \ve{R}{r} \vdd{R}{s} \vd{L}{t} \left(\bar{e}_{p} P_R 
e_{r}\right) \left(\bar{d}_{s} P_L     d_{t}\right) \bigg]\Bigg]\nn\\
&=& 
\sum_{p,r,s,t}\Bigg[ \cledq{\nu}{e}{d}{u}
\left(\bar{\nu}_{p} P_R e_{r}\right) 
\left(\bar{d}_{s} P_L u_{t}\right) + 
 \cledq{e}{e}{d}{d} \left(\bar{e}_{p} P_R 
e_{r}\right) \left(\bar{d}_{s} P_L     d_{t}\right)\Bigg], \label{op2-massbasis}
\end{eqnarray}
where
\begin{eqnarray}
\sum_{p',r',s',t'} [C_{ledq}]^\prime_{p'r's't'} \vnd{L}{p} \ve{R}{r} \vdd{R}{s} 
\vu{L}{t}  &=& \cledq{\nu}{e}{d}{u} \nn\\
\sum_{p',r',s',t'} [C_{ledq}]^\prime_{p'r's't'} \ved{L}{p} \ve{R}{r} \vdd{R}{s} 
\vd{L}{t}  &=&  \cledq{e}{e}{d}{d} \label{op2-massbasis-def}
\end{eqnarray}

\subsubsection*{$\boxed{ [C_{lequ}^{(1)}]^{\prime}_{p'r's't'}\left(\bar{l^\prime}_{p'}^j e^\prime_{r'} 
\right) \epsilon_{jk} \left( \bar{q^\prime}_{s'}^k u^\prime_{t'} \right) }$ : }

\begin{eqnarray}
&& \sum_{p',r',s',t'} 
[C_{lequ}^{(1)}]^{\prime}_{p'r's't'}\left(\bar{l^\prime}_{p'}^j e^\prime_{r'} 
\right) \epsilon_{jk} \left( 
\bar{q^\prime}_{s'}^k u^\prime_{t'} \right)\nn\\
&=& \sum_{p,r,s,t}\Bigg[ \sum_{p',r',s',t'} 
[C_{lequ}^{(1)}]^{\prime}_{p'r's't'} \bigg[
\vnd{L}{p} \ve{R}{r} \vdd{L}{s} \vu{R}{t} 
\left(\bar{\nu}_{p} P_R e_{r}\right) \left(\bar{d}_{s} P_R u_{t}\right) \nn\\
&-& \ved{L}{p} \ve{R}{r} \vud{L}{s} \vu{R}{t} 
\left(\bar{e}_{p} P_R e_{r}\right) \left(\bar{u}_{s} P_R    u_{t}\right) 
\bigg]\Bigg]  \nn\\
&& \hspace{-1.2cm} = \sum_{p,r,s,t}\Bigg[\cledqO {\nu}{e}{d}{u}
\left(\bar{\nu}_{p} P_R e_{r}\right) \left(\bar{d}_{s} P_R u_{t}\right) -
\cledqO{e}{e}{d}{u}
\left(\bar{e}_{p} P_R e_{r}\right) \left(\bar{u}_{s} P_R    u_{t}\right) \Bigg],
\end{eqnarray}
where
\begin{eqnarray}
  \sum_{p',r',s',t'} [C_{lequ}^{(1)}]^{\prime}_{p'r's't'} \vnd{L}{p} \ve{R}{r} 
\vdd{L}{s} \vu{R}{t}  &=& \cledqO{\nu}{e}{d}{u} \nn\\
\sum_{p',r',s',t'} [C_{lequ}^{(1)}]^{\prime}_{p'r's't'} \ved{L}{p} \ve{R}{r} 
\vud{L}{s} \vu{R}{t}  &=& \cledqO{e}{e}{d}{u}
\end{eqnarray}

Thus, 
\bea
&&\Delta C_{\rm SL}^{cb\tau\nu_3} = \frac12 \frac{\Lambda_{\rm SM}^2}{\Lambda^2}
\left( [\tilde C_{ledq}^{ \nu e d u}]_{3332}   +  [\tilde C_{le qu}^{ (1) \nu e d u}]_{3332}  \right)^* \, , \\
&& \Delta C_{\rm PL}^{cb\tau\nu_3} = \frac12 \frac{\Lambda_{\rm SM}^2}{\Lambda^2}
\left( [\tilde C_{ledq}^{ \nu e d u}]_{3332}   -  [\tilde C_{le qu}^{ (1) \nu e d u}]_{3332}  \right)^*  \, . 
\eea

\subsubsection*{$\boxed{ [C_{lequ}^{(3)}]^{\prime}_{p'r's't'} 
\left(\bar{l^\prime}_{p'}^j \sigma^{\mu\nu} e^\prime_{r'} \right) \epsilon_{jk} 
\left( \bar{q^\prime}_{s'}^k \sigma_{\mu\nu} u^\prime_{t'} \right)}$ : }
\begin{eqnarray}
&&  \sum_{p',r',s',t'} [C_{lequ}^{(3)}]^{\prime}_{p'r's't'} 
\left(\bar{l^\prime}_{p'}^j \sigma^{\mu\nu} e^\prime_{r'} \right) \epsilon_{jk} 
\left( \bar{q^\prime}_{s'}^k \sigma_{\mu\nu} u^\prime_{t'} \right)\nn\\
&= & \sum_{p,r,s,t}\Bigg[  \sum_{p',r',s',t'} 
[C_{lequ}^{(3)}]^{\prime}_{p'r's't'} \bigg[
\vnd{L}{p} \ve{R}{r} \vdd{L}{s} \vu{R}{t} 
\left(\bar{\nu}_{p} \sigma^{\mu\nu} P_R e_{r}\right) 
\left(\bar{d}_{s}   \sigma_{\mu\nu} P_R u_{t}\right) \nn\\
&- & \ved{L}{p} \ve{R}{r} \vud{L}{s} \vu{R}{t} 
\left(\bar{e}_{p} \sigma^{\mu\nu} P_R e_{r}\right) 
\left(\bar{u}_{s} 
\sigma_{\mu\nu} P_R    u_{t}\right) \bigg]\Bigg]\nn\\
&= & \sum_{p,r,s,t}\Bigg[  \cledqT{\nu}{e}{d}{u}
\left(\bar{\nu}_{p} \sigma^{\mu\nu} P_R e_{r}\right) 
\left(\bar{d}_{s}   \sigma_{\mu\nu} P_R u_{t}\right) -  \cledqT{e}{e}{d}{u}
\left(\bar{e}_{p} \sigma^{\mu\nu} P_R e_{r}\right) 
\left(\bar{u}_{s} 
\sigma_{\mu\nu} P_R    u_{t}\right) \Bigg], \nn \\
\end{eqnarray}
where
\begin{eqnarray}
  \sum_{p',r',s',t'} [C_{lequ}^{(3)}]^{\prime}_{p'r's't'} \vnd{L}{p} \ve{R}{r} 
\vdd{L}{s} \vu{R}{t}  &=& \cledqT{\nu}{e}{d}{u} \nn\\
\sum_{p',r',s',t'} [C_{lequ}^{(3)}]^{\prime}_{p'r's't'} \ved{L}{p} \ve{R}{r} 
\vud{L}{s} \vu{R}{t}  &=& \cledqT{e}{e}{d}{u}
\end{eqnarray}

Thus,
\bea
\Delta C_{\rm TL}^{cb\tau\nu_3} = \frac12  
\frac{\Lambda_{\rm SM}^2}{\Lambda^2} 
\left( [\tilde C_{le qu}^{ (3) \nu e d u}]_{3332}  \right)^* \, .
\eea

\section{Mixing of $[C_{l q}^{(3,1)}]^\prime$ and $[C_{\phi l}^{(3,1)}]^\prime$}
\label{Ztautau-running}

The $\beta$-functions of $[C_{\phi l}^{(3)}]^\prime_{33}$ and $[C_{l q}^{(3)}]^\prime_{3333}$ can be approximately 
written as (assuming that no other couplings are generated at the matching scale $\Lambda$) \cite{Alonso:2013hga,Jenkins:2013wua}
\bea
16
\pi^2\frac{d }{d \log\mu}  [C_{\phi l}^{(3)}]^\prime_{33}  &=&
\left( -5 g_2^2 +6 y_t^2+6 y_b^2+4 y_\tau^2\right) [C_{\phi l}^{(3)}]^\prime_{33} + 3y _\tau^2  \, [C_{\phi l}^{(1)}]^\prime_{33} \nonumber\\
&& + \left( 2 g_2^2 -6 y_b^2 -6 y_t^2 \right) [C_{l q}^{(3)}]^\prime_{3333} \\
16
\pi^2\frac{d }{d \log\mu}  [C_{\phi l}^{(1)}]^\prime_{33}  &=&
\left( \frac{1}{3} g_1^2 +6 y_t^2+6 y_b^2+6 y_\tau^2\right) [C_{\phi l}^{(1)}]^\prime_{33} + 9 y _\tau^2  \, [C_{\phi l}^{(3)}]^\prime_{33} \nonumber\\
&& +  \left( \frac{2}{3} g_1^2 -6 y_b^2 +6 y_t^2 \right)  [C_{l q}^{(1)}]^\prime_{3333}
\eea
This gives
\begin{eqnarray}
[C_{\phi l}^{(3)}]'_{33} (m_t) \simeq  \, 0.027  \, [C_{l q}^{(3)}]'_{3333} (\Lambda) \, \log(\Lambda/m_t) \, ,
\end{eqnarray}
\begin{eqnarray}
[C_{\phi l}^{(1)}]'_{33} (m_t) \simeq  \, -0.034 \, [C_{l q}^{(1)}]'_{3333} (\Lambda) \, \log(\Lambda/m_t) \, . 
\end{eqnarray}

Thus, we get 
\bea
\Delta g_L^\tau &\simeq&  \frac{1}{2} \,  \left( [C_{\phi l}^{(3)}]^\prime_{33} (m_t) + [C_{\phi l}^{(3)}]^{\prime \, *}_{33} (m_t) 
+ [C_{\phi l}^{(1)}]^\prime_{33} (m_t) + [C_{\phi l}^{(1)}]^{\prime \, *}_{33} (m_t)\right) \, \frac{v^2}{\Lambda^2} \nonumber \\
&\simeq&  \left( 0.0014 \, ([C_{l q}^{(3)}]'_{3333} + [C_{l q}^{(3)}]^{' \, *}_{3333}) - 0.0018 \, ([C_{l q}^{(1)}]'_{3333} + [C_{l q}^{(1)}]^{' \, *}_{3333})\right) 
\nonumber \\
&& \hspace*{5cm} \times \left(\frac{\text{TeV}}{\Lambda}\right)^2 \left(1+ 0.6 \log(\Lambda/\text{TeV}) \right) 
\eea
\bea
\Delta g_L^\nu &\simeq&  \frac{1}{2} \,  \left( - [C_{\phi l}^{(3)}]^\prime_{33} (m_t) - [C_{\phi l}^{(3)}]^{\prime \, *}_{33} (m_t) 
+ [C_{\phi l}^{(1)}]^\prime_{33} (m_t) + [C_{\phi l}^{(1)}]^{\prime \, *}_{33} (m_t)\right) \, \frac{v^2}{\Lambda^2} \nonumber \\
&\simeq&  - \left( 0.0014 \, ([C_{l q}^{(3)}]'_{3333} + [C_{l q}^{(3)}]^{' \, *}_{3333}) + 0.0018 \, ([C_{l q}^{(1)}]'_{3333} + [C_{l q}^{(1)}]^{' \, *}_{3333})\right) 
\nonumber \\
&& \hspace*{5cm} \times \left(\frac{\text{TeV}}{\Lambda}\right)^2 \left(1+ 0.6 \log(\Lambda/\text{TeV}) \right) 
\eea
\bea
\Delta g_W^\tau &\simeq&  - \left( [C_{\phi l}^{(3)}]^\prime_{33} (m_t) + [C_{\phi l}^{(3)}]^{\prime \, *}_{33} (m_t) \right)
\, \frac{v^2}{\Lambda^2} \nonumber \\
&\simeq&  -0.0028 \left( [C_{l q}^{(3)}]'_{3333} + [C_{l q}^{(3)}]^{' \, *}_{3333} \right) 
\left(\frac{\text{TeV}}{\Lambda}\right)^2 \left(1+ 0.6 \log(\Lambda/\text{TeV}) \right)
\eea

Using $|\Delta g_L^\tau| \lesssim 6\times 10^{-4}$, and in the absence of $[C_{l q}^{(1)}]'_{3333}$, we get 
\bea
\left|[C_{l q}^{(3)}]'_{3333} + [C_{l q}^{(3)}]^{' \, *}_{3333}\right| &\lesssim&  
\frac{0.43}{\left(1+ 0.6 \log(\Lambda/\text{TeV}) \right)} \left(\frac{\Lambda}{\text{TeV}}\right)^2 \\
\eea
 In the presence of both $[C_{l q}^{(1)}]'_{3333}$ and $[C_{l q}^{(3)}]'_{3333}$, combining all the constraints on $\Delta g_W^\tau$,
$\Delta g_L^\tau$ and $|\Delta g_L^\nu|< 1.2 \times 10^{-3}$ we get
\bea
\label{eq:running}
&\left | [C_{l q}^{(1)}]'_{3333} + [C_{l q}^{(1)}]^{' \, *}_{3333}  \right| \lesssim  
\frac{0.5}{\left(1+ 0.6 \log(\Lambda/\text{TeV}) \right)} \left(\frac{\Lambda}{\text{TeV}}\right)^2 \nonumber\\
&\frac{-0.63}{\left(1+ 0.6 \log(\Lambda/\text{TeV}) \right)} \left(\frac{\Lambda}{\text{TeV}}\right)^2\lesssim  [C_{l q}^{(3)}]'_{3333} + [C_{l q}^{(3)}]^{' \, *}_{3333}  \lesssim 
\frac{0.14}{\left(1+ 0.6 \log(\Lambda/\text{TeV}) \right)} \left(\frac{\Lambda}{\text{TeV}}\right)^2 \, .
\eea


\section{Constraints from $Z$ interactions with fermions}
\label{sec:Zint}
%
One of the advantages of the MCHM5 model is that it can provide protection for some of the
 $g_{Z}^\tau,g_{Z}^b, g_{Z}^\nu$ couplings.
Indeed, discrete  $P_{LR}$ symmetry \cite{Agashe:2006at} protects $g_{Z}^\tau$ ,  however it cannot protect $(g_Z^\nu, g_Z^\tau,g_W^\tau)$
at the same time.  
Indeed let us consider the leptonic part of the lagrangian in Eq.\ref{Fmixing} 
and allow the splitting of the mixing parameters defined in the Eq.\ref{splitting}:
\bea
{\cal L}&=&i  \overline{\tilde{\mathcal{O}}}_{l_1}\left( \slash{D}+i \slash E \right)\tilde{\mathcal{O}}_{l_1}+
i \overline{\tilde{\cal O}}_{l_2}\left( \slash{D}+i \slash E \right){\cal O}_{l_2}+\left (i c_1  \overline{\tilde{\cal O}}_{l_1}^i \slash d_i \tilde{\cal O}_N+ 
i c_2 \overline{\tilde{\cal O}}_{l_2}^i \slash d_i\tilde{\mathcal{O}}_e+ h.c.\right)\nonumber\\
&-& m_4^{(1)}   \overline{\tilde{\mathcal{O}}}_{l_1}\tilde{\cal O}_{l_1}- m_4^{(2)}   
\overline{\tilde{\mathcal{O}}}_{l_2}\tilde{\cal O}_{l_2}-m_1^{(e)}   \overline{\tilde{\mathcal{O}}}_{e}\tilde{\cal O}_{e}- m_1^{(N)}   
\overline{\tilde{\mathcal{O}}}_{N}\tilde{\cal O}_{N}\nonumber\\
&+&\lambda^{(4)}_l \overline {\tilde {l}}_L U(h)_{I i} {\cal O}_{l_1}+\lambda^{(1)}_l \overline {\tilde {l}}_L U(h)_{I 5} {\cal O}_{N}
+\tilde\lambda^{(4)}_l \overline {\tilde {l}}_L U(h)_{I i} {\cal O}_{l_2}+\tilde\lambda^{(1)}_l \overline {\tilde {l}}_L U(h)_{I 5} {\cal O}_{e} \, . 
\eea
Then the modifications to $g_{Z}^\tau,\, g_{Z}^\nu$ can be read-off from the Refs.~\cite{Grojean:2013qca,Azatov:2016xik}, where
analogous discussion was applied to the top quark, so that 
\bea
\delta g_Z^\tau=-\frac{v^2}{4 f^2}\frac{M_*^2\left[
\left(\tilde \lambda_l^{(4)} m_1^{(e)}\right)^2 +\left(\tilde \lambda_l^{(1)} m_4^{(2)} \right)^2 -2 \sqrt 2 c_2\tilde \lambda_l^{(4)}  \tilde \lambda_l^{(1)} m_1^{(e)}  m_4^{(2)}
\right]}{\left(m_1^{(e)}\right)^2\left(   \left(m_4^{(2)}\right)^2+\left(\tilde \lambda_l^{(4)} M_*\right)^2 \right)},\nonumber\\
\delta g_Z^\nu=-\frac{v^2}{4 f^2}\frac{M_*^2\left[
\left( \lambda_l^{(4)} m_1^{(N)}\right)^2 +\left( \lambda_l^{(1)} m_4^{(1)} \right)^2 -2 \sqrt 2 c_1 \lambda_l^{(4)}   \lambda_l^{(1)} m_1^{(N)}  m_4^{(2)}
\right]}{\left(m_1^{(N)}\right)^2\left(   \left(m_4^{(1)}\right)^2+\left( \lambda_l^{(4)} M_*\right)^2 \right)}.\nonumber\\
\eea
We can see that $P_{LR}$ symmetry forces the $\delta g_{Z}^\tau$ to depend only on $\tilde \lambda_l^{(1,4)}$ and $\delta g_{Z}^\nu$ on $ \lambda_l^{(1,4)}$.
Since the bound on $g_{Z}^\tau$ is a bit stronger, it is natural to assume that $ \lambda_{l}> \tilde \lambda_{l}$  and the contribution to 
$R_{D,D*}$ is dominated by $\lambda_{l}$. Note that this coupling does not enter the leading expression of the $\tau$ mass which scales as
\bea
m_\tau \propto \lambda_e \tilde \lambda_l^{(1,4)} \, .
\eea
Then in order to pass the constraints from $g_{Z}^\nu$ we will have to tune additionally the parameter $c_1$ as was suggested in
\cite{Barbieri:2017tuq}. 

\end{appendices}


\begin{thebibliography}{100}

\bibitem{Aoki:2016frl}
S.~Aoki et~al., \emph{{Review of lattice results concerning low-energy particle
  physics}}, \href{http://dx.doi.org/10.1140/epjc/s10052-016-4509-7}{\emph{Eur.
  Phys. J.} {\bf C77} (2017) 112}, [\href{http://arxiv.org/abs/1607.00299}{{\tt
  1607.00299}}].

\bibitem{Na:2015kha}
{\scshape HPQCD} collaboration, H.~Na, C.~M. Bouchard, G.~P. Lepage, C.~Monahan
  and J.~Shigemitsu, \emph{{$B \rightarrow D l \nu$ form factors at nonzero
  recoil and extraction of $|V_{cb}|$}},
  \href{http://dx.doi.org/10.1103/PhysRevD.93.119906,
  10.1103/PhysRevD.92.054510}{\emph{Phys. Rev.} {\bf D92} (2015) 054510},
  [\href{http://arxiv.org/abs/1505.03925}{{\tt 1505.03925}}]. [Erratum: Phys.
  Rev.D93,no.11,119906(2016)].

\bibitem{Amhis:2016xyh}
Y.~Amhis et~al., \emph{{Averages of $b$-hadron, $c$-hadron, and $\tau$-lepton
  properties as of summer 2016, \textnormal{Online update at}
  \url{http://www.slac.stanford.edu/xorg/hflav/semi/fpcp17/RDRDs.html}}},
  \href{http://arxiv.org/abs/1612.07233}{{\tt 1612.07233}}.

\bibitem{Bigi:2016mdz}
D.~Bigi and P.~Gambino, \emph{{Revisiting $B\to D \ell \nu$}},
  \href{http://dx.doi.org/10.1103/PhysRevD.94.094008}{\emph{Phys. Rev.} {\bf
  D94} (2016) 094008}, [\href{http://arxiv.org/abs/1606.08030}{{\tt
  1606.08030}}].

\bibitem{Fajfer:2012vx}
S.~Fajfer, J.~F. Kamenik and I.~Nisandzic, \emph{{On the $B \to D^* \tau \bar
  \nu_{\tau}$ Sensitivity to New Physics}},
  \href{http://dx.doi.org/10.1103/PhysRevD.85.094025}{\emph{Phys. Rev.} {\bf
  D85} (2012) 094025}, [\href{http://arxiv.org/abs/1203.2654}{{\tt
  1203.2654}}].

\bibitem{Bigi:2017jbd}
D.~Bigi, P.~Gambino and S.~Schacht, \emph{{$R(D^*)$, $|V_{cb}|$, and the Heavy
  Quark Symmetry relations between form factors}},
  \href{http://dx.doi.org/10.1007/JHEP11(2017)061}{\emph{JHEP} {\bf 11} (2017)
  061}, [\href{http://arxiv.org/abs/1707.09509}{{\tt 1707.09509}}].

\bibitem{Hirose:2016wfn}
{\scshape Belle} collaboration, S.~Hirose et~al., \emph{{Measurement of the
  $\tau$ lepton polarization and $R(D^*)$ in the decay $\bar{B} \to D^* \tau^-
  \bar{\nu}_\tau$}},
  \href{http://dx.doi.org/10.1103/PhysRevLett.118.211801}{\emph{Phys. Rev.
  Lett.} {\bf 118} (2017) 211801}, [\href{http://arxiv.org/abs/1612.00529}{{\tt
  1612.00529}}].

\bibitem{Hirose:2017dxl}
{\scshape Belle} collaboration, S.~Hirose et~al., \emph{{Measurement of the
  $\tau$ lepton polarization and $R(D^*)$ in the decay $\bar{B} \rightarrow D^*
  \tau^- \bar{\nu}_\tau$ with one-prong hadronic $\tau$ decays at Belle}},
  \href{http://arxiv.org/abs/1709.00129}{{\tt 1709.00129}}.

\bibitem{Aaij:2017tyk}
{\scshape LHCb} collaboration, R.~Aaij et~al., \emph{{Measurement of the ratio
  of branching fractions
  $\mathcal{B}(B_c^+\,\to\,J/\psi\tau^+\nu_\tau)$/$\mathcal{B}(B_c^+\,\to\,J/\psi\mu^+\nu_\mu)$}},
  \href{http://arxiv.org/abs/1711.05623}{{\tt 1711.05623}}.

\bibitem{Lees:2012xj}
{\scshape BaBar} collaboration, J.~P. Lees et~al., \emph{{Evidence for an
  excess of $\bar{B} \to D^{(*)} \tau^-\bar{\nu}_\tau$ decays}},
  \href{http://dx.doi.org/10.1103/PhysRevLett.109.101802}{\emph{Phys. Rev.
  Lett.} {\bf 109} (2012) 101802}, [\href{http://arxiv.org/abs/1205.5442}{{\tt
  1205.5442}}].

\bibitem{Lees:2013uzd}
{\scshape BaBar} collaboration, J.~P. Lees et~al., \emph{{Measurement of an
  Excess of $\bar{B} \to D^{(*)}\tau^- \bar{\nu}_\tau$ Decays and Implications
  for Charged Higgs Bosons}},
  \href{http://dx.doi.org/10.1103/PhysRevD.88.072012}{\emph{Phys. Rev.} {\bf
  D88} (2013) 072012}, [\href{http://arxiv.org/abs/1303.0571}{{\tt
  1303.0571}}].

\bibitem{Huschle:2015rga}
{\scshape Belle} collaboration, M.~Huschle et~al., \emph{{Measurement of the
  branching ratio of $\bar{B} \to D^{(\ast)} \tau^- \bar{\nu}_\tau$ relative to
  $\bar{B} \to D^{(\ast)} \ell^- \bar{\nu}_\ell$ decays with hadronic tagging
  at Belle}}, \href{http://dx.doi.org/10.1103/PhysRevD.92.072014}{\emph{Phys.
  Rev.} {\bf D92} (2015) 072014}, [\href{http://arxiv.org/abs/1507.03233}{{\tt
  1507.03233}}].

\bibitem{Aaij:2015yra}
{\scshape LHCb} collaboration, R.~Aaij et~al., \emph{{Measurement of the ratio
  of branching fractions $\mathcal{B}(\bar{B}^0 \to
  D^{*+}\tau^{-}\bar{\nu}_{\tau})/\mathcal{B}(\bar{B}^0 \to
  D^{*+}\mu^{-}\bar{\nu}_{\mu})$}},
  \href{http://dx.doi.org/10.1103/PhysRevLett.115.159901,
  10.1103/PhysRevLett.115.111803}{\emph{Phys. Rev. Lett.} {\bf 115} (2015)
  111803}, [\href{http://arxiv.org/abs/1506.08614}{{\tt 1506.08614}}].
  [Erratum: Phys. Rev. Lett.115,no.15,159901(2015)].

\bibitem{Sato:2016svk}
{\scshape Belle} collaboration, Y.~Sato et~al., \emph{{Measurement of the
  branching ratio of $\bar{B}^0 \rightarrow D^{*+} \tau^- \bar{\nu}_{\tau}$
  relative to $\bar{B}^0 \rightarrow D^{*+} \ell^- \bar{\nu}_{\ell}$ decays
  with a semileptonic tagging method}},
  \href{http://dx.doi.org/10.1103/PhysRevD.94.072007}{\emph{Phys. Rev.} {\bf
  D94} (2016) 072007}, [\href{http://arxiv.org/abs/1607.07923}{{\tt
  1607.07923}}].

\bibitem{fpcp}
A.~Rapha{\"e}l.
  \url{https://indico.cern.ch/event/586719/contributions/2531261/attachments/1470695/2275576/2_fpcp_talk_wormser.pdf}{,
  Talk given at FPCP Conference, June 5 2017, Prague}.

\bibitem{Wen-Fei:2013uea}
W.-F. Wang, Y.-Y. Fan and Z.-J. Xiao, \emph{{Semileptonic decays
  $B_c\to(\eta_c,J/\Psi)l\nu$ in the perturbative QCD approach}},
  \href{http://dx.doi.org/10.1088/1674-1137/37/9/093102}{\emph{Chin. Phys.}
  {\bf C37} (2013) 093102}, [\href{http://arxiv.org/abs/1212.5903}{{\tt
  1212.5903}}].

\bibitem{Barbieri:2015yvd}
R.~Barbieri, G.~Isidori, A.~Pattori and F.~Senia, \emph{{Anomalies in
  $B$-decays and $U(2)$ flavour symmetry}},
  \href{http://dx.doi.org/10.1140/epjc/s10052-016-3905-3}{\emph{Eur. Phys. J.}
  {\bf C76} (2016) 67}, [\href{http://arxiv.org/abs/1512.01560}{{\tt
  1512.01560}}].

\bibitem{Gripaios:2014tna}
B.~Gripaios, M.~Nardecchia and S.~A. Renner, \emph{{Composite leptoquarks and
  anomalies in $B$-meson decays}},
  \href{http://dx.doi.org/10.1007/JHEP05(2015)006}{\emph{JHEP} {\bf 05} (2015)
  006}, [\href{http://arxiv.org/abs/1412.1791}{{\tt 1412.1791}}].

\bibitem{Niehoff:2015iaa}
C.~Niehoff, P.~Stangl and D.~M. Straub, \emph{{Direct and indirect signals of
  natural composite Higgs models}},
  \href{http://dx.doi.org/10.1007/JHEP01(2016)119}{\emph{JHEP} {\bf 01} (2016)
  119}, [\href{http://arxiv.org/abs/1508.00569}{{\tt 1508.00569}}].

\bibitem{Barbieri:2016las}
R.~Barbieri, C.~W. Murphy and F.~Senia, \emph{{B-decay Anomalies in a Composite
  Leptoquark Model}},
  \href{http://dx.doi.org/10.1140/epjc/s10052-016-4578-7}{\emph{Eur. Phys. J.}
  {\bf C77} (2017) 8}, [\href{http://arxiv.org/abs/1611.04930}{{\tt
  1611.04930}}].

\bibitem{Niehoff:2016zso}
C.~Niehoff, P.~Stangl and D.~M. Straub, \emph{{Electroweak symmetry breaking
  and collider signatures in the next-to-minimal composite Higgs model}},
  \href{http://dx.doi.org/10.1007/JHEP04(2017)117}{\emph{JHEP} {\bf 04} (2017)
  117}, [\href{http://arxiv.org/abs/1611.09356}{{\tt 1611.09356}}].

\bibitem{Megias:2016bde}
E.~Megias, G.~Panico, O.~Pujolas and M.~Quiros, \emph{{A Natural origin for the
  LHCb anomalies}},
  \href{http://dx.doi.org/10.1007/JHEP09(2016)118}{\emph{JHEP} {\bf 09} (2016)
  118}, [\href{http://arxiv.org/abs/1608.02362}{{\tt 1608.02362}}].

\bibitem{Buttazzo:2016kid}
D.~Buttazzo, A.~Greljo, G.~Isidori and D.~Marzocca, \emph{{Toward a coherent
  solution of diphoton and flavor anomalies}},
  \href{http://dx.doi.org/10.1007/JHEP08(2016)035}{\emph{JHEP} {\bf 08} (2016)
  035}, [\href{http://arxiv.org/abs/1604.03940}{{\tt 1604.03940}}].

\bibitem{Barbieri:2017tuq}
R.~Barbieri and A.~Tesi, \emph{{$B$-decay anomalies in Pati-Salam SU(4)}},
  \href{http://arxiv.org/abs/1712.06844}{{\tt 1712.06844}}.

\bibitem{DAmbrosio:2017wis}
G.~D'Ambrosio and A.~M. Iyer, \emph{{Flavour issues in warped custodial models:
  $B$ anomalies and rare $K$ decays}},
  \href{http://arxiv.org/abs/1712.08122}{{\tt 1712.08122}}.

\bibitem{Sannino:2017utc}
F.~Sannino, P.~Stangl, D.~M. Straub and A.~E. Thomsen, \emph{{Flavor Physics
  and Flavor Anomalies in Minimal Fundamental Partial Compositeness}},
  \href{http://arxiv.org/abs/1712.07646}{{\tt 1712.07646}}.

\bibitem{Carmona:2017fsn}
A.~Carmona and F.~Goertz, \emph{{Recent $\boldsymbol{B}$ Physics Anomalies - a
  First Hint for Compositeness?}},  \href{http://arxiv.org/abs/1712.02536}{{\tt
  1712.02536}}.

\bibitem{Marzocca:2018wcf}
D.~Marzocca, \emph{{Addressing the B-physics anomalies in a fundamental
  Composite Higgs Model}},  \href{http://arxiv.org/abs/1803.10972}{{\tt
  1803.10972}}.

\bibitem{Asadi:2018wea}
P.~Asadi, M.~R. Buckley and D.~Shih, \emph{{It's all right(-handed neutrinos):
  a new $W'$ model for the $R_{D^{(*)}}$ anomaly}},
  \href{http://arxiv.org/abs/1804.04135}{{\tt 1804.04135}}.

\bibitem{Greljo:2018ogz}
A.~Greljo, D.~J. Robinson, B.~Shakya and J.~Zupan, \emph{{$R(D^{(*)})$ from
  $W'$ and right-handed neutrinos}},
  \href{http://arxiv.org/abs/1804.04642}{{\tt 1804.04642}}.

\bibitem{Datta:2012qk}
A.~Datta, M.~Duraisamy and D.~Ghosh, \emph{{Diagnosing New Physics in $b \to c
  \, \tau \, \nu_\tau$ decays in the light of the recent BaBar result}},
  \href{http://dx.doi.org/10.1103/PhysRevD.86.034027}{\emph{Phys. Rev.} {\bf
  D86} (2012) 034027}, [\href{http://arxiv.org/abs/1206.3760}{{\tt
  1206.3760}}].

\bibitem{Tanaka:2012nw}
M.~Tanaka and R.~Watanabe, \emph{{New physics in the weak interaction of $\bar
  B\to D^{(*)}\tau\bar\nu$}},
  \href{http://dx.doi.org/10.1103/PhysRevD.87.034028}{\emph{Phys. Rev.} {\bf
  D87} (2013) 034028}, [\href{http://arxiv.org/abs/1212.1878}{{\tt
  1212.1878}}].

\bibitem{Choudhury:2012hn}
D.~Choudhury, D.~K. Ghosh and A.~Kundu, \emph{{B decay anomalies in an
  effective theory}},
  \href{http://dx.doi.org/10.1103/PhysRevD.86.114037}{\emph{Phys. Rev.} {\bf
  D86} (2012) 114037}, [\href{http://arxiv.org/abs/1210.5076}{{\tt
  1210.5076}}].

\bibitem{Sakaki:2013bfa}
Y.~Sakaki, M.~Tanaka, A.~Tayduganov and R.~Watanabe, \emph{{Testing leptoquark
  models in $\bar B \to D^{(*)} \tau \bar\nu$}},
  \href{http://dx.doi.org/10.1103/PhysRevD.88.094012}{\emph{Phys. Rev.} {\bf
  D88} (2013) 094012}, [\href{http://arxiv.org/abs/1309.0301}{{\tt
  1309.0301}}].

\bibitem{Bhattacharya:2014wla}
B.~Bhattacharya, A.~Datta, D.~London and S.~Shivashankara, \emph{{Simultaneous
  Explanation of the $R_K$ and $R(D^{(*)})$ Puzzles}},
  \href{http://dx.doi.org/10.1016/j.physletb.2015.02.011}{\emph{Phys. Lett.}
  {\bf B742} (2015) 370--374}, [\href{http://arxiv.org/abs/1412.7164}{{\tt
  1412.7164}}].

\bibitem{Cata:2015lta}
O.~Cata and M.~Jung, \emph{{Signatures of a nonstandard Higgs boson from flavor
  physics}}, \href{http://dx.doi.org/10.1103/PhysRevD.92.055018}{\emph{Phys.
  Rev.} {\bf D92} (2015) 055018}, [\href{http://arxiv.org/abs/1505.05804}{{\tt
  1505.05804}}].

\bibitem{Freytsis:2015qca}
M.~Freytsis, Z.~Ligeti and J.~T. Ruderman, \emph{{Flavor models for $\bar{B}
  \to D^{(*)} \tau \bar{\nu}$}},
  \href{http://dx.doi.org/10.1103/PhysRevD.92.054018}{\emph{Phys. Rev.} {\bf
  D92} (2015) 054018}, [\href{http://arxiv.org/abs/1506.08896}{{\tt
  1506.08896}}].

\bibitem{Choudhury:2016ulr}
D.~Choudhury, A.~Kundu, S.~Nandi and S.~K. Patra, \emph{{Unified resolution of
  the $R(D)$ and $R(D^*)$ anomalies and the lepton flavor violating decay
  $h\to\mu\tau$}},
  \href{http://dx.doi.org/10.1103/PhysRevD.95.035021}{\emph{Phys. Rev.} {\bf
  D95} (2017) 035021}, [\href{http://arxiv.org/abs/1612.03517}{{\tt
  1612.03517}}].

\bibitem{Celis:2016azn}
A.~Celis, M.~Jung, X.-Q. Li and A.~Pich, \emph{{Scalar contributions to $b\to c
  (u) \tau \nu$ transitions}},
  \href{http://dx.doi.org/10.1016/j.physletb.2017.05.037}{\emph{Phys. Lett.}
  {\bf B771} (2017) 168--179}, [\href{http://arxiv.org/abs/1612.07757}{{\tt
  1612.07757}}].

\bibitem{Choudhury:2017qyt}
D.~Choudhury, A.~Kundu, R.~Mandal and R.~Sinha, \emph{{Minimal unified
  resolution to $R_{K^{(*)}}$ and $R(D^{(*)})$ anomalies with lepton mixing}},
  \href{http://dx.doi.org/10.1103/PhysRevLett.119.151801}{\emph{Phys. Rev.
  Lett.} {\bf 119} (2017) 151801}, [\href{http://arxiv.org/abs/1706.08437}{{\tt
  1706.08437}}].

\bibitem{Crivellin:2017zlb}
A.~Crivellin, D.~Muller and T.~Ota, \emph{{Simultaneous explanation of
  R(D$^{(?)}$) and b?s?$^{+}$ ?$^{?}$: the last scalar leptoquarks standing}},
  \href{http://dx.doi.org/10.1007/JHEP09(2017)040}{\emph{JHEP} {\bf 09} (2017)
  040}, [\href{http://arxiv.org/abs/1703.09226}{{\tt 1703.09226}}].

\bibitem{Colangelo:2018cnj}
P.~Colangelo and F.~De~Fazio, \emph{{Scrutinizing $\bar B \to D^*(D \pi) \ell^-
  \bar \nu_\ell$ and $\bar B \to D^*(D \gamma) \ell^- \bar \nu_\ell$ in search
  of new physics footprints}},  \href{http://arxiv.org/abs/1801.10468}{{\tt
  1801.10468}}.

\bibitem{Lattice:2015rga}
{\scshape MILC} collaboration, J.~A. Bailey et~al., \emph{{B?D?? form factors
  at nonzero recoil and |V$_{cb}$| from 2+1-flavor lattice QCD}},
  \href{http://dx.doi.org/10.1103/PhysRevD.92.034506}{\emph{Phys. Rev.} {\bf
  D92} (2015) 034506}, [\href{http://arxiv.org/abs/1503.07237}{{\tt
  1503.07237}}].

\bibitem{Melikhov:2000yu}
D.~Melikhov and B.~Stech, \emph{{Weak form-factors for heavy meson decays: An
  Update}}, \href{http://dx.doi.org/10.1103/PhysRevD.62.014006}{\emph{Phys.
  Rev.} {\bf D62} (2000) 014006},
  [\href{http://arxiv.org/abs/hep-ph/0001113}{{\tt hep-ph/0001113}}].

\bibitem{Atoui:2013mqa}
M.~Atoui, D.~Becirevic, V.~Morenas and F.~Sanfilippo, \emph{{Lattice QCD study
  of $B_s\to D_s \ell \bar\nu_\ell$ decay near zero recoil}},
  \href{http://dx.doi.org/10.22323/1.187.0384}{\emph{PoS} {\bf LATTICE2013}
  (2014) 384}, [\href{http://arxiv.org/abs/1311.5071}{{\tt 1311.5071}}].

\bibitem{Caprini:1997mu}
I.~Caprini, L.~Lellouch and M.~Neubert, \emph{{Dispersive bounds on the shape
  of anti-B ---> D(*) lepton anti-neutrino form-factors}},
  \href{http://dx.doi.org/10.1016/S0550-3213(98)00350-2}{\emph{Nucl. Phys.}
  {\bf B530} (1998) 153--181}, [\href{http://arxiv.org/abs/hep-ph/9712417}{{\tt
  hep-ph/9712417}}].

\bibitem{Amhis:2014hma}
{\scshape Heavy Flavor Averaging Group (HFAG)} collaboration, Y.~Amhis et~al.,
  \emph{{Averages of $b$-hadron, $c$-hadron, and $\tau$-lepton properties as of
  summer 2014}},  \href{http://arxiv.org/abs/1412.7515}{{\tt 1412.7515}}.

\bibitem{Bailey:2014tva}
{\scshape Fermilab Lattice, MILC} collaboration, J.~A. Bailey et~al.,
  \emph{{Update of $|V_{cb}|$ from the $\bar{B}\to D^*\ell\bar{\nu}$ form
  factor at zero recoil with three-flavor lattice QCD}},
  \href{http://dx.doi.org/10.1103/PhysRevD.89.114504}{\emph{Phys. Rev.} {\bf
  D89} (2014) 114504}, [\href{http://arxiv.org/abs/1403.0635}{{\tt
  1403.0635}}].

\bibitem{Bernlochner:2017jka}
F.~U. Bernlochner, Z.~Ligeti, M.~Papucci and D.~J. Robinson, \emph{{Combined
  analysis of semileptonic $B$ decays to $D$ and $D^*$: $R(D^{(*)})$,
  $|V_{cb}|$, and new physics}},
  \href{http://dx.doi.org/10.1103/PhysRevD.95.115008,
  10.1103/PhysRevD.97.059902}{\emph{Phys. Rev.} {\bf D95} (2017) 115008},
  [\href{http://arxiv.org/abs/1703.05330}{{\tt 1703.05330}}]. [Erratum: Phys.
  Rev.D97,no.5,059902(2018)].

\bibitem{Jung:2018lfu}
M.~Jung and D.~M. Straub, \emph{{Constraining new physics in $b\to c\ell\nu$
  transitions}},  \href{http://arxiv.org/abs/1801.01112}{{\tt 1801.01112}}.

\bibitem{Bardhan:2016uhr}
D.~Bardhan, P.~Byakti and D.~Ghosh, \emph{{A closer look at the R$_{D}$ and
  R$_{D^*}$ anomalies}},
  \href{http://dx.doi.org/10.1007/JHEP01(2017)125}{\emph{JHEP} {\bf 01} (2017)
  125}, [\href{http://arxiv.org/abs/1610.03038}{{\tt 1610.03038}}].

\bibitem{Alonso:2013hga}
R.~Alonso, E.~E. Jenkins, A.~V. Manohar and M.~Trott, \emph{{Renormalization
  Group Evolution of the Standard Model Dimension Six Operators III: Gauge
  Coupling Dependence and Phenomenology}},
  \href{http://dx.doi.org/10.1007/JHEP04(2014)159}{\emph{JHEP} {\bf 04} (2014)
  159}, [\href{http://arxiv.org/abs/1312.2014}{{\tt 1312.2014}}].

\bibitem{Alonso:2016oyd}
R.~Alonso, B.~Grinstein and J.~Martin~Camalich, \emph{{The lifetime of the
  $B_c^-$ meson and the anomalies in $B\to D^{(*)}\tau\nu$}},
  \href{http://dx.doi.org/10.1103/PhysRevLett.118.081802}{\emph{Phys. Rev.
  Lett.} {\bf 118} (2017) 081802}, [\href{http://arxiv.org/abs/1611.06676}{{\tt
  1611.06676}}].

\bibitem{Akeroyd:2017mhr}
A.~G. Akeroyd and C.-H. Chen, \emph{{Constraint on the branching ratio of $B_c
  \to \tau \bar{\nu}$ from LEP1 and consequences for $R(D^{(*)})$ anomaly}},
  \href{http://dx.doi.org/10.1103/PhysRevD.96.075011}{\emph{Phys. Rev.} {\bf
  D96} (2017) 075011}, [\href{http://arxiv.org/abs/1708.04072}{{\tt
  1708.04072}}].

\bibitem{Dorsner:2013tla}
I.~Dorsner, S.~Fajfer, N.~Kosnik and I.~Nisandzic, \emph{{Minimally flavored
  colored scalar in $\bar B \to D^{(*)} \tau \bar \nu$ and the mass matrices
  constraints}}, \href{http://dx.doi.org/10.1007/JHEP11(2013)084}{\emph{JHEP}
  {\bf 11} (2013) 084}, [\href{http://arxiv.org/abs/1306.6493}{{\tt
  1306.6493}}].

\bibitem{Becirevic:2018afm}
D.~Becirevic, I.~Dorsner, S.~Fajfer, N.~Kosnik, D.~A. Faroughy and
  O.~Sumensari, \emph{{Scalar leptoquarks from grand unified theories to
  accommodate the $B$-physics anomalies}},
  \href{http://dx.doi.org/10.1103/PhysRevD.98.055003}{\emph{Phys. Rev.} {\bf
  D98} (2018) 055003}, [\href{http://arxiv.org/abs/1806.05689}{{\tt
  1806.05689}}].

\bibitem{Nierste:2008qe}
U.~Nierste, S.~Trine and S.~Westhoff, \emph{{Charged-Higgs effects in a new B
  ---> D tau nu differential decay distribution}},
  \href{http://dx.doi.org/10.1103/PhysRevD.78.015006}{\emph{Phys. Rev.} {\bf
  D78} (2008) 015006}, [\href{http://arxiv.org/abs/0801.4938}{{\tt
  0801.4938}}].

\bibitem{Duraisamy:2013kcw}
M.~Duraisamy and A.~Datta, \emph{{The Full $B \to D^{*} \tau^{-}
  \bar{\nu_\tau}$ Angular Distribution and CP violating Triple Products}},
  \href{http://dx.doi.org/10.1007/JHEP09(2013)059}{\emph{JHEP} {\bf 09} (2013)
  059}, [\href{http://arxiv.org/abs/1302.7031}{{\tt 1302.7031}}].

\bibitem{Ligeti:2016npd}
Z.~Ligeti, M.~Papucci and D.~J. Robinson, \emph{{New Physics in the Visible
  Final States of $B\to D^{(*)}\tau\nu$}},
  \href{http://dx.doi.org/10.1007/JHEP01(2017)083}{\emph{JHEP} {\bf 01} (2017)
  083}, [\href{http://arxiv.org/abs/1610.02045}{{\tt 1610.02045}}].

\bibitem{Alonso:2016gym}
R.~Alonso, A.~Kobach and J.~Martin~Camalich, \emph{{New physics in the
  kinematic distributions of $\bar B\to
  D^{(*)}\tau^-(\to\ell^-\bar\nu_\ell\nu_\tau)\bar\nu_\tau$}},
  \href{http://dx.doi.org/10.1103/PhysRevD.94.094021}{\emph{Phys. Rev.} {\bf
  D94} (2016) 094021}, [\href{http://arxiv.org/abs/1602.07671}{{\tt
  1602.07671}}].

\bibitem{Alok:2016qyh}
A.~K. Alok, D.~Kumar, S.~Kumbhakar and S.~U. Sankar, \emph{{$D^{*}$
  polarization as a probe to discriminate new physics in $\bar{B}\to D^{*} \tau
  \bar{\nu}$}}, \href{http://dx.doi.org/10.1103/PhysRevD.95.115038}{\emph{Phys.
  Rev.} {\bf D95} (2017) 115038}, [\href{http://arxiv.org/abs/1606.03164}{{\tt
  1606.03164}}].

\bibitem{Grzadkowski:2010es}
B.~Grzadkowski, M.~Iskrzynski, M.~Misiak and J.~Rosiek, \emph{{Dimension-Six
  Terms in the Standard Model Lagrangian}},
  \href{http://dx.doi.org/10.1007/JHEP10(2010)085}{\emph{JHEP} {\bf 10} (2010)
  085}, [\href{http://arxiv.org/abs/1008.4884}{{\tt 1008.4884}}].

\bibitem{ALEPH:2005aa}
{\scshape DELPHI, OPAL, ALEPH, LEP Electroweak Working Group, L3}
  collaboration, \emph{{A Combination of preliminary electroweak measurements
  and constraints on the standard model}},
  \href{http://arxiv.org/abs/hep-ex/0511027}{{\tt hep-ex/0511027}}.

\bibitem{ALEPH:2005ab}
{\scshape SLD Electroweak Group, DELPHI, ALEPH, SLD, SLD Heavy Flavour Group,
  OPAL, LEP Electroweak Working Group, L3} collaboration, S.~Schael et~al.,
  \emph{{Precision electroweak measurements on the $Z$ resonance}},
  \href{http://dx.doi.org/10.1016/j.physrep.2005.12.006}{\emph{Phys. Rept.}
  {\bf 427} (2006) 257--454}, [\href{http://arxiv.org/abs/hep-ex/0509008}{{\tt
  hep-ex/0509008}}].

\bibitem{Pich:2013lsa}
A.~Pich, \emph{{Precision Tau Physics}},
  \href{http://dx.doi.org/10.1016/j.ppnp.2013.11.002}{\emph{Prog. Part. Nucl.
  Phys.} {\bf 75} (2014) 41--85}, [\href{http://arxiv.org/abs/1310.7922}{{\tt
  1310.7922}}].

\bibitem{Grygier:2017tzo}
{\scshape Belle} collaboration, J.~Grygier et~al., \emph{{Search for
  $\boldsymbol{B\to h\nu\bar{\nu}}$ decays with semileptonic tagging at
  Belle}}, \href{http://dx.doi.org/10.1103/PhysRevD.96.091101}{\emph{Phys.
  Rev.} {\bf D96} (2017) 091101}, [\href{http://arxiv.org/abs/1702.03224}{{\tt
  1702.03224}}].

\bibitem{Buras:2014fpa}
A.~J. Buras, J.~Girrbach-Noe, C.~Niehoff and D.~M. Straub, \emph{{$ B\to
  {K}^{\left(\ast \right)}\nu \overline{\nu} $ decays in the Standard Model and
  beyond}}, \href{http://dx.doi.org/10.1007/JHEP02(2015)184}{\emph{JHEP} {\bf
  02} (2015) 184}, [\href{http://arxiv.org/abs/1409.4557}{{\tt 1409.4557}}].

\bibitem{Hambrock:2015wka}
C.~Hambrock, A.~Khodjamirian and A.~Rusov, \emph{{Hadronic effects and
  observables in $B\to \pi\ell^{+}\ell^{-}$ decay at large recoil}},
  \href{http://dx.doi.org/10.1103/PhysRevD.92.074020}{\emph{Phys. Rev.} {\bf
  D92} (2015) 074020}, [\href{http://arxiv.org/abs/1506.07760}{{\tt
  1506.07760}}].

\bibitem{Manoni:2017lxj}
{\scshape Belle II} collaboration, E.~Manoni, \emph{{Studies of missing energy
  decays of B meson at Belle II}},
  \href{http://dx.doi.org/10.22323/1.314.0226}{\emph{PoS} {\bf EPS-HEP2017}
  (2017) 226}.

\bibitem{Capdevila:2017iqn}
B.~Capdevila, A.~Crivellin, S.~Descotes-Genon, L.~Hofer and J.~Matias,
  \emph{{Searching for New Physics with $b\to s\tau^+\tau^-$ processes}},
  \href{http://arxiv.org/abs/1712.01919}{{\tt 1712.01919}}.

\bibitem{Faroughy:2016osc}
D.~A. Faroughy, A.~Greljo and J.~F. Kamenik, \emph{{Confronting lepton flavor
  universality violation in B decays with high-$p_T$ tau lepton searches at
  LHC}}, \href{http://dx.doi.org/10.1016/j.physletb.2016.11.011}{\emph{Phys.
  Lett.} {\bf B764} (2017) 126--134},
  [\href{http://arxiv.org/abs/1609.07138}{{\tt 1609.07138}}].

\bibitem{Feruglio:2016gvd}
F.~Feruglio, P.~Paradisi and A.~Pattori, \emph{{Revisiting Lepton Flavor
  Universality in B Decays}},
  \href{http://dx.doi.org/10.1103/PhysRevLett.118.011801}{\emph{Phys. Rev.
  Lett.} {\bf 118} (2017) 011801}, [\href{http://arxiv.org/abs/1606.00524}{{\tt
  1606.00524}}].

\bibitem{Buttazzo:2017ixm}
D.~Buttazzo, A.~Greljo, G.~Isidori and D.~Marzocca, \emph{{B-physics anomalies:
  a guide to combined explanations}},
  \href{http://dx.doi.org/10.1007/JHEP11(2017)044}{\emph{JHEP} {\bf 11} (2017)
  044}, [\href{http://arxiv.org/abs/1706.07808}{{\tt 1706.07808}}].

\bibitem{Feruglio:2017rjo}
F.~Feruglio, P.~Paradisi and A.~Pattori, \emph{{On the Importance of
  Electroweak Corrections for B Anomalies}},
  \href{http://dx.doi.org/10.1007/JHEP09(2017)061}{\emph{JHEP} {\bf 09} (2017)
  061}, [\href{http://arxiv.org/abs/1705.00929}{{\tt 1705.00929}}].

\bibitem{Bobeth:2011st}
C.~Bobeth and U.~Haisch, \emph{{New Physics in $\Gamma_{12}^s$: ($\bar{s}
  b$)$(\bar{\tau} \tau)$ Operators}},
  \href{http://dx.doi.org/10.5506/APhysPolB.44.127}{\emph{Acta Phys. Polon.}
  {\bf B44} (2013) 127--176}, [\href{http://arxiv.org/abs/1109.1826}{{\tt
  1109.1826}}].

\bibitem{Dighe:2012df}
A.~Dighe and D.~Ghosh, \emph{{How large can the branching ratio of $B_s \to
  \tau^+ \tau^-$ be ?}},
  \href{http://dx.doi.org/10.1103/PhysRevD.86.054023}{\emph{Phys. Rev.} {\bf
  D86} (2012) 054023}, [\href{http://arxiv.org/abs/1207.1324}{{\tt
  1207.1324}}].

\bibitem{Kaplan:1983fs}
D.~B. Kaplan and H.~Georgi, \emph{{SU(2) x U(1) Breaking by Vacuum
  Misalignment}},
  \href{http://dx.doi.org/10.1016/0370-2693(84)91177-8}{\emph{Phys. Lett.} {\bf
  136B} (1984) 183--186}.

\bibitem{Kaplan:1991dc}
D.~B. Kaplan, \emph{{Flavor at SSC energies: A New mechanism for dynamically
  generated fermion masses}},
  \href{http://dx.doi.org/10.1016/S0550-3213(05)80021-5}{\emph{Nucl. Phys.}
  {\bf B365} (1991) 259--278}.

\bibitem{Contino:2003ve}
R.~Contino, Y.~Nomura and A.~Pomarol, \emph{{Higgs as a holographic
  pseudoGoldstone boson}},
  \href{http://dx.doi.org/10.1016/j.nuclphysb.2003.08.027}{\emph{Nucl. Phys.}
  {\bf B671} (2003) 148--174}, [\href{http://arxiv.org/abs/hep-ph/0306259}{{\tt
  hep-ph/0306259}}].

\bibitem{Contino:2010rs}
R.~Contino, \emph{{The Higgs as a Composite Nambu-Goldstone Boson}},  in
  \emph{{Physics of the large and the small, TASI 09, proceedings of the
  Theoretical Advanced Study Institute in Elementary Particle Physics, Boulder,
  Colorado, USA, 1-26 June 2009}}, pp.~235--306, 2011.
\newblock \href{http://arxiv.org/abs/1005.4269}{{\tt 1005.4269}}.
\newblock \href{http://dx.doi.org/10.1142/9789814327183_0005}{DOI}.

\bibitem{Panico:2015jxa}
G.~Panico and A.~Wulzer, \emph{{The Composite Nambu-Goldstone Higgs}},
  \href{http://dx.doi.org/10.1007/978-3-319-22617-0}{\emph{Lect. Notes Phys.}
  {\bf 913} (2016) pp.1--316}, [\href{http://arxiv.org/abs/1506.01961}{{\tt
  1506.01961}}].

\bibitem{Contino:2006nn}
R.~Contino, T.~Kramer, M.~Son and R.~Sundrum, \emph{{Warped/composite
  phenomenology simplified}},
  \href{http://dx.doi.org/10.1088/1126-6708/2007/05/074}{\emph{JHEP} {\bf 05}
  (2007) 074}, [\href{http://arxiv.org/abs/hep-ph/0612180}{{\tt
  hep-ph/0612180}}].

\bibitem{Coleman:1969sm}
S.~R. Coleman, J.~Wess and B.~Zumino, \emph{{Structure of phenomenological
  Lagrangians. 1.}},
  \href{http://dx.doi.org/10.1103/PhysRev.177.2239}{\emph{Phys. Rev.} {\bf 177}
  (1969) 2239--2247}.

\bibitem{Callan:1969sn}
C.~G. Callan, Jr., S.~R. Coleman, J.~Wess and B.~Zumino, \emph{{Structure of
  phenomenological Lagrangians. 2.}},
  \href{http://dx.doi.org/10.1103/PhysRev.177.2247}{\emph{Phys. Rev.} {\bf 177}
  (1969) 2247--2250}.

\bibitem{Contino:2011np}
R.~Contino, D.~Marzocca, D.~Pappadopulo and R.~Rattazzi, \emph{{On the effect
  of resonances in composite Higgs phenomenology}},
  \href{http://dx.doi.org/10.1007/JHEP10(2011)081}{\emph{JHEP} {\bf 10} (2011)
  081}, [\href{http://arxiv.org/abs/1109.1570}{{\tt 1109.1570}}].

\bibitem{Ecker:1989yg}
G.~Ecker, J.~Gasser, H.~Leutwyler, A.~Pich and E.~de~Rafael, \emph{{Chiral
  Lagrangians for Massive Spin 1 Fields}},
  \href{http://dx.doi.org/10.1016/0370-2693(89)91627-4}{\emph{Phys. Lett.} {\bf
  B223} (1989) 425--432}.

\bibitem{Bona:2007vi}
{\scshape UTfit} collaboration, M.~Bona et~al., \emph{{Model-independent
  constraints on $\Delta F=2$ operators and the scale of new physics}},
  \href{http://dx.doi.org/10.1088/1126-6708/2008/03/049}{\emph{JHEP} {\bf 03}
  (2008) 049}, [\href{http://arxiv.org/abs/0707.0636}{{\tt 0707.0636}}].

\bibitem{Carrasco:2013zta}
{\scshape ETM} collaboration, N.~Carrasco et~al., \emph{{B-physics from $N_f$ =
  2 tmQCD: the Standard Model and beyond}},
  \href{http://dx.doi.org/10.1007/JHEP03(2014)016}{\emph{JHEP} {\bf 03} (2014)
  016}, [\href{http://arxiv.org/abs/1308.1851}{{\tt 1308.1851}}].

\bibitem{Csaki:2008zd}
C.~Csaki, A.~Falkowski and A.~Weiler, \emph{{The Flavor of the Composite
  Pseudo-Goldstone Higgs}},
  \href{http://dx.doi.org/10.1088/1126-6708/2008/09/008}{\emph{JHEP} {\bf 09}
  (2008) 008}, [\href{http://arxiv.org/abs/0804.1954}{{\tt 0804.1954}}].

\bibitem{Agashe:2008uz}
K.~Agashe, A.~Azatov and L.~Zhu, \emph{{Flavor Violation Tests of
  Warped/Composite SM in the Two-Site Approach}},
  \href{http://dx.doi.org/10.1103/PhysRevD.79.056006}{\emph{Phys. Rev.} {\bf
  D79} (2009) 056006}, [\href{http://arxiv.org/abs/0810.1016}{{\tt
  0810.1016}}].

\bibitem{Barbieri:2011ci}
R.~Barbieri, G.~Isidori, J.~Jones-Perez, P.~Lodone and D.~M. Straub,
  \emph{{$U(2)$ and Minimal Flavour Violation in Supersymmetry}},
  \href{http://dx.doi.org/10.1140/epjc/s10052-011-1725-z}{\emph{Eur. Phys. J.}
  {\bf C71} (2011) 1725}, [\href{http://arxiv.org/abs/1105.2296}{{\tt
  1105.2296}}].

\bibitem{Barbieri:2012tu}
R.~Barbieri, D.~Buttazzo, F.~Sala, D.~M. Straub and A.~Tesi, \emph{{A 125 GeV
  composite Higgs boson versus flavour and electroweak precision tests}},
  \href{http://dx.doi.org/10.1007/JHEP05(2013)069}{\emph{JHEP} {\bf 05} (2013)
  069}, [\href{http://arxiv.org/abs/1211.5085}{{\tt 1211.5085}}].

\bibitem{Sirunyan:2018omb}
{\scshape CMS} collaboration, A.~M. Sirunyan et~al., \emph{{Search for
  vector-like T and B quark pairs in final states with leptons at $\sqrt{s} =$
  13 TeV}}, \href{http://dx.doi.org/10.1007/JHEP08(2018)177}{\emph{JHEP} {\bf
  08} (2018) 177}, [\href{http://arxiv.org/abs/1805.04758}{{\tt 1805.04758}}].

\bibitem{Aaboud:2018pii}
{\scshape ATLAS} collaboration, M.~Aaboud et~al., \emph{{Combination of the
  searches for pair-produced vector-like partners of the third-generation
  quarks at $\sqrt{s} =$ 13 TeV with the ATLAS detector}},
  \href{http://arxiv.org/abs/1808.02343}{{\tt 1808.02343}}.

\bibitem{Ciuchini:2013pca}
M.~Ciuchini, E.~Franco, S.~Mishima and L.~Silvestrini, \emph{{Electroweak
  Precision Observables, New Physics and the Nature of a 126 GeV Higgs Boson}},
  \href{http://dx.doi.org/10.1007/JHEP08(2013)106}{\emph{JHEP} {\bf 08} (2013)
  106}, [\href{http://arxiv.org/abs/1306.4644}{{\tt 1306.4644}}].

\bibitem{Baak:2014ora}
{\scshape Gfitter Group} collaboration, M.~Baak, J.~Cuth, J.~Haller,
  A.~Hoecker, R.~Kogler, K.~Monig et~al., \emph{{The global electroweak fit at
  NNLO and prospects for the LHC and ILC}},
  \href{http://dx.doi.org/10.1140/epjc/s10052-014-3046-5}{\emph{Eur. Phys. J.}
  {\bf C74} (2014) 3046}, [\href{http://arxiv.org/abs/1407.3792}{{\tt
  1407.3792}}].

\bibitem{deBlas:2016ojx}
J.~de~Blas, M.~Ciuchini, E.~Franco, S.~Mishima, M.~Pierini, L.~Reina et~al.,
  \emph{{Electroweak precision observables and Higgs-boson signal strengths in
  the Standard Model and beyond: present and future}},
  \href{http://dx.doi.org/10.1007/JHEP12(2016)135}{\emph{JHEP} {\bf 12} (2016)
  135}, [\href{http://arxiv.org/abs/1608.01509}{{\tt 1608.01509}}].

\bibitem{Azatov:2013ura}
A.~Azatov, R.~Contino, A.~Di~Iura and J.~Galloway, \emph{{New Prospects for
  Higgs Compositeness in $h \to Z\gamma$}},
  \href{http://dx.doi.org/10.1103/PhysRevD.88.075019}{\emph{Phys. Rev.} {\bf
  D88} (2013) 075019}, [\href{http://arxiv.org/abs/1308.2676}{{\tt
  1308.2676}}].

\bibitem{Grojean:2013qca}
C.~Grojean, O.~Matsedonskyi and G.~Panico, \emph{{Light top partners and
  precision physics}},
  \href{http://dx.doi.org/10.1007/JHEP10(2013)160}{\emph{JHEP} {\bf 10} (2013)
  160}, [\href{http://arxiv.org/abs/1306.4655}{{\tt 1306.4655}}].

\bibitem{Ghosh:2015wiz}
D.~Ghosh, M.~Salvarezza and F.~Senia, \emph{{Extending the Analysis of
  Electroweak Precision Constraints in Composite Higgs Models}},
  \href{http://dx.doi.org/10.1016/j.nuclphysb.2016.11.013}{\emph{Nucl. Phys.}
  {\bf B914} (2017) 346--387}, [\href{http://arxiv.org/abs/1511.08235}{{\tt
  1511.08235}}].

\bibitem{Redi:2011zi}
M.~Redi and A.~Weiler, \emph{{Flavor and CP Invariant Composite Higgs Models}},
  \href{http://dx.doi.org/10.1007/JHEP11(2011)108}{\emph{JHEP} {\bf 11} (2011)
  108}, [\href{http://arxiv.org/abs/1106.6357}{{\tt 1106.6357}}].

\bibitem{Chala:2018igk}
M.~Chala and M.~Spannowsky, \emph{{On the behaviour of composite resonances
  breaking lepton flavour universality}},
  \href{http://arxiv.org/abs/1803.02364}{{\tt 1803.02364}}.

\bibitem{DAmico:2017mtc}
G.~D'Amico, M.~Nardecchia, P.~Panci, F.~Sannino, A.~Strumia, R.~Torre et~al.,
  \emph{{Flavour anomalies after the $R_{K^*}$ measurement}},
  \href{http://dx.doi.org/10.1007/JHEP09(2017)010}{\emph{JHEP} {\bf 09} (2017)
  010}, [\href{http://arxiv.org/abs/1704.05438}{{\tt 1704.05438}}].

\bibitem{Capdevila:2017bsm}
B.~Capdevila, A.~Crivellin, S.~Descotes-Genon, J.~Matias and J.~Virto,
  \emph{{Patterns of New Physics in $b\to s\ell^+\ell^-$ transitions in the
  light of recent data}},
  \href{http://dx.doi.org/10.1007/JHEP01(2018)093}{\emph{JHEP} {\bf 01} (2018)
  093}, [\href{http://arxiv.org/abs/1704.05340}{{\tt 1704.05340}}].

\bibitem{Ciuchini:2017mik}
M.~Ciuchini, A.~M. Coutinho, M.~Fedele, E.~Franco, A.~Paul, L.~Silvestrini
  et~al., \emph{{On Flavourful Easter eggs for New Physics hunger and Lepton
  Flavour Universality violation}},
  \href{http://dx.doi.org/10.1140/epjc/s10052-017-5270-2}{\emph{Eur. Phys. J.}
  {\bf C77} (2017) 688}, [\href{http://arxiv.org/abs/1704.05447}{{\tt
  1704.05447}}].

\bibitem{Geng:2017svp}
L.-S. Geng, B.~Grinstein, S.~Jager, J.~Martin~Camalich, X.-L. Ren and R.-X.
  Shi, \emph{{Towards the discovery of new physics with lepton-universality
  ratios of $b\to s\ell\ell$ decays}},
  \href{http://dx.doi.org/10.1103/PhysRevD.96.093006}{\emph{Phys. Rev.} {\bf
  D96} (2017) 093006}, [\href{http://arxiv.org/abs/1704.05446}{{\tt
  1704.05446}}].

\bibitem{Altmannshofer:2017yso}
W.~Altmannshofer, P.~Stangl and D.~M. Straub, \emph{{Interpreting Hints for
  Lepton Flavor Universality Violation}},
  \href{http://dx.doi.org/10.1103/PhysRevD.96.055008}{\emph{Phys. Rev.} {\bf
  D96} (2017) 055008}, [\href{http://arxiv.org/abs/1704.05435}{{\tt
  1704.05435}}].

\bibitem{Ghosh:2017ber}
D.~Ghosh, \emph{{Explaining the $R_K$ and $R_{K^*}$ anomalies}},
  \href{http://dx.doi.org/10.1140/epjc/s10052-017-5282-y}{\emph{Eur. Phys. J.}
  {\bf C77} (2017) 694}, [\href{http://arxiv.org/abs/1704.06240}{{\tt
  1704.06240}}].

\bibitem{Bardhan:2017xcc}
D.~Bardhan, P.~Byakti and D.~Ghosh, \emph{{Role of Tensor operators in $R_K$
  and $R_{K^*}$}},
  \href{http://dx.doi.org/10.1016/j.physletb.2017.08.062}{\emph{Phys. Lett.}
  {\bf B773} (2017) 505--512}, [\href{http://arxiv.org/abs/1705.09305}{{\tt
  1705.09305}}].

\bibitem{Patrignani:2016xqp}
{\scshape Particle Data Group} collaboration, C.~Patrignani et~al.,
  \emph{{Review of Particle Physics}},
  \href{http://dx.doi.org/10.1088/1674-1137/40/10/100001}{\emph{Chin. Phys.}
  {\bf C40} (2016) 100001}.

\bibitem{Colquhoun:2015oha}
{\scshape HPQCD} collaboration, B.~Colquhoun, C.~T.~H. Davies, R.~J. Dowdall,
  J.~Kettle, J.~Koponen, G.~P. Lepage et~al., \emph{{B-meson decay constants: a
  more complete picture from full lattice QCD}},
  \href{http://dx.doi.org/10.1103/PhysRevD.91.114509}{\emph{Phys. Rev.} {\bf
  D91} (2015) 114509}, [\href{http://arxiv.org/abs/1503.05762}{{\tt
  1503.05762}}].

\bibitem{Colquhoun:2016osw}
{\scshape HPQCD} collaboration, B.~Colquhoun, C.~Davies, J.~Koponen, A.~Lytle
  and C.~McNeile, \emph{{$B_c$ decays from highly improved staggered quarks and
  NRQCD}}, {\emph{PoS} {\bf LATTICE2016} (2016) 281},
  [\href{http://arxiv.org/abs/1611.01987}{{\tt 1611.01987}}].

\bibitem{Watanabe:2017mip}
R.~Watanabe, \emph{{New Physics effect on $B_c \to J/\psi \tau\bar\nu$ in
  relation to the $R_{D^{(*)}}$ anomaly}},
  \href{http://arxiv.org/abs/1709.08644}{{\tt 1709.08644}}.

\bibitem{Bordone:2017bld}
M.~Bordone, C.~Cornella, J.~Fuentes-Martin and G.~Isidori, \emph{{A three-site
  gauge model for flavor hierarchies and flavor anomalies}},
  \href{http://dx.doi.org/10.1016/j.physletb.2018.02.011}{\emph{Phys. Lett.}
  {\bf B779} (2018) 317--323}, [\href{http://arxiv.org/abs/1712.01368}{{\tt
  1712.01368}}].

\bibitem{Jenkins:2013wua}
E.~E. Jenkins, A.~V. Manohar and M.~Trott, \emph{{Renormalization Group
  Evolution of the Standard Model Dimension Six Operators II: Yukawa
  Dependence}}, \href{http://dx.doi.org/10.1007/JHEP01(2014)035}{\emph{JHEP}
  {\bf 01} (2014) 035}, [\href{http://arxiv.org/abs/1310.4838}{{\tt
  1310.4838}}].

\bibitem{Agashe:2006at}
K.~Agashe, R.~Contino, L.~Da~Rold and A.~Pomarol, \emph{{A Custodial symmetry
  for $Zb \bar b$}},
  \href{http://dx.doi.org/10.1016/j.physletb.2006.08.005}{\emph{Phys. Lett.}
  {\bf B641} (2006) 62--66}, [\href{http://arxiv.org/abs/hep-ph/0605341}{{\tt
  hep-ph/0605341}}].

\bibitem{Azatov:2016xik}
A.~Azatov, C.~Grojean, A.~Paul and E.~Salvioni, \emph{{Resolving gluon fusion
  loops at current and future hadron colliders}},
  \href{http://dx.doi.org/10.1007/JHEP09(2016)123}{\emph{JHEP} {\bf 09} (2016)
  123}, [\href{http://arxiv.org/abs/1608.00977}{{\tt 1608.00977}}].

\end{thebibliography}

\providecommand{\href}[2]{#2}\begingroup\raggedright\endgroup

\end{document}